\documentclass[12pt]{iopart}
\usepackage{graphicx}
 \usepackage{lscape}
\usepackage{url}            
\usepackage{txfonts}
\usepackage{natbib}
\bibpunct{(}{)}{;}{a}{}{,}  
\usepackage{color}
\usepackage[percent]{overpic} 

\def\OII{[O\,\textsc{ii}]}

\def\microns{$\textrm{$\mu$m}$}                             

\def\sun{\hbox{$_\odot$}}                                    
\def\arcmin{\hbox{$^\prime$}}                               
\def\arcsec{\hbox{$^{\prime\prime}$}}                       
\def\degr{\hbox{$^\circ$}}                                  

\def\la{\mathrel{\mathchoice {\vcenter{\offinterlineskip\halign{\hfil
$\displaystyle##$\hfil\cr<\cr\sim\cr}}}
{\vcenter{\offinterlineskip\halign{\hfil$\textstyle##$\hfil\cr
<\cr\sim\cr}}}
{\vcenter{\offinterlineskip\halign{\hfil$\scriptstyle##$\hfil\cr
<\cr\sim\cr}}}
{\vcenter{\offinterlineskip\halign{\hfil$\scriptscriptstyle##$\hfil\cr
<\cr\sim\cr}}}}}                                            
\def\ga{\mathrel{\mathchoice {\vcenter{\offinterlineskip\halign{\hfil
$\displaystyle##$\hfil\cr>\cr\sim\cr}}}
{\vcenter{\offinterlineskip\halign{\hfil$\textstyle##$\hfil\cr
>\cr\sim\cr}}}
{\vcenter{\offinterlineskip\halign{\hfil$\scriptstyle##$\hfil\cr
>\cr\sim\cr}}}
{\vcenter{\offinterlineskip\halign{\hfil$\scriptscriptstyle##$\hfil\cr
>\cr\sim\cr}}}}}                                            

\def\zg1{$z\!>\!1$}                                         
\def\zga1{$z\!\ga\!1$}                                      
\def\zsim1{$z\!\sim\!1$}                                    
\def\reds{red-sequence }                               
\def\h70{$h_{70}$}                                          
\def\hinv70{$h^{-1}_{70}$}                                  

\newcommand\newblock{\hskip .11em\@plus.33em\@minus.07em} 

\begin{document}

\title[The X-ray luminous galaxy cluster population at $0.9\!<\!z\!\le\!1.6$ as revealed by the XDCP]{The X-ray luminous galaxy cluster population at $\bf 0.9\!<\!z\!\la\!1.6$ as revealed by the XMM-{\it \bf Newton} Distant Cluster Project\footnote{Based on observations under program IDs 079.A-0634 and 085.A-0647
 collected at the European Organisation for Astronomical Research in the Southern Hemisphere, Chile, and observations collected at the Centro Astron\'omico Hispano Alem\'an (CAHA) at Calar Alto, operated jointly by the Max-Planck Institut f\"ur Astronomie and the Instituto de Astrof\'isica de Andaluc\'ia (CSIC).}}

\author{R Fassbender$^1$, H B\"ohringer$^1$, A Nastasi$^1$, R Suhada$^1$, M M\"uhlegger$^1$,  A de Hoon$^2$, J Kohnert$^2$, G Lamer$^2$, J J Mohr$^{1,3,4}$,  D Pierini\footnote{Visiting astronomer at MPE.}, G W Pratt$^5$,  H Quintana$^6$, P Rosati$^7$, J S Santos$^8$, A D Schwope$^2$}

\address{$^1$ Max-Planck-Institut f\"ur extraterrestrische Physik (MPE),
              Giessenbachstrasse~1, 85748 Garching, Germany}
\address{$^2$  Leibniz-Institut f\"ur Astrophysik Potsdam (AIP), An der Sternwarte~16, 14482 Potsdam, Germany}
\address{$^3$  Department of Physics, Ludwigs-Maximilians Universit\"at M\"unchen, Scheinerstr. 1, 81679  Munich, Germany}
\address{$^4$  Excellence Cluster Universe, Boltzmannstr.~2, 85748 Garching, Germany }
\address{$^5$  CEA \/ Saclay, Service d'Astrophysique, L'Orme des Merisiers, B\^at. 709, 91191 Gif-sur-Yvette Cedex, France}
\address{$^6$  Departamento de Astronom\'ia y Astrof\'isica, Pontificia Universidad Cat\'olica de Chile, Casilla 306, Santiago 22, Chile}
\address{$^7$  European Southern Observatory (ESO), Karl-Scharzschild-Str.~2, 85748 Garching, Germany}
\address{$^8$  European Space Astronomy Centre (ESAC), 7828691 Villanueva de la Canada, Madrid, Spain}
\eads{\mailto{rfassben@mpe.mpg.de}}


\begin{abstract}
We present the largest sample of spectroscopically confirmed X-ray luminous high-redshift galaxy clusters to date comprising   22 systems in the 
range  $0.9\!<\!z\!\la\!1.6$ as part of the XMM-{\it Newton} Distant Cluster Project (XDCP). All systems were initially selected as extended X-ray sources over 76.1\,deg$^2$ of non-contiguous deep archival XMM-{\it Newton} 
coverage, of which 49.4\,deg$^2$ are part of the core survey with a quantifiable selection function and 17.7\,deg$^2$ are classified as `gold' coverage as starting point for upcoming cosmological applications. 
Distant cluster candidates were followed-up with moderately deep optical and near-infrared imaging 
 in at least two bands to photometrically identify the cluster galaxy populations and obtain redshift estimates based on 
  colors of simple stellar population models.
We test and calibrate the most promising redshift estimation  techniques based on the R$-$z and z$-$H colors  for efficient distant cluster identifications and find a good 
redshift accuracy performance of the z$-$H color out to at least $z$$\sim$$1.5$, while the redshift evolution of the  R$-$z color leads to increasingly large 
uncertainties at $z\!\ga\!0.9$. Photometrically identified high-$z$ systems are spectroscopically confirmed with VLT/FORS\,2 with a  minimum of three concordant cluster member redshifts.
We present first details of two newly identified clusters, XDCP\,J0338.5+0029 at $z$=0.916 and  XDCP\,J0027.2+1714 at $z$=0.959, and 
investigate the X-ray properties of SpARCS\,J003550-431224 at $z$=1.335, which shows evidence for ongoing major merger activity along the line-of-sight. 
We provide X-ray properties and luminosity-based total mass estimates for the full sample of 22 high-$z$ clusters, of which 17 are at  $z\!\ge\!1.0$ and 7 populate the highest redshift bin at $z\!>\!1.3$.  
The median system mass of the sample is M$_{200}\!\simeq\!2 \times 10^{14}$\,M$_{\sun}$, while the probed mass range for the distant clusters spans approximately (0.7-7)$\times$$10^{14}$\,M$_{\sun}$. The majority ($>$70\%) of the X-ray selected clusters 
show rather regular 
X-ray morphologies, albeit in most cases with a discernible elongation along one axis. In contrast to local clusters, the $z\!>\!0.9$ systems do mostly  not harbor  
central dominant galaxies coincident with the X-ray centroid position,  but rather exhibit significant BCG offsets from the X-ray center with a median value of about 50\,kpc in projection and a smaller median luminosity gap to the second-ranked galaxy of  $\Delta m_{12}\!\simeq\!0.3$\,mag.  
We estimate a fraction of cluster-associated NVSS 1.4\,GHz radio sources of about 30\%, preferentially located within 1\arcmin \ from the X-ray center. 
This value suggests an increase of the fraction of very luminous cluster-associated radio sources by about a factor of 2.5-5 relative to low-$z$ systems. 
The galaxy populations in $z\!\ga\!1.5$ cluster environments show first evidence for drastic changes 
on the high-mass end of galaxies 
and signs for a gradual disappearance of a well-defined cluster \reds as strong star formation activity is observed in an increasing fraction of massive galaxies down to the densest core regions.
The presented XDCP high-$z$ sample will allow first detailed studies of the cluster population  during the critical cosmic epoch at lookback times of 7.3-9.5\,Gyr on the aggregation and evolution of baryons in the cold and hot phases as a function of redshift {\em and} system mass. 
\end{abstract}

\maketitle



\section{Introduction}

The most extreme mass peaks in the primordial matter density field have developed into the present day galaxy cluster population through gravitational amplification and more than 13\,Gyrs of hierarchical structure formation at work. As such, clusters of galaxies form the top level of the hierarchy  and are the latecomers on the  stage of cosmic structures with the most extreme masses and dimensions for gravitationally bound objects. Besides their role as key tracers of the cosmic large-scale structure, clusters  
are also intriguing multi-component astrophysical systems for the study of dark matter, baryons in the hot and cold phases, and a multitude of resulting interaction processes  between them. 

However, one of the major observational challenges is to provide sizable samples of galaxy clusters at high redshift ($z\!>\!0.8$) in order to trace the evolution of the cluster population and their matter components back to the first half of cosmic time, corresponding to lookback times of  7-10\,Gyrs. Bona fide clusters of galaxies with total masses of $M\!\ga\!10^{14}\,\mathrm{M_{\sun}}$ are rare objects, in particular at high $z$, 
which requires large survey areas (tens of square degrees) on one hand and a high observational sensitivity for the identification and investigation of the galaxy- and intracluster medium (ICM) components on the other hand. 
Examples of successful high-$z$ galaxy cluster surveys based on optical/infrared observations of the galaxy populations include 
\citet{Gonzalez2001a}, \citet{Gladders2005a}, \citet{Olsen2007a}, \citet{Eisenhardt2008a},
 \citet{Muzzin2009a},  \citet{Grove2009a}, \citet{Erben2009a}, \citet{Adami2010a}, \citet{Roeser2010a}, and \citet{Gilbank2011a}.
X-ray selected distant  cluster searches include the work of \citet{Rosati1998a}, \citet{Pacaud2007a}, \citet{Suhada2010a}, and 
\citet{Mehrtens2011a}, while detected $z\!>\!0.8$  systems based on the Sunyaev-Zeldovich effect (SZE) are reported e.g.~in 
\citet{Marriage2010a} and \citet{Williamson2011a}.
For a general overview of different survey techniques and an updated status report of distant galaxy cluster research we refer to the accompanying review of  Rosati and Fassbender (in prep.) 

In this paper we provide a comprehensive overview of the XMM-{\it Newton} Distant Cluster Project (XDCP), a serendipitous X-ray survey specifically designed for finding and studying distant X-ray luminous galaxy clusters at $z\!\ge\!0.8$.
The main aims of this article are a description of the cluster sample construction in the XDCP and a report on the status of the compilation of the largest distant X-ray luminous galaxy cluster sample to date.
The paper follows  and combines a series of previous multi-wavelength studies of individual high-$z$ clusters discovered in the XDCP\footnote{An updated list of XDCP publications can be found at \\ \url{http://www.xray.mpe.mpg.de/theorie/cluster/XDCP/xdcp_publications.html} .}. 
To this end, we start with the general goals and design of the survey 
in Sect.\,\ref{c2_XDCP}, followed by an overview of the observational techniques in Sect.\,\ref{c3_Observations}. New results are discussed in Sect.\,\ref{c4_Results}, the current sample of 22 X-ray clusters at $z\!>\!0.9$ is presented in  Sect.\,\ref{c5_Sample}, and 
Sect.\,\ref{c7_Summary} summarizes our findings and conclusions.

Throughout this work we use a standard $\Lambda$CDM cosmological model with parameters ($H_0$, $\Omega_{\mathrm{m}}$, $\Omega_{\mathrm{DE}}$, w)=(70\,km\,s$^{-1}$Mpc$^{-1}$, 0.3, 0.7, -1), physical quantities (e.g.~$R_{500}$, $M_{200}$) are derived for radii for which the mean total mass 
density of the cluster is 500 or 200 
times the critical energy density of the Universe $\rho_{\mathrm{cr}}(z)$ at the
given redshift $z$, and all reported magnitudes are given in the Vega system.



\section{The XMM-{\it \bf Newton} Distant Cluster Project}
\label{c2_XDCP}

The XMM-{\it Newton} Distant Cluster Project was initiated in 
2003 with the main objective of a systematic search for distant X-ray luminous galaxy clusters, with a special focus on the $z\!>\!1$ regime \citep{HxB2005a}. 
Before 2005 only five confirmed clusters at redshifts beyond unity were known \citep{Stanford2002a,Rosati2004a,Hashimoto2005a} up to a maximum redshift for clusters with an X-ray detection from the ROSAT era of $z\!=\!1.26/1.27$ for the two Lynx systems RX\,J0848.9+4452 and CIG\,J0848.6+4453 \citep{Rosati1999a,Stanford1997a}.
However, the rapid growth of data in the XMM-{\it Newton} archive offered the possibility for a new generation of 
serendipitous  X-ray  galaxy cluster surveys with an order of magnitude 
better sensitivity and greatly improved resolution capabilities  \citep[e.g.][]{Romer2001a}.

\subsection{Science objectives}

From the very start, the XDCP focussed  on the galaxy cluster population in the first half of the present age of the Universe, i.e.~at redshifts $z\!\ga\!0.8$. This specialization made the survey manageable in terms of the 
required follow-up resources and, moreover, allowed  the deployment of optimized observational techniques and instrumentation for high-$z$ studies as discussed in Sect.\,\ref{c3_Observations}. The final aim of the XDCP survey is the compilation of an X-ray selected distant galaxy cluster sample with a minimum of 50 test objects at  $z\!>\!0.8$  (30 at $z\!>\!1$) to allow statistically meaningful  evolution studies of the cluster population in at least three mass and redshift bins. 

With such a sample 
numerous open questions on the formation and early evolution of the most massive bound structures in the Universe can be addressed observationally. Some of the key areas include:
\vspace{-0.5ex}
\begin{enumerate}
\item Galaxy evolution in the densest high-$z$ environments 
\item Redshift evolution of the X-ray scaling relations
\item Evolution of the thermal structure and the metal enrichment of the intracluster medium
\item Number density evolution of massive clusters at $z\!>\!0.8$ for cosmological tests
\end{enumerate}
\vspace{-0.5ex}
\noindent
For the cosmological applications (iv) a well controlled selection function is a crucial prerequisite, which will be further discussed in Sect.\,\ref{s3_Xsimulations}.
Some first 
results on the galaxy populations in high-$z$  clusters are shown in Sects.\,\ref{c4_Results}\,\&\,\ref{c5_Sample} and in publications on individual systems (e.g.~Santos et al. \citeyear{Santos2009a}; Strazzullo et al. \citeyear{Strazzullo2010a}; Fassbender et al. \citeyear{Fassbender2011b}b). 
Combining
the existing literature data on the scaling relations of cluster X-ray properties 
with  recent deep X-ray observations of new distant systems from our and other projects we obtained tighter constraints on the evolution of scaling relations with redshift as presented in \citet{Reichert2011a}. 
These results support the picture of an early energy input into the intracluster medium as advocated in preheating models \citep[e.g.~][]{Stanek2010a,Short2010a} rather than a late energy input from interactions with a central AGN at low redshift ($z\!\la\!1$). Since the cluster mass function evolves very rapidly on the massive end and the degree of evolutions depends sensitively on the cosmological parameters, an X-ray selected sample of distant massive clusters is particularly well suited to test cosmological models. Notably, the effect of Dark Energy on structure growth is expected to be most pronounced in the redshift range $0\!<\!z\!\la\!2$.
 The competitive cosmological and Dark Energy  constraints of Vikhlinin et al. (\citeyear{Vik2009a}a; \citeyear{Vik2009b}b) based on the observed evolution of the cluster mass function  with only 37 moderate redshift systems ($0.35\!<\!z\!<\!0.9$)
clearly demonstrated the high potential of distant X-ray clusters as  Dark Energy probes. Therefore the XDCP survey will be ideally suited to extend this test to the next higher redshift regime soon once 
 a sizable subsample of the survey is completed.



\subsection{Survey strategy} 
\label{s2_SurveyStrategy}

The XDCP survey  is based on the following four stage strategy: 
\vspace{-0.5ex}
\begin{description}
\item[X-ray source detection and candidate selection:]  Deep, extragalactic\footnote{The extragalactic sky is defined here as the sky region with galactic latitudes $|b|\!\geq\!20\degr$ that avoids the large extinction and dense stellar fields of the galactic band.} XMM-{\it Newton} archival fields are screened for serendipitous {\em extended} X-ray sources, which are in their vast majority associated with galaxy clusters. The positions of the detected extended X-ray sources are cross-correlated with available optical data and extragalactic database information to test for the existence of a detectable optical cluster counterpart. For about 30\% of the X-ray sources  no optical counterpart could be identified.
These sources are selected as {\em distant} cluster candidates for further follow-up.

\item[Follow-up imaging and redshift estimation:]  The selected distant cluster candidates 
are targeted with sufficiently deep imaging data in at least two suitable optical or near-infrared (NIR) bands. The data allow as a first identification step to probe the existence of an overdensity of (red) galaxies coincident with the extended X-ray source and in a second step a  
cluster redshift estimate based on the comparison of the color of red-ridgeline galaxies with simple stellar population (SSP) evolution models for passive galaxies.

\item[Spectroscopic confirmation:]  Photometrically identified systems at  $z\!>\!0.8$ are further targeted with deep optical spectroscopy in order to confirm the gravitationally bound nature of the systems and to determine the final accurate redshifts of the newly discovered galaxy clusters. 

\item[Multi-wavelength follow-up of selected systems:]  The most interesting and intriguing distant systems are further studied in more detail in different wavelength regimes, e.g.~with deeper X-ray data or multi-band imaging observations in the optical and infrared. 

\end{description}

\noindent
The first $z\!>\!1$ cluster 
discovered with this strategy was XDCP\,J2235.3-2557 at $z\!=\!1.39$ 
\citep{Mullis2005a}, which started the ongoing era of distant cluster detections with XMM-{\it Newton}.



\section{Observational techniques and reduction pipelines}
\label{c3_Observations}

The following section introduces and discusses  the different relevant observational techniques for the first three XDCP survey stages  in more detail. A full comprehensive description of observational aspects and reduction pipelines can be found in \citet{RF2007PhD}.

\subsection{X-ray data}
\label{s3_Xdata}

The XMM-{\it Newton} observatory currently provides  by far the best capabilities for 
detecting the typically faint extended X-ray sources associated with distant galaxy clusters. The most important  key features of  XMM-{\it Newton} for this task are (i) the large effective collecting area ($\sim$2500\,cm$^2$ on-axis at 1\,keV), (ii) the 30\arcmin \ diameter field-of-view (FoV, $\sim$0.2\,deg$^2$), and (iii) a sufficiently good spatial resolution of 5\arcsec-15\arcsec \ (FWHM) to 
identify distant clusters as extended sources.

\begin{figure}[t]
\centering
\includegraphics[angle=0,clip,width=0.6\textwidth]{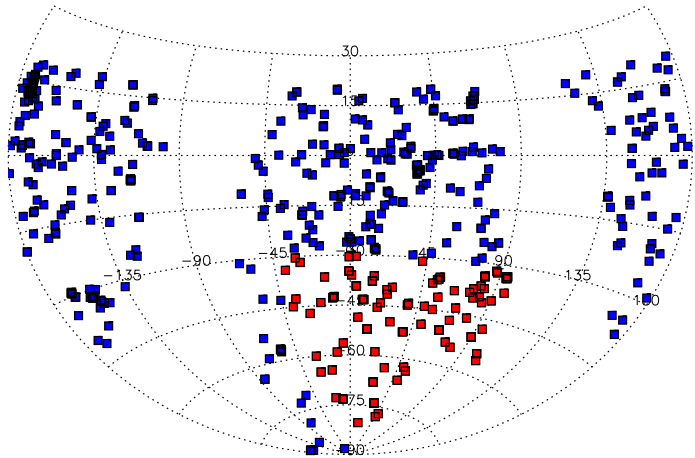}
\vspace{-1ex}
\caption{Sky distribution of the 575 Southern XMM-{\it Newton} fields considered for the XDCP survey, 469 of which were successfully processed and analyzed. The red squares indicate fields  within the original footprint of the South Pole Telescope survey. Square symbols are not to scale. 
}
\label{f6_Sky_Coverage}       
\end{figure}

The XMM-{\it Newton}  data archive is a very rich resource to start a systematic search for distant clusters
based on their characteristic X-ray signature, the {\em extended} thermal ICM emission, which clearly discriminates these sources from the point-like AGN population that dominates the X-ray sky in extragalactic fields.  For the definition of the XDCP survey fields, the public XMM-{\it Newton} archive as of 2 November 2004 was considered, i.e.~the public data of the first 5 years of the mission.
Out of the 2960 observed fields available at that time with a combined nominal exposure time\footnote{The exposure time listed in the XMM-{\it Newton} archive.} of 72.3\,Msec, 1109 fields remained after applying the conditions of (i) imaging mode  observations of at least one of the three cameras, (ii) a minimum nominal exposure time of 10\,ksec, and (iii) field positions outside the galactic plane ($|b|\!\geq\!20\degr$) and away from the Magellanic Clouds and M31\footnote{The minimal angular distances for the field positions were 10.8\degr \ for the LMC, 5.3\degr \ for the SMC, and 3.2\degr \ for M31 \citep[see e.g.][]{Kim2004a}.}. After a further removal of (iv)  major dedicated survey fields (e.g.~COSMOS) and (v)  constraining the area 
 to the VLT-accessible part of the sky  (DEC$\leq\!+20$\degr) for the follow-up program, 575 archival observations remained as input for the survey (see Fig.\,\ref{f6_Sky_Coverage}). Out of these fields, 29  were discarded as non-usable for the survey after a visual screening of all fields.

The 
remaining 546 XMM-{\it Newton}  archival  fields with a nominal total exposure time of 17.5\,Msec were processed and analyzed as detailed below. The final XDCP sample of successfully processed  and analyzed fields amounts to  469 individual XMM-{\it Newton} pointings (29 fields had corrupted data and 48 were flared) comprising 15.2\,Msec of  X-ray data with a total sky coverage of 76.1\,deg$^2$ (see Table\,\ref{tab_survey area}). The initial XDCP pilot study (de Hoon et al., in prep.) for testing and qualifying the survey strategy of Sect.\,\ref{s2_SurveyStrategy} was based on an earlier processing and candidate selection of about 20\% of these fields.

\subsubsection{X-ray processing.} 
\label{s3_Xprocessing}
The task of processing several hundred XMM-{\it Newton} archival fields requires an efficient automated X-ray reduction pipeline with minimized manual interaction.  To this end, a designated, distant cluster optimized XDCP reduction and source detection pipeline was developed based on the XMM Science Analysis Software\footnote[1]{\url{http://xmm.esac.esa.int/sas/}} (SAS).
All selected XMM-{\it Newton}  data sets were homogeneously processed with this pipeline  using the  version SAS 6.5 released in August 2005.

The data processing starts with the Observation Data File (ODF) for each archival field. In a first reduction step the SAS tasks
{\tt cifbuild}, {\tt odfingest}, {\tt emchain}, and {\tt epchain} are run to set up the appropriate calibration files for the field, ingest the housekeeping data, and produce calibrated photon event files for the PN and the two MOS X-ray imaging instruments of XMM-{\it Newton}.

In a second step, periods of increased background levels, most notably due to solar soft proton flares, are removed from the data in a strict two level flare cleaning process \citep[see e.g.][]{Pratt2003a}. This task is of crucial importance for the detectability of faint extended X-ray sources.
Due to the flat nature of the flare spectrum, time periods with background levels significantly higher than the quiescent count rates are in the first cleaning stage efficiently identified in the hardest energy band of 12-14\,keV (10-12\,keV) for the PN (MOS) detector and removed from the data with an automated 3-$\sigma$ clipping algorithm. However, residual soft flare peaks can still remain in the data, which are subsequently removed  by applying a second soft-band cleaning stage to the full 0.3-10\,keV band with a similar clipping procedure. 
The resulting cleaned photon event lists for each detector contain now only the selected science usable time periods, which is on 
 average about two thirds of the nominal field exposure time, i.e.~one third of the observation is typically lost due to flares and instrumental overheads. 

We define the {\em clean effective exposure time} as the period during which all three instruments in imaging operation would collect the equivalent number of soft science photons for the particular observation. 
The 48 fields with a resulting clean effective exposure time of $<$5\,ksec were declared as flared and discarded from further processing. In addition,  29 archival fields with corrupted data files were not considered. The resulting 469 XDCP survey fields comprise a total of 8.8\,Msec of clean effective exposure time, with an average (median) clean field depth of 18.78\,ksec (15.71\,ksec).

In a third step, 
images with a pixel scale of 4\arcsec/pixel are generated for different X-ray energy bands  from the clean event lists for each of the three instruments. The  redshifted spectra of distant clusters with ICM temperatures of  2-6\,keV  
have their observed bulk emission in the soft X-ray band. Images are hence generated for the standard XMM bands  0.3-0.5\,keV, 0.5-2.0\,keV, 2.0-4.5\,keV,  and a very broad band with  0.5-7.5\,keV. Moreover, it is possible to define a single energy band which maximizes the expected signal-to-noise ratio (SNR) for $z\!>\!0.8$  systems following the work of \citet{Scharf2002a}, which leads to the definition of an additional optimized XDCP detection band for the energy range  0.35-2.4\,keV.

For all images, corresponding exposure maps are generated with the SAS task {\tt eexpmap}, which contain the effective local integration times associated with each detector pixel scaled to the on-axis exposure. These X-ray exposure maps, similarly to the concept of  flatfields in optical and NIR imaging,  contain the calibration information on the radial vignetting function, the energy dependent detector quantum efficiency, chip gaps, dead detector columns, the transmission function of the used optical blocking filter, and the field-of-view of the detectors.

Exposure corrected, i.e.~flatfielded,  images are obtained by dividing the photon images of each detector by the corresponding exposure map. The full data stack for each energy band is obtained by combining the PN, MOS1, and MOS2 images weighted with the corresponding effective collecting area of each telescope-camera system. For visual inspection purposes the combined and exposure corrected X-ray images in each energy band are smoothed with a 4\arcsec \ Gaussian filter, from which  logarithmically spaced X-ray flux contours are generated to be overlaid on optical images for the source identification process (Sects.\,\ref{s3_SourceScreening}\,\&\,\ref{s3_FollowUpImaging}).

\subsubsection{X-ray source detection.}
\label{s3_Xdetection}
 
The X-ray source detection is run on each field individually, even in case of multiple observations of the same target or overlapping fields. In the event of multiple detections of the same extended X-ray source in overlapping fields, the highest significance source is retained on the cluster candidate list, while the others are flagged as duplicate detections.
The main technical reason for this field-by-field approach is that the X-ray point-spread-function (PSF) at each detector position has to be known as accurately  as possible in order to allow a robust 
determination of extent likelihoods. Since the PSF\footnote[2]{The source detection relies on the tabulated energy and position dependent PSF model as provided in the calibration database for SAS 6.5.} of XMM-{\it Newton}'s telescopes varies considerably across the field, in particular as a function of increasing off-axis angle, detections in combined mosaic  fields would have added significant systematic uncertainty to the results based on the available PSF calibration and SAS status at the time of data reduction.    

 The main XDCP source detection method relies on a sliding box detection with the SAS task {\tt eboxdetect} followed by a maximum likelihood fitting and source evaluation with  {\tt emldetect}. As a preparatory step, detection masks are created with {\tt emask} that define the area over which the source detection is to be performed. Regions contaminated by bright sources in the FoV, i.e.~in general the targets of the observation such as nearby clusters or luminous AGN, are excised from the survey area at this point by defining circular exclusion regions for the detection mask, which also takes into account detector artifacts of the individual instruments, such as chip gaps and dead columns.
 
The next crucial step is the determination of robust background maps in each field for all instruments and energy bands with {\tt esplinemap}. For robustness and to avoid possible artificial background fluctuations from spline fits with many degrees of freedom, we make use of the smooth two-component background model option, which is based on the
linear combination of a spatially constant background contribution (quiescent particle induced
background and instrumental noise) and a vignetted component (CXB and residuals of
the soft proton particle background). Background maps are produced by first running {\tt eboxdetect}  with a local background determination around the detection cell in order
to produce a preliminary list of X-ray sources, which are subsequently  excised from the field before performing the two-component fit for the global background map. 

The sliding box source detection is then repeated with {\tt eboxdetect}  using the previously determined global background maps for each detector and varying detection cell sizes to account for the extended sources. This way, a  list of  positions of X-ray source candidates is produced, which serves as input list for the subsequent detailed analysis
and source characterization via maximum likelihood (ML) fitting with {\tt emldetect}. The maximum likelihood fitting for the source evaluation and parameter estimation is performed simultaneously for all used energy 
bands and the three individual detectors, with the associated global background maps, exposure maps, and detection masks provided to the task.  

The maximum likelihood PSF fitting procedure applied to the photon images evaluates the significance for the detection ({\tt DET\,ML\/}) and the extent ({\tt EXT\,ML\/}) of an X-ray source expressed in terms of the  likelihood
$L\!=\!-\ln p_{\mathrm{Pois}}$ \citep{Cruddace1988a}, where $p_{\mathrm{Pois}}$ is the probability of a Poissonian random background fluctuation of counts in the detection cell, which would result in at least the number of observed counts.
X-ray sources are flagged as extended  with  core radius $r_{\mathrm{c}}\!>\!0$
if a King profile fit\footnote[3]{Radial surface brightness profile with functional form $S(r)=S_0\cdot[1+(r/r_{\mathrm{c}})^2]^{-3/2}$.} with a fixed $\beta\!=\!2/3$ returns a significantly improved likelihood above a minimum threshold value compared to a local point source model. 
Moreover, the 
extended nature of a source is only accepted if the model likelihood  of the fit to the X-ray photon distribution supersedes the probability of a model with two overlapping  point sources (i.e.~point source confusion). 
According to this likelihood evaluation, sources are either characterized as point sources with detection likelihood {\tt DET\,ML\/} and the free parameters position and count rate in each band, or as extended source with extent likelihood {\tt EXT\,ML\/} and the additional core radius parameter $r_{\mathrm{c}}$.

The inherent thresholding procedure used in {\tt emldetect} and the test performed  for source confusion of two PSF-like components does not allow a subsequent evaluation of the extent probabilities of all sources, but rather divides the populations into point sources with, by definition, zero core radius and extent probability, and extended sources above a minimum extent likelihood threshold  for {\tt EXT\,ML\/}. This implies that the critical thresholding parameters for the detection of extended X-ray sources have to be optimized prior to the actual detection run. To this end, source detection tests with various input parameter combinations were performed on the XMM-{\it Newton} data set in the COSMOS field, which were compared to the actual extended X-ray source catalog of confirmed galaxy  groups and clusters of  \citet{Alexis2006a}.

The XDCP source detection procedure follows two main objectives: (i) the construction of a quantifiable extended X-ray source sample ({\em survey sample}) with an accurately characterizable selection function over a suitable part of the X-ray coverage (Sects.\,\ref{s3_Xsimulations}\,\&\,\ref{s3_XSurveyArea}), and (ii) a supplementary X-ray selected cluster candidate sample ({\em supplementary sample}) from the full XDCP sky coverage and down to the faintest feasible X-ray flux levels that still allow the blind detection of extended sources. The scientific applications of the first objective are statistical and cosmological studies of a well-controlled high-$z$ galaxy cluster sample with quantified detection characteristics drawn from a known survey volume. 
To this end, the final {\em survey sample} is selected from the inner parts ($\Theta\!\le\!12\arcmin$) of the detector area (survey level 2 in Table\,\ref{tab_survey area} and Sect.\,\ref{s3_XSurveyArea}) based on significant extended X-ray sources above a minimum flux cut-off, which is determined through extensive simulations (Sect.\,\ref{s3_Xsimulations}).
The second objective for the compilation of the additional {\em supplementary sample} aims at an extended coverage of the accessible range of cluster parameters by considering also sources of lower significance and at large off-axis angles at the expense of higher impurity levels. Applications for this  {\em supplementary sample} include (i) new rare massive clusters found in the additional larger survey area covered by the outer parts of the detectors (survey level 1 in Table\,\ref{tab_survey area} and Sect.\,\ref{s3_XSurveyArea}), (ii) the detection of lower mass and higher redshift systems at lower flux levels, and (iii) the general 
 exploration of the feasibility limits of the source detection and X-ray cluster surveys.  

 
The adopted XDCP source detection procedure for the construction of the survey sample rests upon the conceptually simplest 
detection strategy by deploying the single, distant cluster optimized, detection band for the energy range 0.35-2.4\,keV. This choice is expected to yield optimal signal-to-noise ratios for the X-ray sources associated with  the targeted distant cluster population with ICM temperatures $T_{\mathrm{X}}\!\ga\!2$\,keV. This primary XDCP detection scheme is also the easiest to characterize through simulations (Sects.\,\ref{s3_Xsimulations}). The critical thresholding parameters for the detection of X-ray sources are set to  
 {\tt DET\,ML}$\ge$6 ($p_{\mathrm{real}}\!\ge\!0.998$, significance\,$\ga$$3.1\,\sigma$) as the minimum likelihood for the existence of a source, and  {\tt EXT\,ML}$\ge$5 ($p_{\mathrm{ext}}\!\ge\!0.993$,  significance\,$\ga$$2.7\,\sigma$) as lower threshold for the extent probability.
 
For the supplementary sample, additional cluster candidates down to lower extent likelihoods of  {\tt EXT\,ML}$>$3 ($p_{\mathrm{ext}}\!\ge\!0.95$,  significance\,$\ga$$2\,\sigma$) are considered by re-running the source detection two more times using
different detection schemes. The first one is the basic XMM  `standard scheme'  covering the energy range 0.3-4.5\,keV with three input bands (0.3-0.5\,keV, 0.5-2.0\,keV, 2.0-4.5\,keV). The second setup is an experimental  `spectral matched filter scheme'
that covers the broader energy range 0.3-7.5\,keV with an increased weight on the lower energy range by using five overlapping bands  (0.3-0.5\,keV, 0.5-2.0\,keV, 2.0-4.5\,keV, 0.35-2.4\,keV, 0.5-7.5\,keV).
Detection results based on the complementary wavelet detection method with the 
SAS task {\tt ewavelet} were additionally used as a qualitative cross-check of detected extended sources at low significance levels.

This redundancy strategy with  source detection results from different detection schemes 
offers cross-comparison possibilities that are particularly advantageous when evaluating flagged extended sources very close to the threshold of detectability. To this end, the supplementary schemes add extra information to the source lists from the primary detection band scheme, such as standard 0.5-2.0\,keV flux estimates and several hardness ratios. Furthermore, the stability of the best fitting extended source model can be evaluated by cross-comparing the core radius measurements and extent likelihoods obtained with the different detection schemes, which allows a more reliable identification of spurious sources from background fluctuations and spurious extent flags associated with point sources.  

The XDCP source detection run based on the discussed schemes and applied to the 469 XMM-{\it Newton} archival survey fields resulted in about 2000 flagged extended source candidates as raw input list for the  combined survey and supplementary samples.  
These flagged sources are further evaluated in the three stage screening process detailed below.

\subsubsection{Source screening.}
\label{s3_SourceScreening}
\label{s3_Xscreening}

A visual inspection and screening of candidate extended X-ray sources detected in XMM-{\it Newton} data is inevitable even at significance threshold levels much higher than for the XDCP scheme. At the first screening stage on the X-ray level, obvious spurious detections of extended sources are removed from the source list. Various calibration and detection method limitations as well as  instrumental artifacts can lead to spurious detections of extended sources. The most obvious false detections originate from (i) secondary detections in wings of large (partially masked out) {\em extended} sources, (ii) artifacts at the edges of the field-of-view, and (iii) PSF residuals in the wings of very bright point sources. These `level 1' spurious extended X-ray sources, totaling about  15\% of the raw 
catalog,   can be readily and safely removed by inspecting the locations of the candidate sources in the FoV of the combined soft-band X-ray image.

For identifying and removing the more subtle `level 2' false detections, a second X-ray screening stage is required that 
is based on a close inspection and evaluation of
every source individually with complementary information on potential contaminations from optical imaging data. For this task, a set of  diagnostic images is produced to evaluate the  source environment based on the X-ray flux contours, the original combined X-ray photon image, the flux distribution in the three individual detectors, and the overlaid X-ray contours on optical imaging data. For the latter X-ray-optical overlays the online  all-sky data base of the Second Digitized Sky Survey\footnote[4]{\url{http://archive.eso.org/dss}} (DSS\,2)  is queried for image cutouts in the red (DSS\,2-red) and NIR (DSS\,2-infrared) bands.  
The aim of the second X-ray screening stage is to identify false detections originating from e.g.~(iv) blends of three or more point sources, (v) spurious sources related to an underestimation of the local background, 
vi) chip boundary effects, (vii) residuals from the correction of the so-called out-of-time event trails, and (viii) `optical loading' residuals caused by bright optical sources. The conservative flagging of such `level 2' false detections reduces the original raw source catalog by an additional 20\% resulting in a double X-ray screened input list of about 1300 extended sources with a remaining impurity level of 10-20\%\footnote[5]{Based on a preliminary empirical evaluation with wide field follow-up imaging data.}. 

\enlargethispage{4ex}

The third and final screening stage aims to identify the optical counterparts associated with the extended X-ray sources based on the X-ray-optical overlays and additional queries to the NASA Extragalactic Data Base\footnote[6]{\url{http://nedwww.ipac.caltech.edu}} (NED) to check for known objects and redshift information. Approximately 100 (8\%) of the extended sources can be readily identified as non-cluster objects, mostly nearby galaxies and galactic sources, e.g.~supernova remnants. From the remaining list of $\sim$1200 galaxy cluster candidates, about 70\% show optical signatures of low and intermediate redshift  clusters or groups, which can be identified typically up to 
$z$$\sim$0.5-0.6. The 
final  fraction of 30\% of the sources with an uncertain or no optical counterpart enter the list of XDCP distant cluster candidates, which are carried over to the dedicated follow-up imaging program of the survey (Sect.\,\ref{s3_FollowUpImaging}). After removing double detections from overlapping XMM-{\it Newton} fields, the final XDCP sample comprises 
990 individual galaxy cluster candidates. From this point on, the further XDCP survey efforts are focussed on the identification and deeper study of the selected $\sim$300 distant cluster candidates, i.e.~the extended X-ray sources without optical counterpart.

\subsubsection{Detection sensitivity.}
\label{s3_Xsimulations}

\begin{figure}[t]
\centering
\includegraphics[angle=0,clip,width=\textwidth]{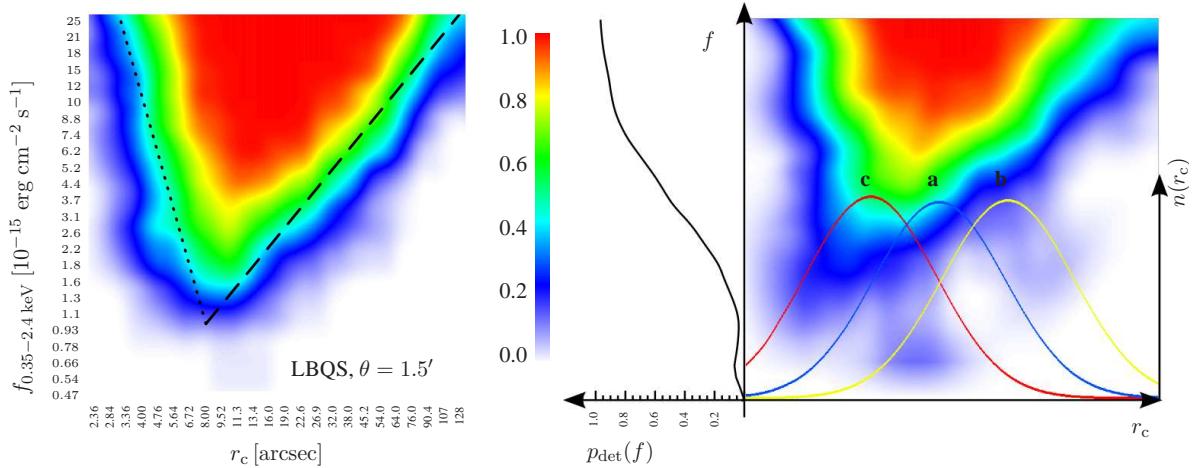}
\vspace{-1ex}
\caption{Source detection simulation results. {\em Left:} Detection sensitivity for the central part of a deep (51.7\,ksec) survey field with color coded recovery fractions according to the 
vertical color bar
across the source flux versus core radius parameter plane.  
The dotted black line indicates the XMM-{\it Newton} resolution limit, whereas the dashed line follows the background limit as a function of source extent. {\em Right:}  Illustration on how a completeness function $p_\mathrm{det}(f)$ is obtained as a function of flux $f$ by weighting the detection probabilities in the source extent direction with an assumed core-radius distribution (a) of the cluster population. Input curve (a) shows the observed local $r_c$ distribution scaled to apparent sizes at $z\!=\!1$, whereas (b) and (c) are up- and down-scaled  distributions by factors of 2.
Plots adapted from \citet{Muehlegger2010}.}
\label{fig_DetSensitivity}       
\end{figure}

One of the main strengths of X-ray cluster surveys is the ability to accurately quantify the 
 detection process and the resulting effective survey volume  through simulations  \citep[see e.g.][]{Pacaud2006a,Burenin2007a,Mantz2010b,Lloyd2010a_aph}.
For the characterization of the XDCP survey sample a dedicated simulation pipeline was developed by \citet{Muehlegger2010} that follows the actual survey data and detection procedure as closely as possible. 

For the background limited regime of deep XMM-{\it Newton} fields, the minimum flux levels $f_{\mathrm{lim}}$ required for the detection of idealized resolvable (i.e.~$r_{\mathrm{c}}\!>\!r_{\mathrm{min}}$) extended sources with angular core radius $r_{\mathrm{c}}$ scale as 
$f_{\mathrm{lim}}(r_{\mathrm{c}}\!>\!r_{\mathrm{min}}) \propto r_{\mathrm{c}}\cdot[B(\Theta,\phi)/t_{\mathrm{eff}}(\Theta)]^{1/2}$, where $t_{\mathrm{eff}}$ is the effective exposure time at off-axis angle $\Theta$ and $B$ is the total local background count rate, which can additionally 
vary with azimuthal angle $\phi$. This strong positional dependence of detection sensitivities for a heterogeneous serendipitous XMM-{\it Newton} survey implies that the accurate reconstruction of the selection function requires a local approach for 
each solid angle element of the X-ray coverage. 

To this end, the  full  XDCP survey area is characterized by analyzing the detection performance of 7.5 million simulated, circularly symmetric mock $\beta$-model\footnote{More complicated (e.g.~double-$\beta$) models are currently not considered owing to the increasing complexity to adequately cover the model parameter space.} cluster sources spanning a wide range of core radii  (2-128\arcsec) and net source counts (20-1280) in 25 logarithmic steps each. Simulated clusters with a poissonized two-dimensional photon distribution are convolved with the  local PSF and then placed directly into the observed XDCP survey fields at various off-axis angles and random azimuthal positions. This approach accounts by design for all local properties at a given position in a survey field, such as local background, exposure time, and possible contamination from surrounding X-ray sources. In order to obtain sufficient statistics for the covered parameter space and the different positions across the FoV, more than 1500 field realizations are generated, each with ten additional inserted mock clusters. 
These mock fields with simulated cluster sources of known flux and position are then analyzed by the XDCP source detection pipeline  for the primary detection scheme with the optimized  0.35-2.4\,keV band.
The detected extended sources in each field realization are subsequently matched to the simulated input catalog, from which the fraction of recovered detected cluster sources can be determined as a function of input flux, core radius, and  off-axis angle.    

Figure\,\ref{fig_DetSensitivity} (left) shows the simulation results for the central part of one of the deepest XDCP survey fields with 51.7\,ksec clean effective exposure time, originally observed as part of the Large Bright Quasar Survey 
\citep[LBQS,][]{Hewett1995a}. The shown detection sensitivity as a function of total source flux versus  angular core radius is representative for the deepest part of XDCP,
while the typical median survey sensitivity 
along the y-axis  is a factor 2.5-3 higher. The figure illustrates well the XMM-{\it Newton} detection capabilities and its limitations. The `shark tooth' shaped colored region of extended source detectability 
is confined by two limits. The dotted black line at small core radii marks the manifestation of the  XMM-{\it Newton} resolution limit and is governed by the extent significance ({\tt EXT\,ML}) determination for the sources. The core radius detection threshold decreases slightly with increasing flux, i.e.~number of source photons, since a smaller core can be compensated with an increased photon statistics of the PSF-convolved 
surface brightness profile in order to yield the same extent significance of the source. 
The dashed line towards large core radii indicates the surface brightness limit and is governed by the detection threshold ({\tt DET\,ML}) in order to identify the presence of a source with low central surface brightness above the background level. This limit prevents the detection of very extended sources and closely follows the expected scaling behavior $f_{\mathrm{lim}}\propto r_{\mathrm{c}}$.

The highest detection sensitivity is achieved at the tip of the `shark tooth' for angular core radii in the range 6-12\arcsec, corresponding to physical core sizes of 50-100\,kpc at  $z\!>\!0.8$. For such relatively compact cores, extended sources down to flux levels of  $\sim$$10^{-15}$\,erg\,s$^{-1}$cm$^{-2}$ in the detection band can be identified  for the deepest parts of the  XDCP survey. Conversely, this implies a  sweet spot for cluster detections at the lowest flux levels (i.e.~the supplementary sample), e.g.~at the highest redshifts of $z\!\ga\!1.4$, where this 
effect introduces a detection preference towards clusters with compact cores. However, by applying a flux cut well above the known extreme tip of detectability the construction of fair and morphologically unbiased cluster samples is still straightforward.

In order to obtain detection probabilities as a function of flux $p_\mathrm{det}(f)$ that 
provide an average over the cluster structures in the survey, the simulation results along the angular core radius axis have to be weighted with the actual core radius distribution of the underlying cluster population, which is illustrated in the right panel of Fig.\,\ref{fig_DetSensitivity}. Ideally, one would like the $z\!>\!0.8$  cluster core radius distribution as input function for this task, which is  observationally not determined at this point. As a starting point, we hence have to revert to the observed local core radius distribution from \citet{Vik1998a}, which follows a log-normal distribution with a central peak at 112\,kpc corresponding to an angular scale of 14\arcsec  \ at  $z\!=\!1$  (curve [a] in Fig.\,\ref{fig_DetSensitivity}), which only varies by $\pm$6\% across our  $0.8\!\la\!z\!\la\!1.6$ redshift interval of interest.  
The weighting procedure according to this input function yields the resulting displayed $p_\mathrm{det}(f)$ function. However,  high-$z$ clusters are expected to exhibit more compact cores due to the higher critical background density at the collapse epoch. The effect on the XDCP selection function can be investigated by downscaling the local distribution by a factor of 2 (curve [c]), resulting in only a moderate change of the median flux limit by about 10\%. The unknown structural properties of the  high-$z$ cluster population are thus only minimally affecting the survey sensitivity characterization, as long as the average  high-$z$ cluster core radii are not decreasing by more than a factor of 2. A more significant decrease of the average detection sensitivity would occur in the unexpected case of increasing core radii (curve [b]).    
In the future, we hope to recover the shape distribution function of distant clusters directly from our survey, once the statistics of systems with good X-ray data is sufficiently large.  

Another important result of the performed simulations is the determination of an optimized  maximal acceptable XMM-{\it Newton} off-axis angle for which the enclosed  detector area is well characterizable, without compromising the detection sensitivity and reliability due to the off-axis PSF characteristics and other instrumental artifacts.  This optimal maximum off-axis angle of the well characterizable detector area was found to be $\Theta_\mathrm{max}\!=\!12\arcmin$, implying an enclosed solid angle of 0.126\,deg$^2$ per XMM-{\it Newton} field.  At $\Theta_\mathrm{max}\!=\!12\arcmin$, the PSF FWHM blurring factor is about 40\% increased and the effective area is decreased to 44\% compared to the on-axis characteristics of XMM-{\it Newton}. 

While the analysis of the full XDCP survey simulations  is still ongoing, the basic global survey characteristics are known and are discussed in the next section.

\subsubsection{Survey area and X-ray cluster candidate sample.}
\label{s3_XSurveyArea}
\label{s3_Xsample}

\begin{table}
\caption{Basic characteristics of the XDCP X-ray coverage for different survey levels. Properties that apply to the full area of a given survey level are indicated by a `yes'.}\label{tab_survey area}
\begin{indented}
\item[]\begin{tabular}{@{}llll} 
\br
& Full X-ray coverage & {\bf Main Survey} & Gold Coverage \\

\mr
Survey Level & SL\,1 & {\bf SL\,2} & SL\,3 \\
Solid Angle [deg$^2$] & 76.1 & {\bf 49.4} & 17.7 \\
Number of Cluster Candidates & 990 & {\bf 752} & 310 \\
Cluster Candidates per deg$^2$ & 13.0 & {\bf 15.2} & 17.5 \\
0.5-2\,keV Sensitivity [$10^{-14}$\,erg\,s$^{-1}$cm$^{-2}$] & $\sim$1.0 & {\bf $\sim$0.8} & 0.6 \\
Analyzed XMM Survey Fields & 469 & {\bf 469} & 160 \\
$|b|\!\geq\!20$\degr & yes & {\bf yes} & yes \\
DEC$\leq\!20$\degr & yes & {\bf yes} & yes \\
XMM Nom. Exp. $\geq\!10$\,ksec  & yes & {\bf yes} & yes \\
XMM Off-axis Angle $\leq\!12$\arcmin &  & {\bf yes} & yes \\
XMM Clean Exp. $\geq\!10$\,ksec  &  &  & yes \\
N$_{\mathrm H}\leq\!6\times 10^{20}$cm$^{-2}$  &  &  & yes \\
Low Background/Contamination  &  &  & yes \\

\br
\end{tabular}
\end{indented}
\end{table}


Table\,\ref{tab_survey area} provides an overview  of the XDCP survey coverage and the basic sample properties for three different subsets of the X-ray data, called survey levels (SL). The full X-ray coverage (SL\,1) comprises a total solid angle of non-overlapping area of 76.1\,deg$^2$ from which a combined galaxy cluster candidate sample of 990 sources was identified. 
This corresponds to a candidate surface density of 13.0  per deg$^2$, which is comparable to the total cluster density in the XMM-LSS survey  \citep{Adami2011a}.
The SL\,1 coverage has an average soft band sensitivity for extended sources of  
$\sim$$10^{-14}$\,erg\,s$^{-1}$cm$^{-2}$ and a sample impurity of up to 20\%. The main aim of this full X-ray coverage sample is to increase the area for the search of  the rarest, most massive  high-$z$ systems, which requires the largest possible survey volume for these sources with flux levels  bright enough to be identified even at large detector off-axis angles. 

A complete and detailed survey characterization will be available for the main survey (SL\,2) which is constrained to the X-ray coverage enclosed within the inner 12\arcmin \ of the detector area, comprising 49.4\,deg$^2$ and 752 cluster candidates. The average sensitivity is $\sim$$0.8\!\times\!10^{-14}$\,erg\,s$^{-1}$cm$^{-2}$, which is also reflected in the increased  surface density  of 15.2  per deg$^2$ and significantly improved purity levels. The coverage of SL\,2 will be the maximum solid angle for studies that require a detailed knowledge of the selection function, i.e.~cosmological applications.

As the starting point for the construction of a first sizable and statistically complete sample of  $z\!\ge\!0.8$ clusters, we added a third survey level comprising the `gold coverage' sample of 310 sources selected from the best 17.7\,deg$^2$ of X-ray data. This  SL\,3 has the additional field  constraints of an effective clean exposure time of $\ge$10\,ksec, upper limits on the Galactic hydrogen absorption column, and stricter field selection  cuts concerning the background levels and contaminating sources in the FoV.  The simulation run for all 160 SL\,3 fields is completed, and yielded an average soft band cluster detection sensitivity of $0.6\times 10^{-14}$\,erg\,s$^{-1}$cm$^{-2}$, with a 
 candidate surface density of 17.5  per deg$^2$ and an expected initial purity level of $>$90\%. The depth of this third XDCP survey level is hence in between the 2\,deg$^2$ coverage in the COSMOS field  \citep{Alexis2006a} and dedicated contiguous cluster surveys with XMM-LSS-like exposure times \citep{Pacaud2006a}.

The initial impurity levels of the different subsamples are based on the selection of extended sources down to the pursued low extent significance threshold of 2-3\,$\sigma$ for the supplementary sample. With the limiting average flux levels for SL\,2 or SL\,3 at hand, it is straightforward to construct 
statistically complete cluster subsamples with negligible impurity levels based on subsequently applied flux cuts.  
As an example for the  `gold coverage' of SL\,3, a minimum flux cut  of $1.5\times 10^{-14}$\,erg\,s$^{-1}$cm$^{-2}$  imposed on the confirmed $z\!\ge\!0.8$ cluster population  is a factor of 2.5 above the average detection sensitivity, and will thus result in a highly complete, morphologically unbiased, and fair census of luminous distant clusters over the SL\,3 solid angle.

Since the XDCP survey is focussed on the distant cluster population that is part of the $\sim$30\% distant candidate subsample without initial optical counterpart identification, most of the initial sample impurity is part of these  $\sim$300 candidates selected for follow-up imaging. Hence, the fraction of spurious sources that  passed the  X-ray screening procedure of Sect.\,\ref{s3_Xscreening} has to be identified as false positives during the follow-up imaging campaigns discussed in Sect.\,\ref{s3_FollowUpImaging}. Preliminary results suggest that about 1/3 of the distant cluster candidate sample are false positives, corresponding to a total number of order 100, or 10\% of the parent sample of SL\,1 cluster candidates. 
In contrast, about the same fraction are photometrically identified  $z\!\ga\!0.8$ candidates for the final spectroscopic confirmation step of  Sect.\,\ref{s3_Spectroscopy}, and about half of these turn out to be bona fide $z\!>\!1$ clusters, corresponding to 5\% of the parent cluster candidate sample or  2.5\% of the first uncleaned source list. 

With about 100 false positives in the XDCP follow up sample, one can ask the question on the odds of finding chance galaxy structures at the sky position of a spurious X-ray source. Assuming random sky positions for spurious sources, we are probing a combined sky area of about 0.022\,deg$^2$ (0.088\,deg$^2$)  for the presence of a galaxy cluster center within 30\arcsec \ (60\arcsec) around initially detected spurious X-ray positions.  This sky area is to be compared to the expected surface density of the objects we are looking for, which is of the order of one $z\!\ge\!0.8$ cluster per deg$^2$, implying a chance of $<$10\% in the case of allowed cluster center offsets of up to 1\arcmin \ and even a factor of 4 lower for the 0.5\arcmin \ offset radius.
This estimate results in the conclusion that the odds of finding even a single `chance cluster' in the full XDCP survey that is randomly associated with a low significance extended X-ray source is very low. However, the situation may look different in case systematic astrophysical effects that can mimic extended X-ray emission are present or enhanced in  high-$z$ group and low-mass cluster environments, such as multiple weak AGN in clusters  that could cause a systematic point source confusion in these systems.   
Answers on the presence and abundance of such systems will come from the XDCP survey itself once the spectroscopic follow-up is completed and {\it Chandra}  X-ray data is available for some of the potentially 
contaminated systems.

\subsubsection{Detailed source characterization.}
\label{s3_Xcharacterization}

The automated XDCP source detection pipeline provides approximate source parameters, such as  estimates for the source flux and the core radius. However, for a detailed characterization of the X-ray properties of spectroscopically confirmed systems a more elaborate  `post detection processing' is required for determining accurate cluster luminosities and other physical parameters.
To this end,  we re-process the archival data with the latest SAS version and calibration database, manually check and optimize the quiescent time periods used for the double flare cleaning process, and check for potential contaminations of the source environment under study.  At this stage, also the combination of overlapping XMM-{\it Newton} fields is considered for cases of significant signal-to-noise gain of the source.
We then apply  an 
extended version of the  growth curve analysis (GCA) method of \citet{HxB2000a} 
to the point-source excised cluster emission in order
to obtain an accurate 0.5-2\,keV flux measurement of the source as a function of cluster-centric radius. Examples of the cumulative background-subtracted source flux for two clusters are shown in 
Sect.\,\ref{c4_Results} in Fig.\,\ref{fig_012_077com_CMDs}.
The total cluster source flux is determined iteratively by fitting a line to the plateau level of the 
flux and measuring the enclosed total source flux 
within the plateau radius. The uncertainty of the flux measurement is determined from the Poisson errors plus a 5\% systematic uncertainty of the background estimation. 

\enlargethispage{4ex}

The soft band  restframe luminosity   L$^{0.5-2\,\mathrm{keV}}_{\mathrm{X,500}}$  and the bolometric luminosity L$^{\mathrm{bol}}_{\mathrm{X,500}}$ are then self-consistently determined within R$_{500}$ by iterating the estimates for the cluster radius and ICM temperature derived from the scaling relations of \citet{Pratt2009a} (for details see Suhada et al., subm.).  
These X-ray luminosity measurements are the physical key properties of the distant XDCP clusters as these are, by survey design, available for all newly detected systems. The application of  the latest calibration of the   L$_{X}$-T$_{X}$ and  L$_{X}$-M  scaling relations out to  high-$z$ allows subsequent robust estimates of the other fundamental properties  ICM temperature T$_{X}$ and total cluster mass M$_{200}$ (Reichert et al., 2011). Tentative first direct $T_\mathrm{X}$ constraints from X-ray spectroscopy are feasible when several hundred source counts are available, which is the case for about 1/3 of the confirmed distant XDCP clusters based on the available archival data (see Sect.\,\ref{s4_Sparcs} and Table\,\ref{tab_masterlist_Xray}). 


\subsection{Follow-up imaging}
\label{s3_FollowUpImaging}

The task for the follow-up imaging of the second XDCP survey stage is quite challenging: the photometric identification of about 300 X-ray selected $z\!>\!0.5$ cluster candidates with imaging data that has to be sufficiently deep to reach the highest accessible cluster redshifts and to reliably flag the unavoidable fraction of false positives. It is obvious that time and telescope efficient imaging strategies are required to tackle this observational challenge. 

After more than 20 dedicated XDCP imaging campaigns, the data acquisition for the imaging follow-up is now close to completion. In total, we applied and tested five different imaging strategies at five telescopes using eight  optical and NIR imaging instruments: (i) R+z band imaging with VLT/FORS\,2, (ii) z+H imaging with OMEGA2000 at the Calar Alto 3.5\,m telescope , (iii) I+H imaging at NTT with SOFI, EMMI, and EFOSC\,2  , (iv) g+r+i+z+H imaging  at the CTIO Blanco 4\,m with MOSAIC\,II and ISPI , (v) and g+r+i+z+J+H+K$_s$  with the 7-band imager GROND at the ESO/MPG 2.2\,m telescope.  

In the following section, we will discuss cluster identification performance predictions for the different methods, 
provide an overview of a NIR data reduction pipeline developed for the project,  introduce the applied redshift estimation method, and finally evaluate and compare the performance of the R$-$z and z$-$H colors based on our spectroscopic sample.

\subsubsection{Imaging strategies.}
\label{s3_Istrategies}

\begin{figure}[t]
\centering
\includegraphics[angle=0,clip,width=0.495\textwidth]{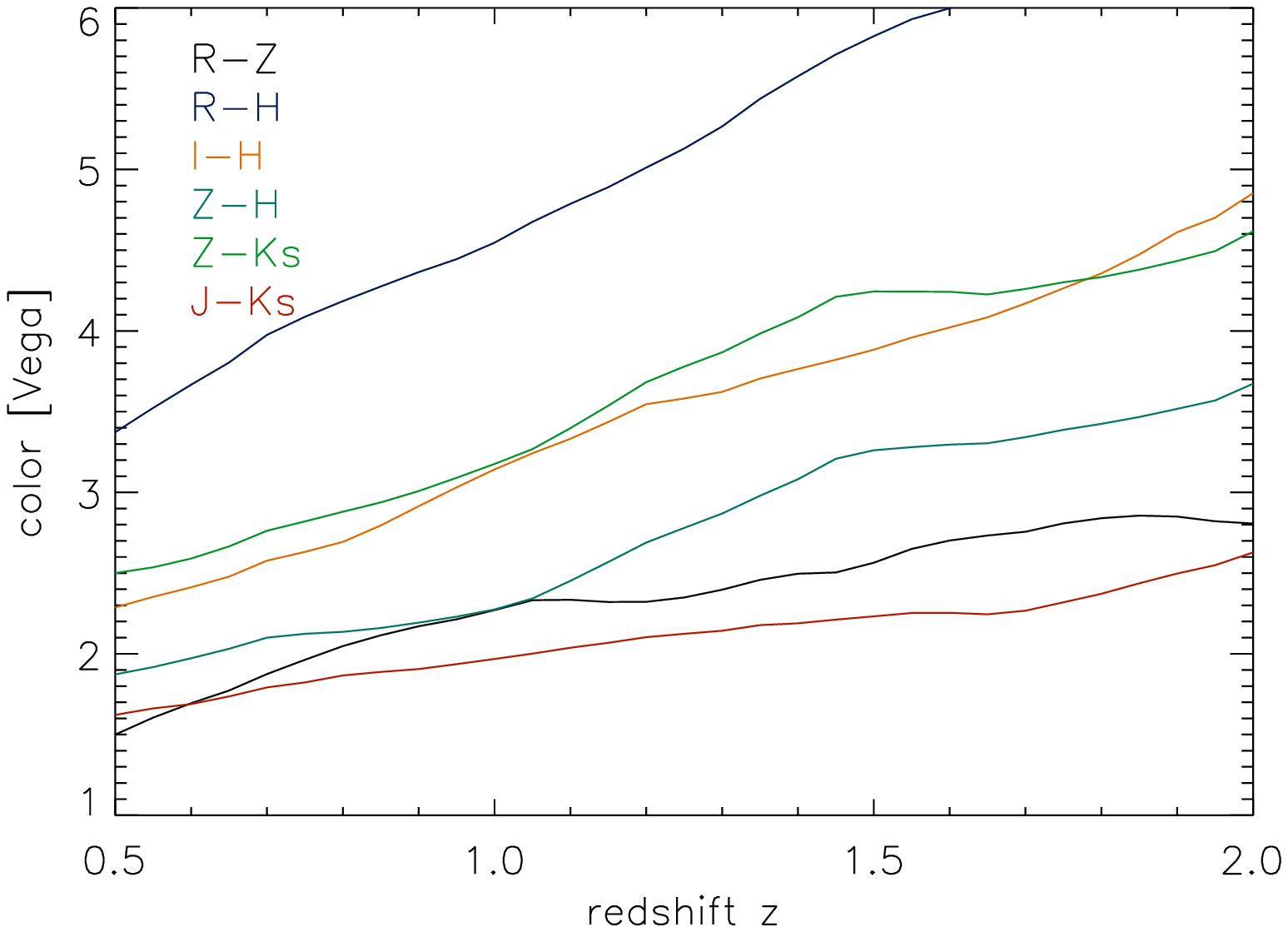}
\includegraphics[angle=0,clip,width=0.495\textwidth]{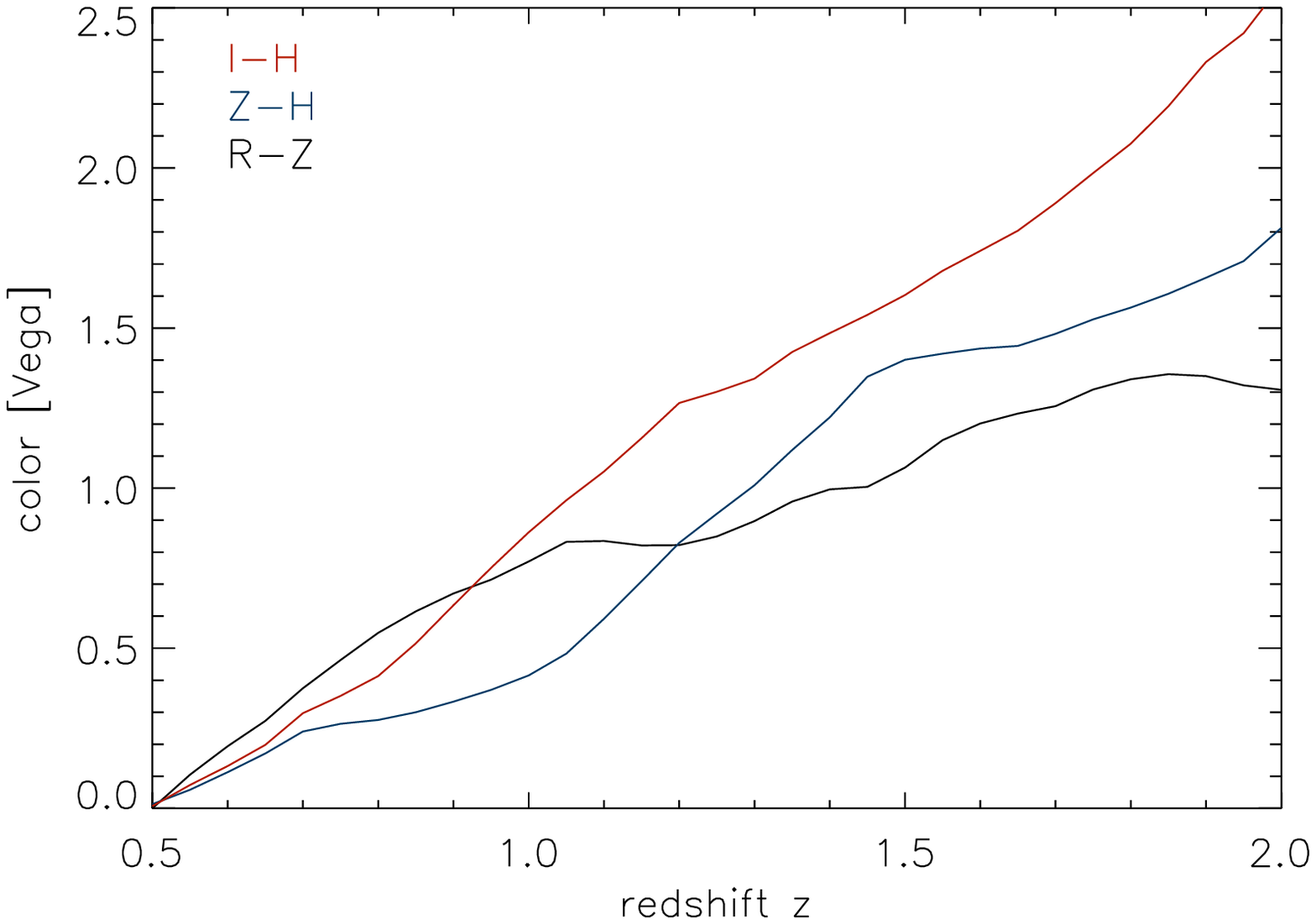}
\includegraphics[angle=0,clip,width=0.495\textwidth]{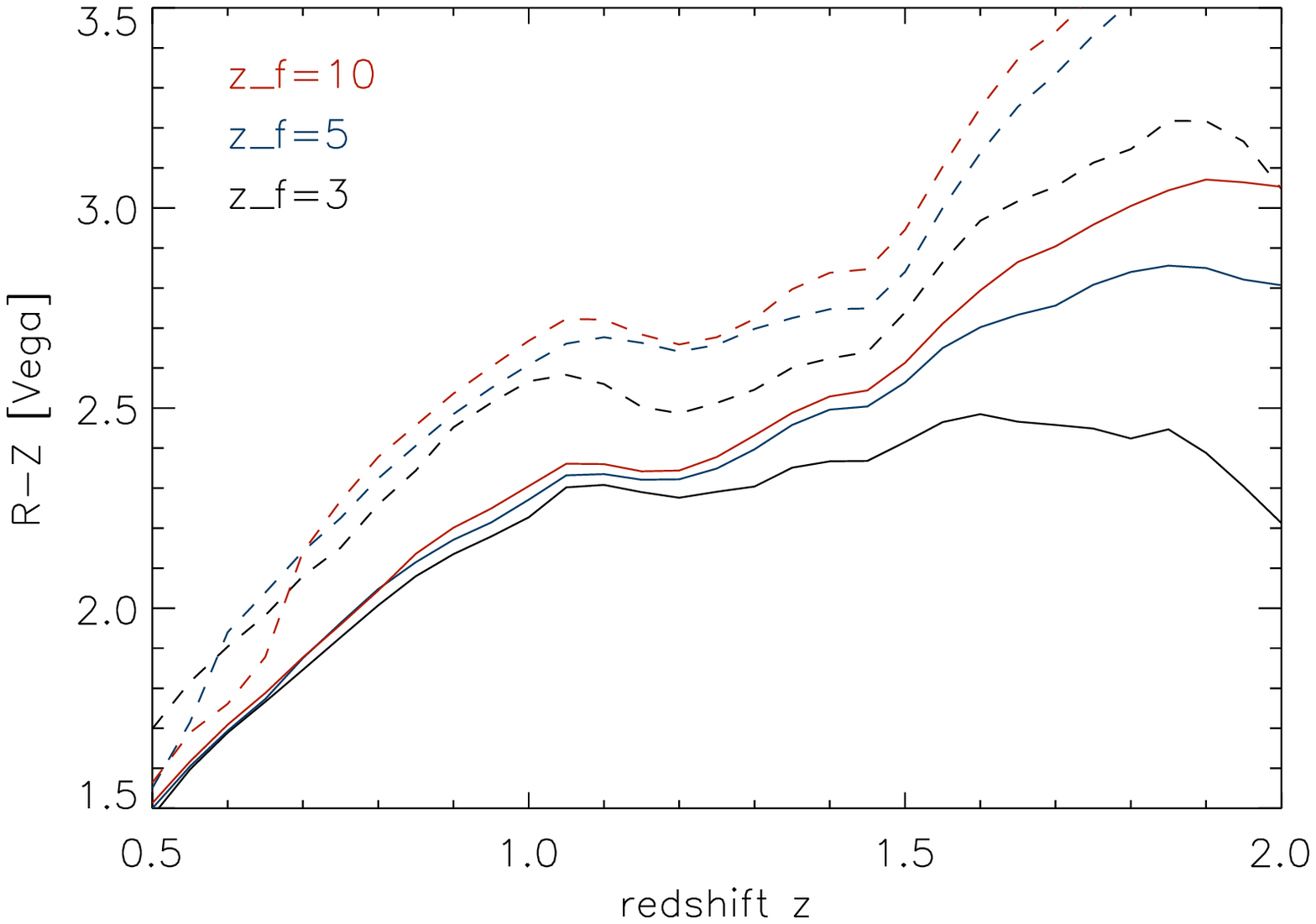}
\includegraphics[angle=0,clip,width=0.495\textwidth]{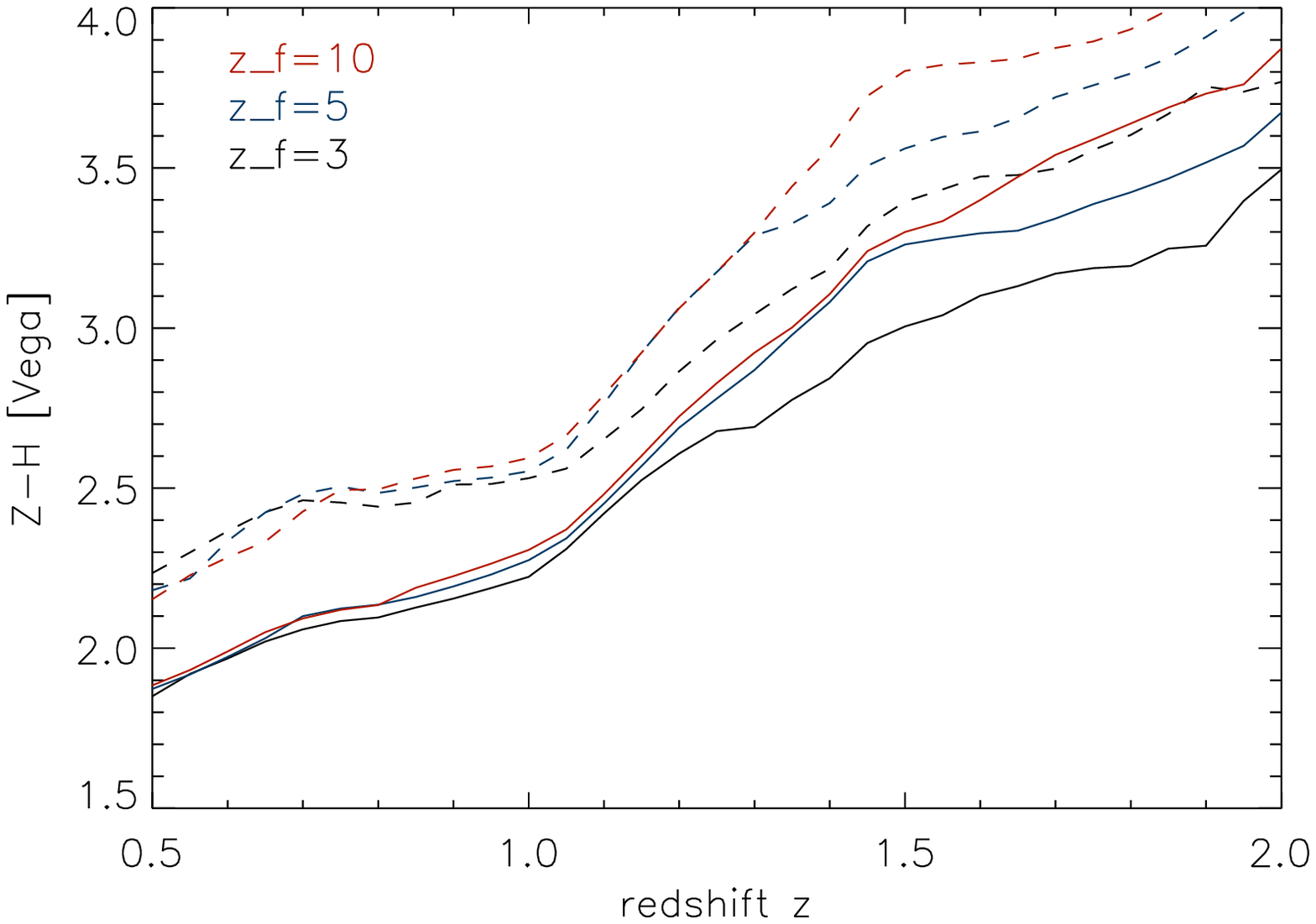}
\caption{
Simple stellar population models for the color evolution of passively evolving galaxies as a function of redshift.
{\em Top left:} Color evolution diagram for a selection of colors based on models with solar metallicity and stellar formation redshift of $z_{\mathrm{f}}\!=\!5$. {\em Top right:} The R$-$z, z$-$H, and I$-$H color shifted to the same origin at $z$=0.5. The slope of the relations determine the redshift sensitivity in the different redshift regimes. {\em Bottom:} SSP model grids of the R$-$z ({\em left}) and z$-$H ({\em right}) color evolution for formation redshifts of three, five, and ten ({\em different colors}) and solar ({\em solid lines}) and three times solar metallicity ({\em dashed lines}). 
}
\label{f7_CMD_models}       
\end{figure}

\begin{table}
\caption{Properties of the main filters used in this section. The second column lists the central wavelengths, columns 3-6 show the expected apparent Vega magnitudes of L* passively evolving galaxies with formation redshift $z_{\mathrm{f}}\!=\!5$ and solar metallicity for four different redshifts, and the last column indicates the additive positive offsets for the conversion to AB magnitudes.}\label{tab_Filters}

\begin{indented}
\item[]\begin{tabular}{@{}lllllll} 
\br

Filter & Center     & m*${(z\!=\!0.5)}$ & m*${(z\!=\!1.0)}$ & m*${(z\!=\!1.5)}$ & m*${(z\!=\!2.0)}$ & m$_{\textrm{AB}}$ (Vega)\\

 &   \microns &  mag  &  mag &  mag  &  mag &  mag \\ 

\mr

R$_{\mathrm{special}}$  & 0.655 & 20.9 & 23.6 & 25.7 & 27.0 & 0.195 \\
I  & 0.798  & 19.8 & 22.2  & 23.8  & 25.4 & 0.440\\
z & 0.90  & 19.4  & 21.4  & 23.3  & 24.2 & 0.521\\
H  & 1.64  & 17.5 & 19.1  & 19.9  & 20.5 & 1.372\\

\br
\end{tabular}
\end{indented}
\end{table}

\begin{figure}[t]
\begin{center}
\includegraphics[angle=0,clip,width=0.66\textwidth]{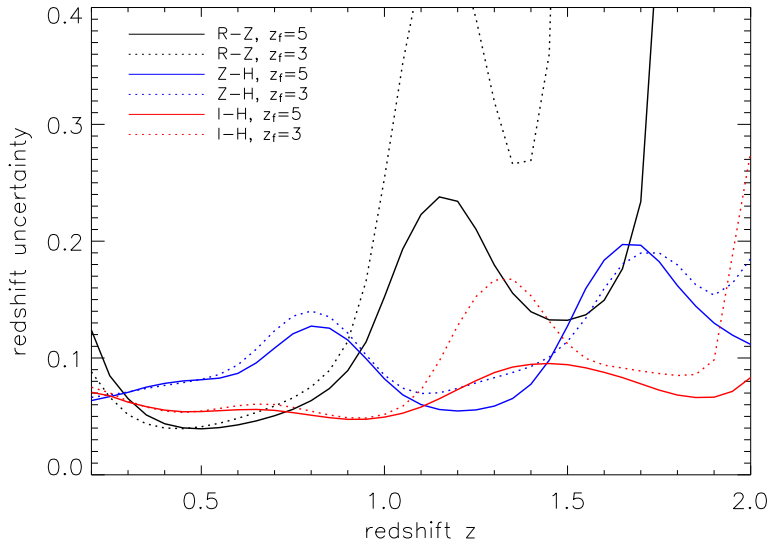}
\end{center}
\vspace{-3.5ex}
\caption{Absolute predicted  redshift uncertainties as a function of $z$ of different \reds \ methods for a 
photometric color error assumption of $\sigma_{\mathrm{color}}\approx 0.05 \cdot (1+z)$\,mag. Error estimates were obtained from the derivatives of the smoothed model colors in Fig.\,\ref{f7_CMD_models} using $\sigma_{z}\approx dz/d(\mathrm{X}-\mathrm{Y}) \cdot \sigma_{\mathrm{color}}$, where $(\mathrm{X}-\mathrm{Y})$ denotes the photometric method. The black solid line illustrates the estimated redshift error for the R$-$z technique under the assumption of a formation redshift of $z_{\mathrm{f}}\!=\!5$, blue shows the z$-$H method, and red I$-$H. The dotted lines use a model formation redshift of $z_{\mathrm{f}}\!=\!3$ for the same methods. 
}
\label{f7_redshift_uncertainties}
\end{figure}


The minimum requirement for the reliable identification of optical counterparts of $z\!>\!0.5$ clusters is a suitable color based on  imaging in two broadband filters (see Table\,\ref{tab_Filters}). Essentially all two-band  imaging techniques used for the identification or selection of galaxy clusters from low to  high-$z$  are variants of the \reds method proposed by    
\citet{Gladders2000a,Gladders2005a}. Based on their predictions, the optical R$-$z color of the cluster \reds was expected to yield reliable redshift estimates out to $z\!\sim\!1.4$. The original XDCP follow-up imaging strategy was based upon this R$-$z method using short snapshot imaging with VLT/FORS\,2 in the z$_\mathrm{Gunn}$ (8\,min) and R$_\mathrm{Special}$ (16\,min) broadband filters. Figure\,\ref{f7_CMD_models} displays simple stellar population (SSP) model predictions based on PEGASE\,2 \citep{Fioc1997a} 
for the observed redshift evolution of the R$-$z color in comparison to other optical/NIR colors (top), and for a model grid of three stellar formation redshifts ($z_f\!=\!3,5,10$) and two metallicities ($Z\!=\!Z_{\sun},3\,Z_{\sun}$).

The limitation that R$-$z  follow-up imaging of targeted  high-$z$ candidates is only efficient with the capabilities and sensitivity of VLT/FORS\,2 led us to develop alternative strategies that are applicable to 4\,m-class telescopes and, at the same time,  provide a higher redshift grasp with the final goal to reach  the $z\!\ga\!1.5$ regime. In order to accomplish both objectives, the used filter combination has to be shifted towards the near-infrared to sample redder parts of the restframe spectral energy distribution (SED) of passive galaxies at redshift beyond unity. The most promising colors with respect to the combination of time efficiency, redshift sensitivity, and redshift limit are z$-$H and I$-$H, which are shown in the upper right panel of Fig.\,\ref{f7_CMD_models} in comparison to R$-$z. Since the achievable accuracy of the \reds based redshift estimate depends on the gradient of the color $d(\mathrm{X}$$-$$\mathrm{Y})$/$dz$   it is immediately evident that  z$-$H and I$-$H are expected to provide significantly improved redshifts at $z\!>\!0.9$. Furthermore, a limiting magnitude of H$_\mathrm{lim}\!\sim\!21$\,mag (Vega) is reachable in less than 1\,h with NIR imagers at 4\,m-class telescopes, corresponding to apparent magnitudes of passive galaxies of m*+1 (m*+0.5) at  $z\!\simeq\!1.5$ ($z\!\simeq\!2$), i.e.~well beyond the characteristic magnitude m* of an L*-galaxy (see Table\,\ref{tab_Filters}).    

To quantify and compare the expected achievable accuracy of \reds  redshift estimates in the presence of photometric and intrinsic color uncertainties 
in an observed color-magnitude-diagram (CMD), we can consider the relation
 $\sigma_{z}\!\approx\!\sigma_{\mathrm{color}} \cdot dz/d(\mathrm{X}\!-\!\mathrm{Y}) $, where $(\mathrm{X}\!-\!\mathrm{Y})$ denotes the photometric method, $\sigma_{\mathrm{color}}$ the observational error of the mean \reds color, and  $\sigma_{z}$ the resulting absolute redshift uncertainty. 
For a realistic photometric color error assumption from good quality data of $\sigma_{\mathrm{color}}\!\approx\! 0.05\cdot(1+z)$\,mag, we obtain the expected absolute redshift uncertainties   shown in Fig.\,\ref{f7_redshift_uncertainties}.
As can be seen, the  z$-$H and I$-$H methods are promising to deliver redshift estimates with uncertainties of    $\sigma_{z}\!\la\!0.1$ all the way to $z\!\sim\!1.5$, while the  high-$z$ uncertainty based on the  R$-$z color is sensitive to the  assumed  stellar formation redshift of the model and  increases dramatically beyond  $z\!\sim\!0.9$, when the 4000\,\AA \ break shifts out of the R$_\mathrm{Special}$  filter.

We implemented and tested the z$-$H imaging technique (Fig.\,\ref{fig_ColorComparison}, right panels) for the identification of  high-$z$ clusters in the year  2006 at the Calar Alto 3.5\,m telescope using the NIR wide field camera OMEGA2000, with results shown throughout this work.
Observations based on the I$-$H method followed from  2007 on at the 3.5\,m NTT with the instrument combination SOFI/EMMI and SOFI/EFOSC\,2. First promising I$-$H results at  $z\!>\!1.5$ are displayed in Fig.\,\ref{fig_HizFrontierClusters} (bottom panels). 
The multi-band approaches pursued at the CTIO Blanco 4\,m with MOSAIC\,II/ISPI and at
the ESO/MPG 2.2\,m telescope with GROND allow the flexibility to make use of all of the discussed colors.

\subsubsection{Near-infrared reduction pipeline.}
\label{s3_NIRpipe}

In the following we provide a brief overview of the  reduction and analysis of the  Calar Alto OMEGA2000 near-infrared data, which is the basis for  many results presented here, in particular for the new systems presented in Sects.\,\ref{s4_012com}\,\&\,\ref{s4_077com}. Details on the reduction of imaging data from other telescopes can be found in \citet{Schwope2010a} for VLT/FORS\,2 (z+R), in  \citet{Santos2011a} for NTT (I+H),  in \citet{Zenteno2011a} for CTIO (griz), and in Pierini et al. (subm.) for GROND data (grizJHK$_s$).  

OMEGA2000 \citep{Bailer2000a} 
is the wide-field NIR prime focus camera at the Calar Alto 3.5\,m telescope with a 15.4\arcmin$\times$15.4\arcmin \ FoV and a pixel scale of 0.45\arcsec \ per pixel. Besides the standard NIR broadband filters J\,H\,K$_s$, the instrument is also equipped with a z-band filter in which the HAWAII-2 detector array still features a high quantum efficiency ($\sim$70\%).
Furthermore, the telescope/instrument system offers an online reduction pipeline \citep{RF2003}, which allows the evaluation of the presence of a distant cluster in real-time (+3\,min) in visitor observing mode. 

The science-grade OMEGA2000 reduction pipeline\footnote[7]{The full OMEGA2000 pipeline is freely available upon request.} was developed for XDCP with a special focus on distant cluster applications, i.e.~faint galaxies.
The full data reduction procedure can be broken up into the
independent processing blocks (i) single image reduction, (ii) image summation, (iii) object mask creation, followed by a second iteration of steps (i)+(ii) with an optimized sky background modeling. 
For the single image reduction the individual 40\,sec (60\,sec) H-band (z-band) exposures are first flatfielded and bad-pixel-corrected. A preliminary, first iteration NIR sky background model is determined from the seven dithered images taken closest to the frame of consideration (i.e.\,$\pm$3 frames) applying a combined median and outlier clipping algorithm in image coordinates for each detector pixel. The modeled background for each image is then subtracted resulting in reduced individual frames.
These individual exposures are then registered in the sky coordinate system by automatically matching reference stars in each image to the underlying masterframe. Deep image stacks in each filter are produced by  co-adding all  
 aligned individual frames with applied fractional pixel offsets while identifying and rejecting cosmic ray events in the process.   
During the stacking process, individual exposures are automatically weighted to yield the optimal SNR in the final stack. Following  \citet{Gabasch2004a}, this  optimal  weight factor in the limit of faint sources scales as $T/(B\cdot\sigma^2)$, where $T$ is the transparency determined from monitoring the fluxes of stars in each frame, $B$ represents the background level, and $\sigma$ denotes the measured seeing .  

The first-iteration summed image stacks in each filter are then used to create an object mask, which flags regions with detectable object flux above the background noise. For the first iteration reduction, the signal of these objects was still in the images used for the sky background model, resulting in determined sky levels which are slightly biased high at these object position,  which in turn translates into a slight background over-subtraction.    
This background bias is overcome in the second iteration of the reduction process, where the object fluxes in each individual exposure are masked out and replaced with the median level of the surrounding unmasked detector area prior to the use of these flux-removed frames as input for the sky modeling process of the time-adjacent  images. This results in the final unbiased reduced single images, which are again stacked in sky coordinates to produce the second iteration final deep image stacks. These final co-added images, based on the discussed double-background subtraction procedure and the optimal weighting process for faint objects, now constitute the basis for the further analysis and distant cluster identification process.

\subsubsection{Redshift estimation.}
\label{s3_Iredshifts}


As a prerequisite for obtaining good distant cluster redshifts estimates, reliable galaxy photometry and color measurements are required as part of the next analysis block. To this end, the final image stacks in the z- and H-band are co-aligned onto identical pixel coordinate grids, i.e.~the same objects in both bands have identical image coordinates.  
As next step, a  deep detection image is created based on the variance-weighted sum of the  z- and H-band stacks, with weighting factors determined from the global background statistics in each frame.
This detection image serves as input frame with maximized depth\footnote[8]{For clusters at $z\!\ga\!1.3$ the detection in H-band may yield improved results.} for the source detection  and for the  green channel  layer of RGB color composites. The  z- and H-band stacks, in which the actual photometric measurements are performed, are then PSF-matched to the 
larger on-frame measured seeing value  of the two bands (typically 0.8-1.5\arcsec) by applying an appropriate Gaussian smoothing kernel (i.e.~$\sigma^{2}_{\mathrm{bad}}\!=\!\sigma^{2}_{\mathrm{good}}\!+\!\sigma^{2}_{\mathrm{smooth}}$) to the frame with better seeing.  As the final preparatory step for the photometry, all frames are  equipped with a
proper equatorial world coordinate system (WCS) from the astrometric plate solution fit with the {\em WCS\,Tools}\footnote[9]{\url{http://tdc-www.cfa.harvard.edu/software/wcstools}} software package.

The actual source photometry is 
performed with {\em SExtractor} \citep{Bertin1996a} run in dual image mode, where the deep detection image is used to find the sources down to faint magnitudes, and the photometric parameters are then extracted directly from the PSF-matched z- and H-band images at the detected source positions. The photometric calibration is achieved in the H-band with stars from the 2-Micron All Sky Survey 
\citep[2MASS,][]{Cutri2003a} directly observed within the large FoV of the science frame.   
The z-band is photometrically calibrated by means of dedicated standard star observations  \citep{Smith2002a} throughout the night, and short photometric overlap observations of the science  field in photometric conditions. 

The  z$-$H  versus H color-magnitude diagram is  constructed from the Galactic extinction-corrected \citep{Schlegel1998a}, i.e.~de-reddened, magnitudes and colors of all galaxies in the FoV, where objects in close proximity (r$<$30\arcsec/60\arcsec) to the X-ray centroid of the candidate source are highlighted (see Fig.\,\ref{fig_012_077com_CMDs}). Total H-band magnitudes ({\tt MAG\,AUTO\/}) are used along the x-axis  since these are directly related to the 
 model predictions. The  z$-$H object  colors are computed from isophotal magnitudes ({\tt MAG\,ISO\/}), which are more accurate for color determinations, since the object flux measurements are restricted to the connected pixels 
above the detection threshold without extrapolations. 

As  the final step, a color-based 
redshift can be estimated from the analysis of  the z$-$H  versus H color-magnitude diagram. Since the location and center of the potential distant cluster is already accurately known from the X-ray centroid, the only unknowns of the candidate system to solve for are the redshift and richness above our limiting magnitude.
One of the main advantages of the X-ray selected XDCP sample is its unbiased nature with respect to the galaxy populations of the systems, which we do not want to give up by requiring a fully developed red-sequence, in particular at the unexplored high-redshift end. Furthermore, the number of detectable cluster galaxies at the limiting depth in $z\!>\!1.3$ systems might be quite small ($<$10) with even fewer accurate color measurements, especially with data taken in poor observing conditions
 or pre-defined exposure times.  A robust redshift estimator for our purposes should thus be able to work with {\em few} cluster galaxies and {\em without requiring} a high signal-to-noise red-sequence.    

For the blind redshift estimation of previously unknown distant cluster candidates, the following objective and reproducible four-step procedure has proven to yield robust results for all used filter combinations: (1) select the third reddest galaxy (3RG) above the magnitude limit detected within 30\arcsec \ from the X-ray centroid, (2) apply a color cut of $\pm$0.3\,mag about the color of 3RG and count the galaxies $N$ within 30\arcsec \ in this color interval and above the magnitude limit, (3) select the central $\sim$68\% percentile of the $N$ galaxies in color space (for $N\!\ge\!4$, otherwise all) yielding  $N_{68}$, (4) determine the minimum (min$_{68}$)  and maximum (max$_{68}$) color of the $N_{68}$ galaxies from which the final best color estimate col$_{68}$= (min$_{68}$+max$_{68}$)/2 is obtained with an associated color uncertainty of $\sigma_{68}$= (max$_{68}$-min$_{68}$)/2. 

The resulting robust color estimator  col$_{68}$$\pm$$\sigma_{68}$ is then compared to the prediction of the input SSP galaxy evolution model to yield the final color-based 
redshift estimate and uncertainty $z_\mathrm{mod}$$\pm$$\sigma_z$ for the candidate system, with a richness estimator  $N_{68}$ and color spread $\sigma_{68}$ that allow conclusions on the existence of a cluster and the presence of a red-sequence in consideration of the magnitude limit of the data set. 
This color estimator is much less demanding in terms of data depth and quality, and with respect to requiring the presence of  an evolved galaxy population for the identification of a distant candidate, compared to actually basing the candidate evaluation on a significant discernible \reds in the CMD of the follow-up data. This is of particular importance when selecting candidate clusters at  $z\!>\!1.4$, for which the observed cluster galaxy population is very faint and the physical \reds is expected to gradually dissolve and eventually disappear (see Sect.\,\ref{s5_HizFrontier}). 

Good redshift estimates for previously unidentified systems based on the color estimator col$_{68}$ rely only on two assumptions: (i) the presence of $\ge$3 red cluster member galaxies brighter than the data limit, whose average color is well represented by the input galaxy evolution model, and (ii) no more than two (background) galaxies located within  30\arcsec \ from the X-ray centroid  (i.e.~the central $\simeq$0.8\,arcmin$^2$), whose colors are significantly redder than the passive member galaxies of the distant cluster candidate.   
The available photometry probes the bright end of the galaxy luminosity function at the cluster redshift, where we expect the SSP models to perform reasonably well.
When extending this  redshift estimation procedure to lower-$z$ calibration clusters at $z\!\la\!0.6$, as in Sect.\,\ref{s3_Rz_zHcomp}, only magnitudes brighter than m*+2 should be 
considered in order to avoid red background galaxies and a too long \reds baseline at the faint end.

By design and purpose, the discussed  color estimation procedure  does naturally not require any available spectroscopic information. The resulting color col$_{68}$ and color uncertainty $\sigma_{68}$ of this empirical approach are hence not necessarily equivalent to the color and scatter of the physical cluster \reds of confirmed early-type passive member galaxies, which are only accessible with high quality data and extensive spectroscopic information. However, in the limit of a discernible, well populated, and tight \reds in the CMD,  col$_{68}$ and $\sigma_{68}$ will converge to the intrinsic physical parameters of the underlying red-sequence.   

Using the extensive spectroscopic information of XDCP (Sect.\,\ref{s3_Spectroscopy}\,\&\,\ref{c5_Sample}), we can now put the most established redshift estimation techniques, R$-$z  and z$-$H, to a critical test and evaluate their performance in practice based on real distant cluster follow-up imaging data.

\subsubsection{Comparison and efficacy of the R$-$z and z$-$H colors.}
\label{s3_Rz_zHcomp}

\begin{figure}[t]
\centering
\includegraphics[angle=0,clip,width=0.49\textwidth]{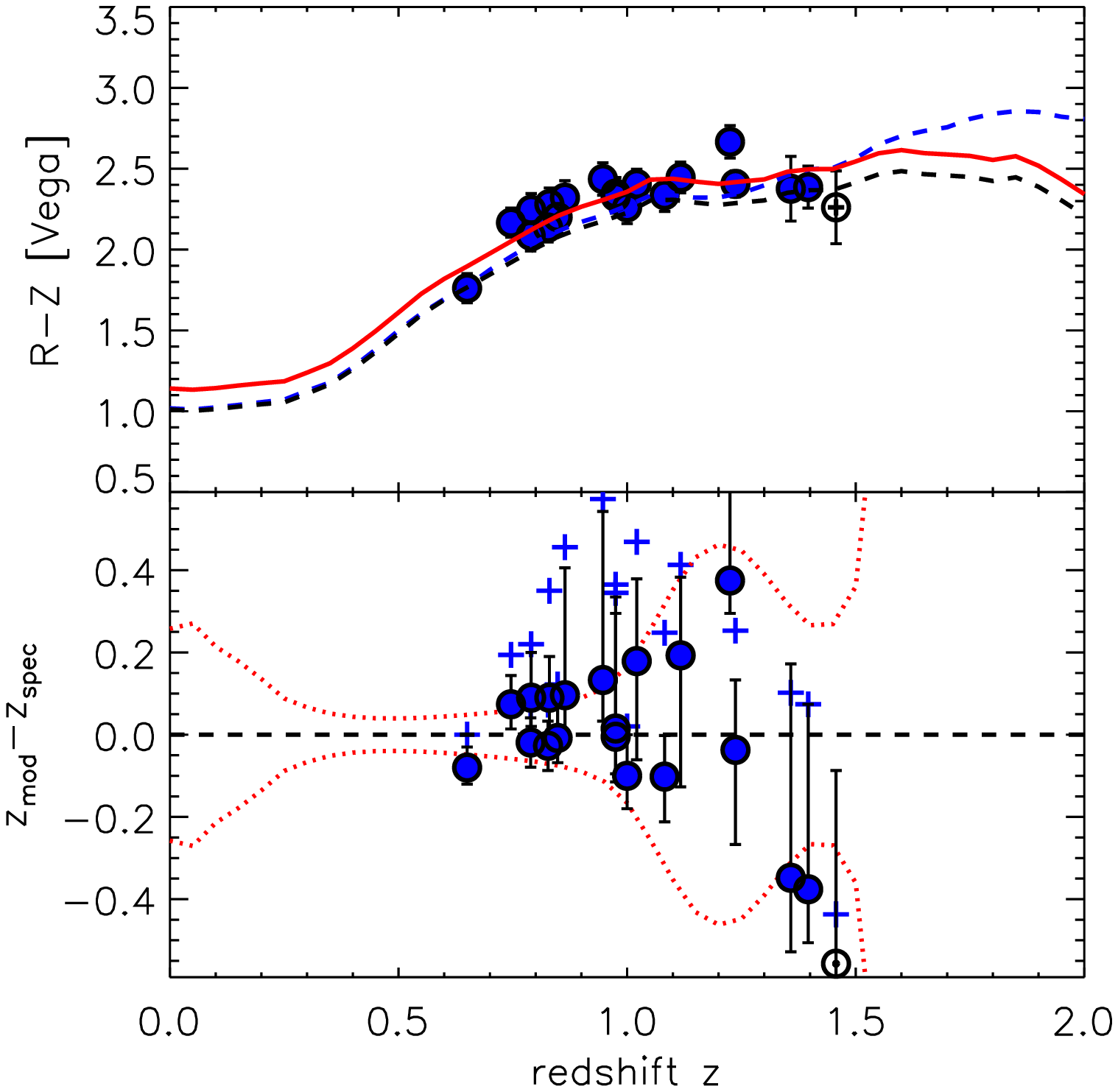}
\includegraphics[angle=0,clip,width=0.49\textwidth]{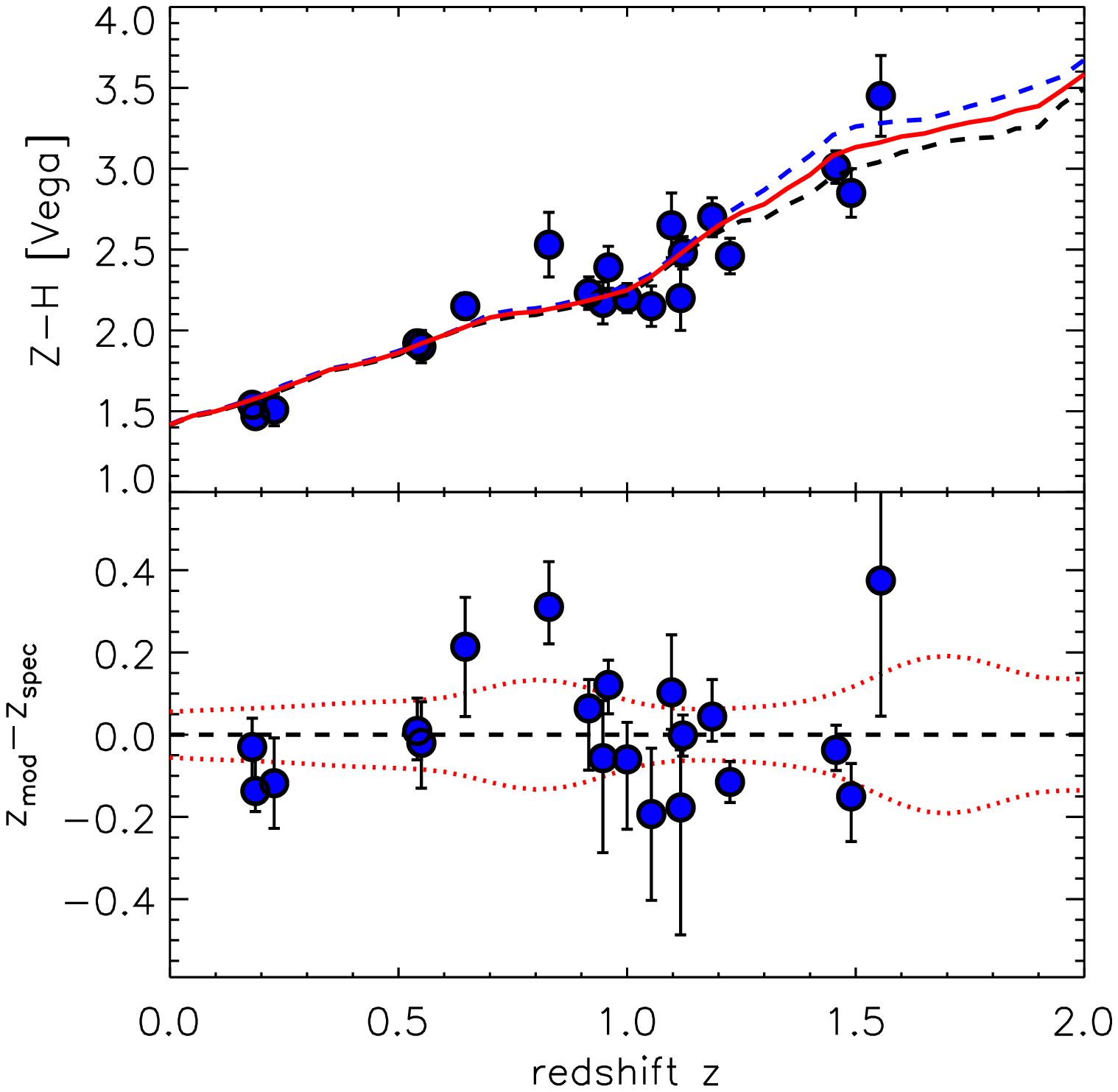}
\vspace{-1ex}
\caption{Comparison of the observed R$-$z (left) and z$-$H (right) colors of spectroscopically confirmed clusters as a function of redshift. The top panels show the measured values col$_{68}$$\pm$$\sigma_{68}$ in comparison to the solar metallicity SSP galaxy evolution models with stellar formation redshift $z_f\!=\!3$ (black dashed line) and $z_f\!=\!5$ (blue dashed line). Red solid lines indicate the best fitting models, which are represented by  a $z_f\!=\!3$ model with a 0.13\,mag positive color offset for R$-$z (left) and the average model for z$-$H (right). The bottom panels show the expected achievable absolute redshift uncertainty based on these models (red dotted line) and the observed redshift offsets of photometric model redshifts $z_\mathrm{mod}$ (with uncertainties) and spectroscopic redshifts $z_\mathrm{spec}$ for each system (blue points).  Blue crosses in the left panel indicate the redshift offsets based on the original model (red dashed line), the open symbol of the highest-z cluster means that no \reds was discernible based on the data.}
\label{fig_ColorComparison}       
\end{figure}

Figure\,\ref{fig_ColorComparison} (top panels)  displays the  measured color col$_{68}$$\pm$$\sigma_{68}$ for spectroscopically confirmed clusters with available   R$-$z (left) or  z$-$H imaging data (right) as a function of the spectroscopic redshift. The 20 confirmed systems with  available FORS\,2 (FoV 6.8\arcmin$\times$6.8\arcmin)  R$-$z data are in their majority targeted XDCP distant cluster candidates, whereas the larger 15.4\arcmin$\times$15.4\arcmin \ FoV of OMEGA2000 enabled the coverage of additional known lower-z calibration clusters at  $z\!\la\!0.6$ in the FoV at no additional observational cost. This yielded also a total of 20 confirmed test objects with z$-$H imaging data, with five overlapping systems present in both data sets  (see Table\,\ref{tab_masterlist_Opt}, column (7)). 

The R$-$z color evolution (col$_{68}$) as a function of redshift (top left panel) shows a low-scatter behavior  with  relatively small photometric uncertainties based on the FORS\,2 data with the targeted exposure depth of 8\,min (16\,min) in z (R). However, as predicted from the SSP models (dashed lines in Fig.\,\ref{fig_ColorComparison} and lower left panel of Fig.\,\ref{f7_CMD_models}) the 
R$-$z color flattens out at $z\!\ga\!0.9$ and eventually turns over  at  high-$z$. As discussed in Sect.\,\ref{s3_Istrategies} and  Fig.\,\ref{f7_redshift_uncertainties} this flattening directly translates into significantly increasing color-based 
redshift uncertainties or even a full model redshift degeneracy.  The observed color-based 
redshift offsets $z_\mathrm{mod}\!-\!z_\mathrm{spec}$ with the derived uncertainty interval are plotted in the lower left panel, together with the discussed predicted absolute uncertainty of the stellar formation redshift  $z_f\!=\!3$ model (dotted lines). The observed widening of the scatter of the redshift difference  with increasing redshift is in very good agreement 
with the prediction, in particular the increasing redshift degeneracy at $z\!\ga\!1$. The open symbol for cluster X2215a at $z\!=\!1.457$  (see Table\,\ref{tab_masterlist_Xray}) indicates that no signature of a \reds (i.e.~$\ge$3 objects at similar red color) could be identified in the  R$-$z versus z CMD, even though the determined  col$_{68}$ from the three reddest galaxies resulted in a reasonable color. Since the physical \reds is present for this system \citep{Hilton2009a}, which is also seen the z$-$H CMD, this indicates that we have surpassed the redshift limit of the  R$-$z observing strategy, which was originally designed to allow cluster identifications up to $z\!\sim\!1.4$
\citep{HxB2005a}.

The best fitting empirical model (red solid line) that yielded the expected behavior for the redshift residuals in the lower  panel is  actually the original solar metallicity, $z_f\!=\!3$ SSP model with an applied R$-$z color offset of +0.13\,mag, empirically determined from the observed data. The original, uncorrected model would (and did) result in the redshift residuals as indicated by the blue crosses in the lower panel, i.e.~a very significant systematic overprediction of the color-based 
redshift estimates of up to 50\%.  
From this it is immediately evident that any applied galaxy evolution model has to be able to predict absolute colors to significantly better than 0.1\,mag in order to yield somewhat reliable redshift estimates at $z\!>\!0.8$ based on the R$-$z color. 
The origin of this observed R$-$z color offset of +0.13\,mag 
could be related to the tabulated transmission functions of the FORS\,2
z$_\mathrm{Gunn}$ and R$_\mathrm{Special}$ filters, or a systematic when calibrating magnitudes observed in  the cut-on z$_\mathrm{Gunn}$ filter to the standard SDSS z-band system by means of SDSS standard star observations. For most of the R$-$z calibration clusters in Fig.\,\ref{fig_ColorComparison} we have results based on
 two independent reduction pipelines, yielding consistent color measurements. Our used SSP color evolution model was also cross-checked with a consistent independent model, providing support for the quality of both the reduction and the model predictions.    
Moreover, a physical explanation for the redder observed colors by invoking super-solar metallicities for the average passive galaxy population (see lower left panel of Fig.\,\ref{f7_CMD_models}) can  be ruled out as well, since such an offset would then also be evident in the right panel for the  z$-$H color.

The observed  z$-$H color evolution on the other hand is in very good agreement with the absolute model prediction over the full probed redshift baseline $0.2\,\la\!z\!\la\!1.55$, fully consistent with both the  $z_f\!=\!3$ (black) and $z_f\!=\!5$ model (blue). As for now, we take the average of these two models as the best fitting model prediction (solid red line). The redshift residuals in the lower panel are in fair agreement  with the predicted general behavior (red dotted lines). 
The z- and H-band observations were taken under average observing conditions significantly worse than for the FORS\,2 data, resulting in photometric uncertainties  which are in some cases larger than the  0.05$\cdot(1+z)$\,mag color error assumed for the redshift uncertainty estimate. Even with these larger observational uncertainties, it is evident that the   z$-$H color clearly outperforms the  R$-$z approach at $z\!\ga\!0.9$, as expected from Fig.\,\ref{f7_redshift_uncertainties}. Reliable z$-$H cluster redshift estimates  have so far been obtained out to  $z\!\sim\!1.5$ and should in principle be extendible  towards even higher redshifts, given the 
presence of such systems with sufficient X-ray brightness  
 within our survey area.  

From the observed color evolution of the tested techniques 
we can confirm that  R$-$z  can indeed  provide accurate color-based
 redshift estimates at $z\!\la\!0.9$, given a sufficiently accurate galaxy evolution model. For clusters at $z\!>\!0.9$ the 
 color-based redshift reliability decreases rapidly due to the flattening of the color function, making spectroscopic follow-up inevitable for the distant cluster candidates we are focussing on. For the aimed at separation of  $z\!<\!0.8$ systems and the identification of $z\!\ga\!0.8$ candidates for spectroscopy based on two-band imaging data, the R$-$z  technique provides an efficient basis up to the limiting redshift of $z\!\sim\!1.4$. 
At $z\!>\!0.9$, the newly established  z$-$H color method provides significantly 
better 
color-based redshift estimates and allows clusters identifications out to  $z\!\ga\!1.5$, whereas the 
uncertainties at the XDCP sample separation redshift of $z\!\simeq\!0.8$ are slightly higher compared to  R$-$z.    
We provide the empirically calibrated,  best fitting R$-$z  and z$-$H  color evolution models as a function of redshift  in text file format as part of  the supplementary material for this paper.

Based on our results, a three-band follow-up approach in R+z+H for future cluster identification projects, e.g.~eROSITA \citep{Predehl2010a},
can provide 
color-based cluster redshift estimates with uncertainties of $\Delta z\!\la\!0.1$ over the full relevant redshift baseline $0.2\,\la\!z\!\la\!1.5$. A similar performance may also be achievable with a two-band approach based on the I$-$H color, which is currently still in the evaluation phase within XDCP.


\subsection{Spectroscopic confirmation}
\label{s3_Spectroscopy}

The third and final stage in the XDCP distant cluster identification process is the spectroscopic confirmation, which is an inevitable and crucial step for all subsequent studies of the $z\!>\!0.8$ galaxy cluster population. The XDCP survey was designed in a way that all potential distant cluster candidate targets are observable with the VLT, with FORS\,2 as the prime instrument of choice for all spectroscopic follow-up work. The excellent red-sensitivity of  FORS\,2 in combination with the multiplexing capabilities with custom-made slit masks allows time-efficient spectroscopic cluster confirmations out to $z\!\ga\!1.5$ with net exposure times of $\sim$3h for the highest-z candidates. 
While the largest possible number of spectroscopic cluster members from single slit mask observations is the  obvious objective,  e.g.~to allow approximate velocity dispersion measurements of the systems, technical and physical limitations  
 make the confirmation process a challenging task, in particular at the highest accessible redshifts. The typical $R_{500}$ radii of the distant cluster candidates are in most cases $\la$1\arcmin \ and the high density core from where the X-ray emission was detected is typically of the order of 30\arcsec. 
 This restricts the slit placement to approximately five within the region of the X-ray emission and ten within  $R_{500}$.
Moreover, the apparent magnitude of cluster galaxies of characteristic luminosity L* is close to the spectroscopic limit for reasonable exposure times once approaching   
$z\!\sim\!1.5$. 

Taking  these challenges into account, we accept a candidate system as a spectroscopically confirmed distant cluster when three conditions are fulfilled: (1) the system was blindly detected as an extended X-ray source, (2) the follow-up imaging revealed a population  of red galaxies (at least 3) coincident ($r\!\la\!1\arcmin$) with the detected X-ray emission, and (3) we find a minimum of three concordant redshifts of associated galaxies.  Since we start with X-ray selected candidates, this strict XDCP definition for confirmed clusters is expected to yield a clean cluster sample concerning the existence of truly gravitationally bound structures.

\subsubsection{Spectroscopic reduction.}
\label{s3_SpecReduction}

The spectroscopic confirmation of newly detected distant X-ray clusters is  one of the prime activities for the current survey phase. In order to allow an efficient and high quality reduction of the spectroscopic data for dozens of systems, we developed a new spectroscopic reduction pipeline called F-VIPGI (Nastasi et al., in prep.), which is the FORS\,2 adaptation of the Vimos Interactive Pipeline Graphical Interface \citep[VIPGI,][]{Scodeggio2005a}.

For the spectroscopic distant cluster confirmations we make use of the 300\,I grism ($\lambda_c\!=\!8\,600$\,\AA), which provides   a resolution of $R\!=\!660$ and a wavelength coverage of
6\,000--10\,500\,\AA. The wavelength coverage on the blue end can be extended down to $\sim$5\,500\,\AA \ when leaving out the standard order sorting filter  OG590, which is in most cases advantageous for the first redshift assessment. Custom made slit masks with  a slit width of 1\arcsec\ and a minimum slit length of 6\arcsec\ allow the placement of about 40 target slits over the 6.8\arcmin$\times$6.8\arcmin \ FORS\,2  FoV. Slits are preferentially placed on color-selected galaxies close to the expected \reds color at the estimated redshift with the highest priority assigned to objects within the detected X-ray emission. Individual exposures are taken with net integration times of 21\,min, whereas the total number of exposures varies from 2 to 10 depending on the estimated system redshift and the faintness of the targeted galaxies. 

The reduction process includes all standard reduction steps, i.e.~bias subtraction, flatfielding, background subtraction, wavelength calibration, extraction of 1-D spectra, and combination of all spectra from the individual exposures including cosmic ray event rejection.  F-VIPGI performs these steps in a semi-automated way with the possibility of interactive quality checks after each process step. The wavelength calibration is achieved by means of a Helium-Argon reference line spectrum observed through the same MXU mask, which allows an absolute calibration with typical rms errors of $\la$\,0.5\,\AA.



\subsubsection{Redshift determination.}
\label{s3_SpecRedshifts}

The final redshifts are obtained by cross-correlating  the reduced spectra with a spectral template library over a wide range of object classes using the 
software packages \texttt{EZ} \citep{Garilli2010a}  and \texttt{IRAF/RVSAO}  \citep{Kurtz1998a}. The best fitting redshift solutions are interactively checked by 
making use of the  graphical VIPGI tools, which allow a simultaneous assessment of the observed spectrum with overplotted redshifted line features, the corresponding sky-subtracted 2-D spectrum, and possible contaminations of observed features related to sky emission lines. 
This way, the final spectroscopic redshift for each galaxy can be determined with typical absolute uncertainties of  $\sigma_z\!\simeq\!(2$-$4)\times10^{-4}$ 
(see e.g.~Table\,\ref{tab_SpecRedshifts_012_077com}), corresponding to a restframe velocity uncertainty of 30-60\,km/s in  $z\!\sim\!1$ systems.

The final system redshift is evaluated by searching for redshift peaks of galaxies in close proximity to the detected center of  X-ray emission. As a first cluster membership classification, galaxies with restframe velocities offsets of $<$3000\,km/s from the redshift peak are considered, corresponding to $\Delta z\!<\!0.01\!\times\!(1\!+\!z_C)$. The systemic cluster redshift  $z_C$ can then be robustly determined as the median of the outlier-clipped redshift distribution of spectroscopic member galaxies.  
For the cases with a sufficiently  high number of identified spectroscopic members ($\ga$8) approximate cluster velocity dispersions are computed by applying the methods of \citet{Danes1980a} and \citet{Beers1990a}.

\section{New distant clusters results}
\label{c4_Results}

In the following section we present results on two newly identified clusters at $z\!\sim\!0.95$ (Sects.\,\ref{s4_012com}\,\&\,\ref{s4_077com}) and 
the  X-ray properties and dynamical state of SpARCS\,J003550-431224  (Sect.\,\ref{s4_Sparcs}).

\subsection{X-ray properties of SpARCS\,J003550-431224 at $z$=1.335}
\label{s4_Sparcs}

\begin{figure}[t]
\centering
\includegraphics[angle=0,clip,width=0.45\textwidth]{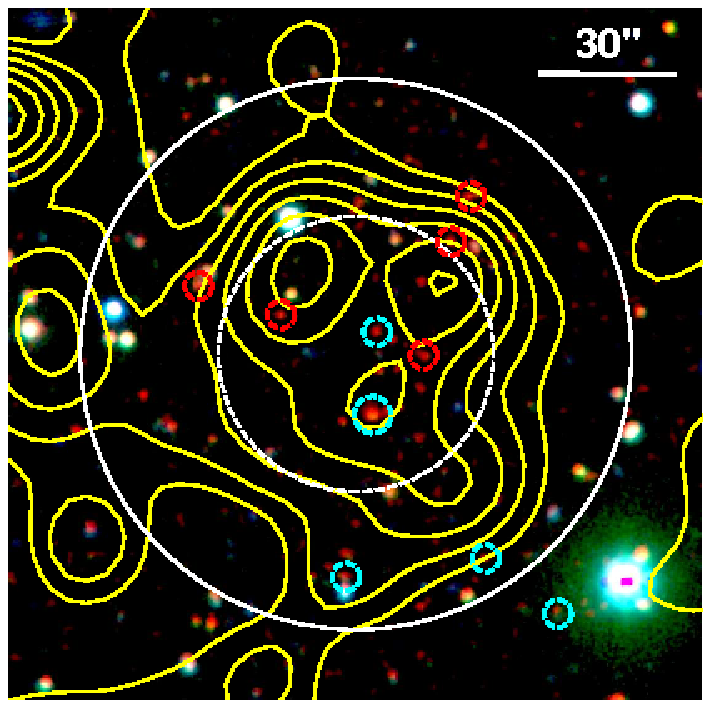}
\includegraphics[angle=0,clip,width=0.54\textwidth]{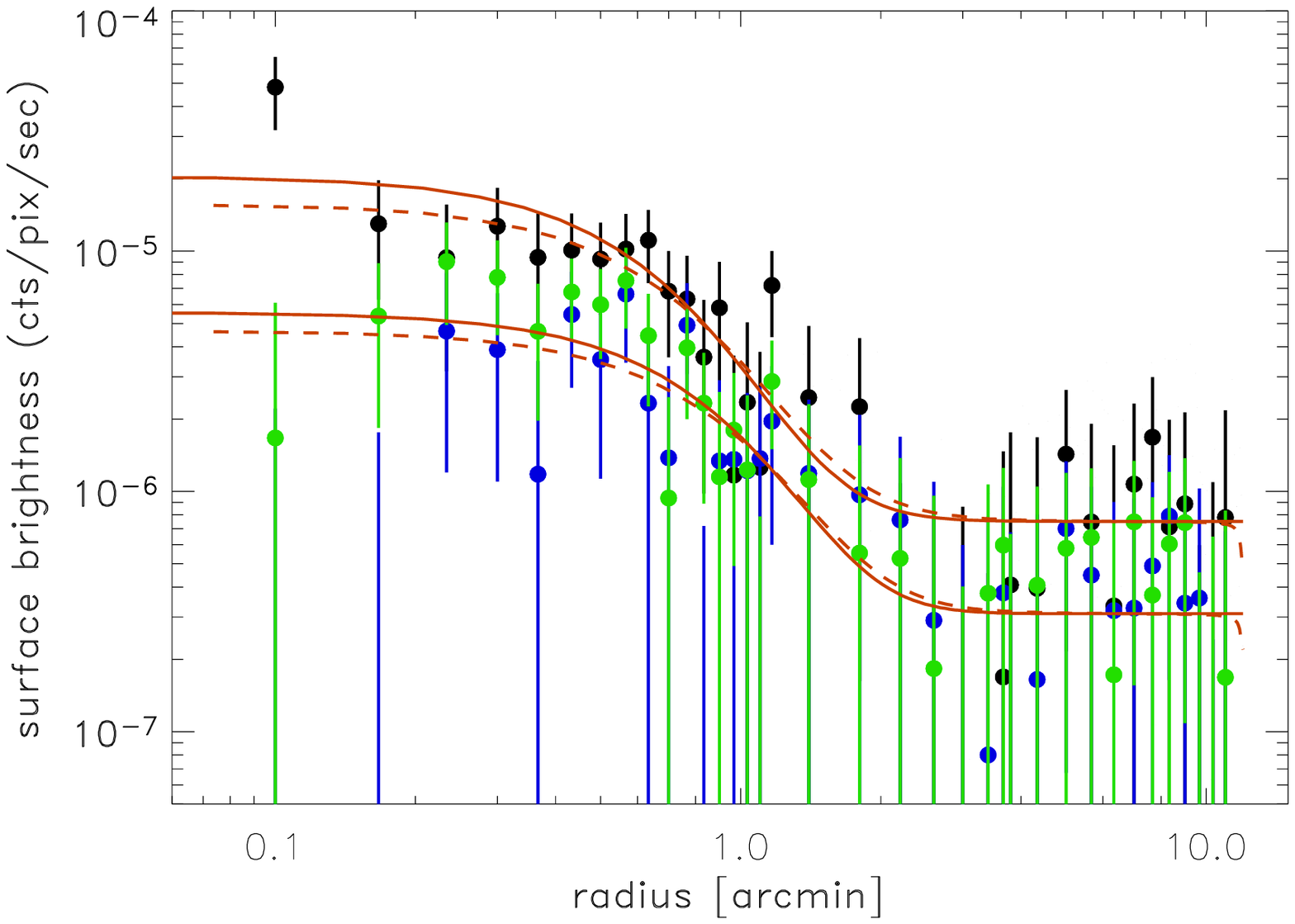}
\vspace{-2ex}
\caption{
Properties of the cluster XMMU\,J0035.8-4312/SpARCS\,J003550-431224  at $z$=1.335. 
{\em Left panel:} Color composite image (gr+iz+H) of the  2.5\arcmin$\times$2.5\arcmin \ cluster environment   with XMM-{\it Newton} X-ray contours overlaid in yellow (North is up, East is to the left). Blue(red)-shifted spectroscopic cluster members with respect to the system redshift are marked by small cyan (red) circles, white circle indicate the 0.5\arcmin \ and 1\arcmin \ radii around the X-ray centroid position. 
{\em Right panel:} XMM-{\it Newton} X-ray surface brightness profile of the cluster's extended emission for the PN (black), MOS1 (green), and MOS2 detectors. Dashed (solid) red lines show the best fit  (PSF-corrected) single $\beta$-model profile for the PN (upper curves) and the combined signal of the MOS instruments (lower curves). 
}
\label{fig_Sparcs_Opt}       
\end{figure}

The galaxy cluster SpARCS\,J003550-431224 was spectroscopically confirmed at a redshift of $z$=1.335 by 
\citet{Wilson2009a} as the currently most distant system within the {\it Spitzer} Adaptation of the Red-sequence Cluster Survey (SpARCS) 
\citep[e.g.][]{Muzzin2009a,Demarco2010a}. This optically rich system was selected within SpARCS based on its \reds in z\arcmin$-$3.6$\mu$m color space and contains 10 spectroscopic members in the range $1.315\!<\!z\!\la\!1.345$ from which a  velocity dispersion value of 1050$\pm$230\,km/s was derived.

Within the XDCP survey, the cluster was independently X-ray selected as the very significantly extended X-ray source XMMU\,J0035.8-4312 at an off-axis angle of 6.3\arcmin \ during the initial  source detection run (Sect.\,\ref{s3_Xdata}) in the XMM-{\it Newton} field with observation ID 0148960101 and an effective clean exposure time of 47.2\,ksec. Owing to the lack of an optical counterpart, the X-ray source was classified as a promising distant cluster candidate and followed-up at the 4\,m CTIO/Blanco telescope with MOSAIC\,II in the g\,r\,i\,z bands on 11 October 2007 and with ISPI in the H-band on 25 October 2007 in good observing conditions. For the visual color composite representation (Figs.\,\ref{fig_Sparcs_Opt}\,\&\,\ref{fig_Gallery}) we co-added the shallower optical images in g (4.2\,min) and r  (10\,min) for the blue channel, and i (20\,min) and z  (11.7\,min) for the green channel, complemented by the 45\,min H-band observation for the red color channel. The core region of the cluster with its rich red galaxy population is depicted in Fig.\,\ref{fig_Gallery}, whereas Fig.\,\ref{fig_Sparcs_Opt} shows the cluster volume to beyond $R_{500}\!\simeq 60\arcsec$ \ (outer white circle) with spectroscopically confirmed members marked as either blue-shifted with respect to the system redshift (cyan circles) or red-shifted (red circles).


Both images have the logarithmically spaced XMM-{\it Newton} X-ray surface brightness contours overlaid in yellow, from which the rather peculiar and irregular 
X-ray  source morphology is evident in comparison to the full distant cluster sample shown in  Fig.\,\ref{fig_Gallery}. Three distinct local surface brightness maxima (left panel of Fig.\,\ref{fig_Sparcs_Opt}) appear to be discernible in the current data within the inner 30\arcsec \ from the X-ray centroid, which is determined as the  `center-of-mass' of the extended X-ray emission. This 
emission is characterized by the most extended surface brightness distribution among all clusters in the presented sample with an effective core radius determined  from the radial profile fit (right panel of Fig.\,\ref{fig_Sparcs_Opt}) of 
r$_{\mathrm{c}}\!\simeq\!34\arcsec^{+15}_{-12}\!\simeq\!280$\,kpc ($\beta$=1.3$^{+1}_{-0.6}$), consistent with the original source detection value of  r$_{\mathrm{c}}\!\simeq\!31\arcsec\!\simeq\!260$\,kpc (for fixed $\beta$=2/3). 
The combination of this large extent with a sufficiently high 
 flux level of f$^{0.5-2\,\mathrm{keV}}_{\mathrm{X,500}}\!\simeq\!(0.80\pm 0.24)\times10^{-14}$\,erg\,s$^{-1}$\,cm$^{-2}$ 
 results in one of the highest extent significances (see Sect.\,\ref{s3_Xsimulations}) for  XMMU\,J0035.8-4312 among the full current distant cluster sample, with  {\tt EXT\,ML}$\simeq$56 and a corresponding formal probability for a spurious extent of $<$$10^{-24}$. 

The observed diffuse and  distorted 
 X-ray morphology of  XMMU\,J0035.8-4312 can best be explained by an ongoing major merger scenario, where at least two main components 
 are in the process of coalescence with bulk flow velocities mainly along the line-of-sight.
This scenario is supported by the bimodal velocity structure of the spectroscopic members reported in \citet{Wilson2009a} with five member galaxies at $z\!<\!1.33$ (cyan circles in Fig.\,\ref{fig_Sparcs_Opt}) and five other members centered around $z\!\sim\!1.34$ (red circles). The median redshifts of the two spectroscopic member bins that are likely associated with different sub-components of the merging process differ by $\Delta z\!=\!0.013$ or a rest-frame velocity offset of $\Delta z\!\simeq\!1700$\,km/s, which is typical for major mergers 
\citep[e.g.][]{Markevitch2007a}.
The velocity sub-structure of XMMU\,J0035.8-4312 is  also visible in the spatial distribution of the spectroscopic members, where the blue-shifted galaxies, i.e.~infalling from behind the cluster, are associated with  the Southern and South-Western extensions of the X-ray emission, whereas the galaxies with positive rest-frame velocities are all located in the Northern half of the system. The brightest cluster galaxy (BCG, larger cyan circle) is associated with one of the local X-ray peaks 13\arcsec \ (109\,kpc) away  from the X-ray centroid and could possibly be the former center of the infalling (blue-shifted) component from the SW radial direction.

\begin{figure}[t]
\centering
\includegraphics[angle=0,clip,width=0.80\textwidth]{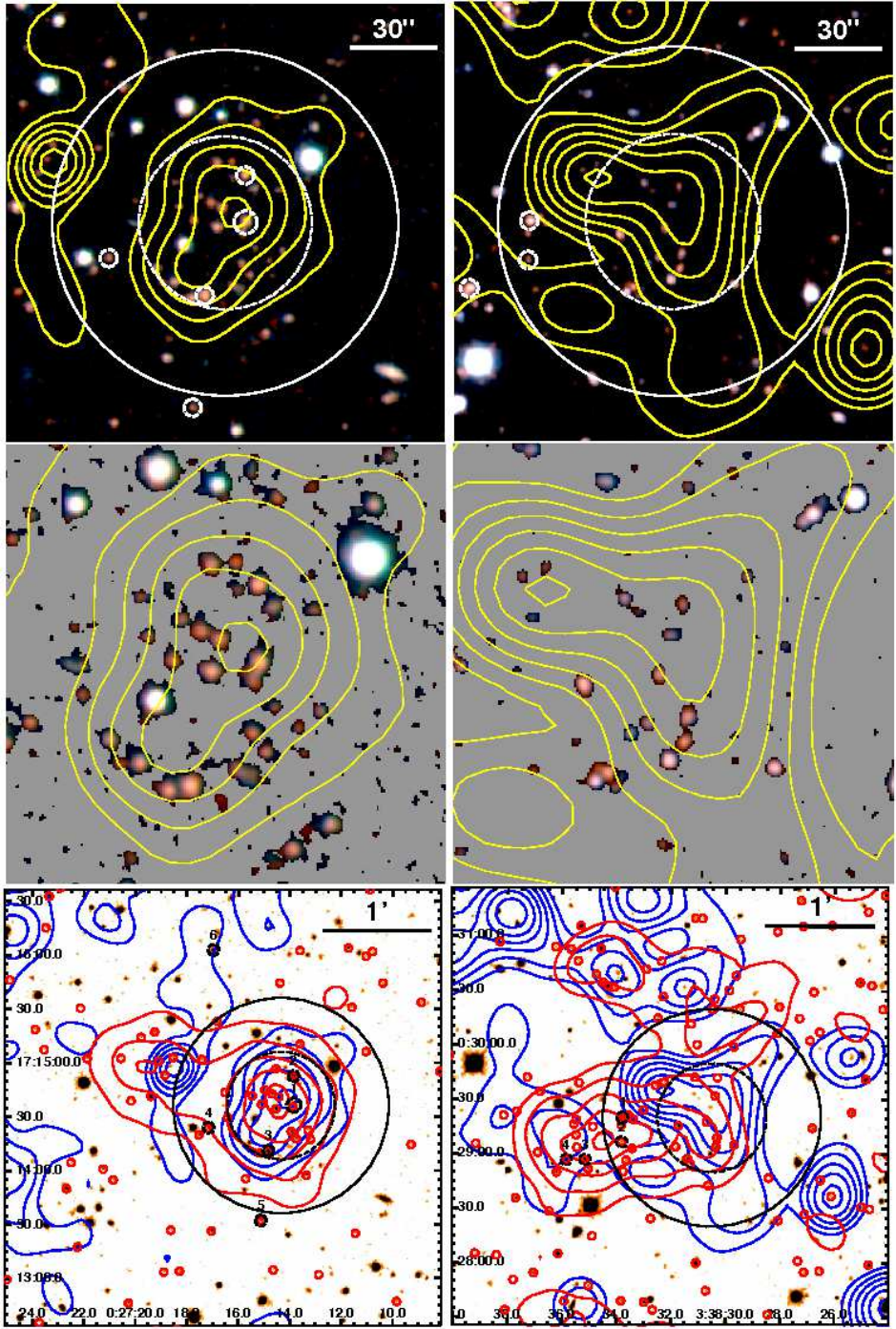} 
\vspace{-1ex}
\caption{Optical and X-ray properties of the clusters XDCP\,J0027.2+1714 at $z$=0.959 (left column) and XDCP\,J0338.5+0029 at $z$=0.916
(right column). {\em Top panels:} 2.5\arcmin$\times$2.5\arcmin \ z+H band color composite images of the clusters with XMM-{\it Newton} X-ray contours overlaid in yellow (North is up, East is to the left).  Spectroscopic member galaxies are indicated by small circles, the two large circles mark the 0.5\arcmin \ and 1\arcmin \ radii around the X-ray centroid position.
 {\em Central panels:} Same as above for a 1.5\arcmin$\times$1.5\arcmin \ zoom on the core region with the black background remapped to gray scale for contrast enhancement. 
 {\em Bottom panels:} 4\arcmin$\times$4\arcmin \ H-band images of the cluster environments with X-ray contours in blue and density contours of color selected galaxies close to the expected \reds color in red, with small red circles indicating the individual galaxies. Black circles have the same meaning as the white ones above.
}
\label{fig_012_077com_Opt}       
\end{figure}

\begin{figure}[t]
\centering
\includegraphics[angle=0,clip,width=0.508\textwidth]{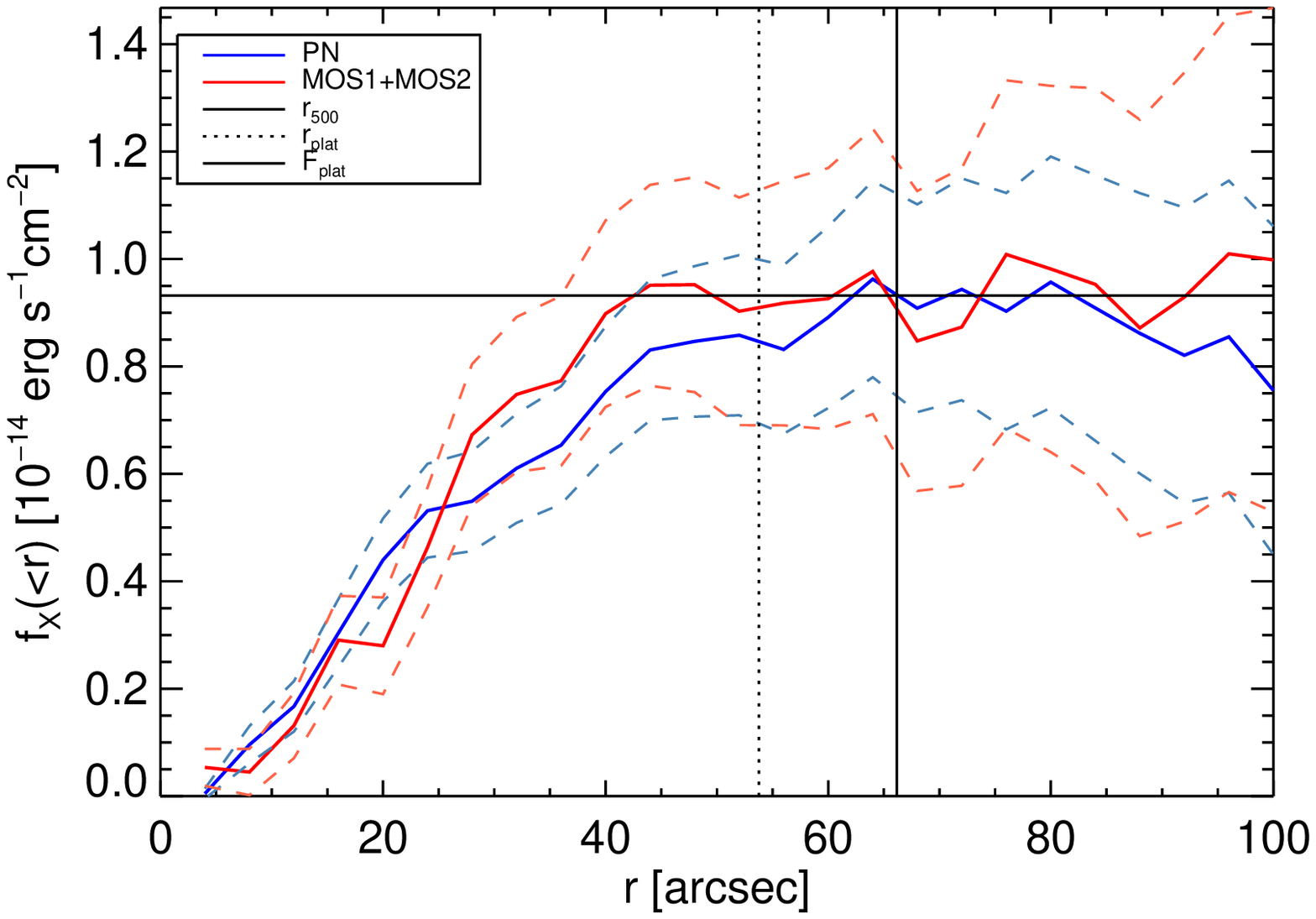} 
\includegraphics[angle=0,clip,width=0.481\textwidth]{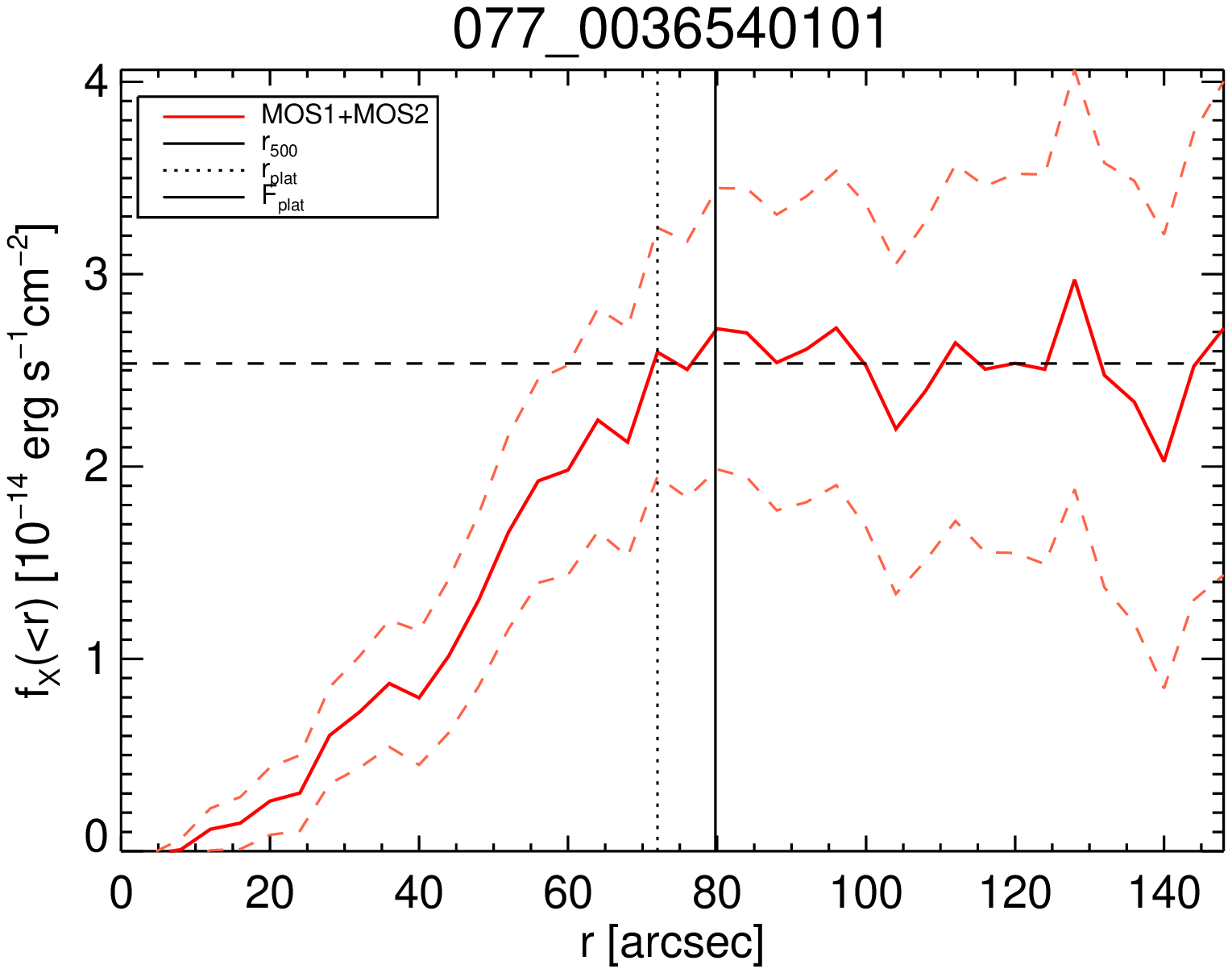} 
\includegraphics[angle=0,clip,width=0.496\textwidth]{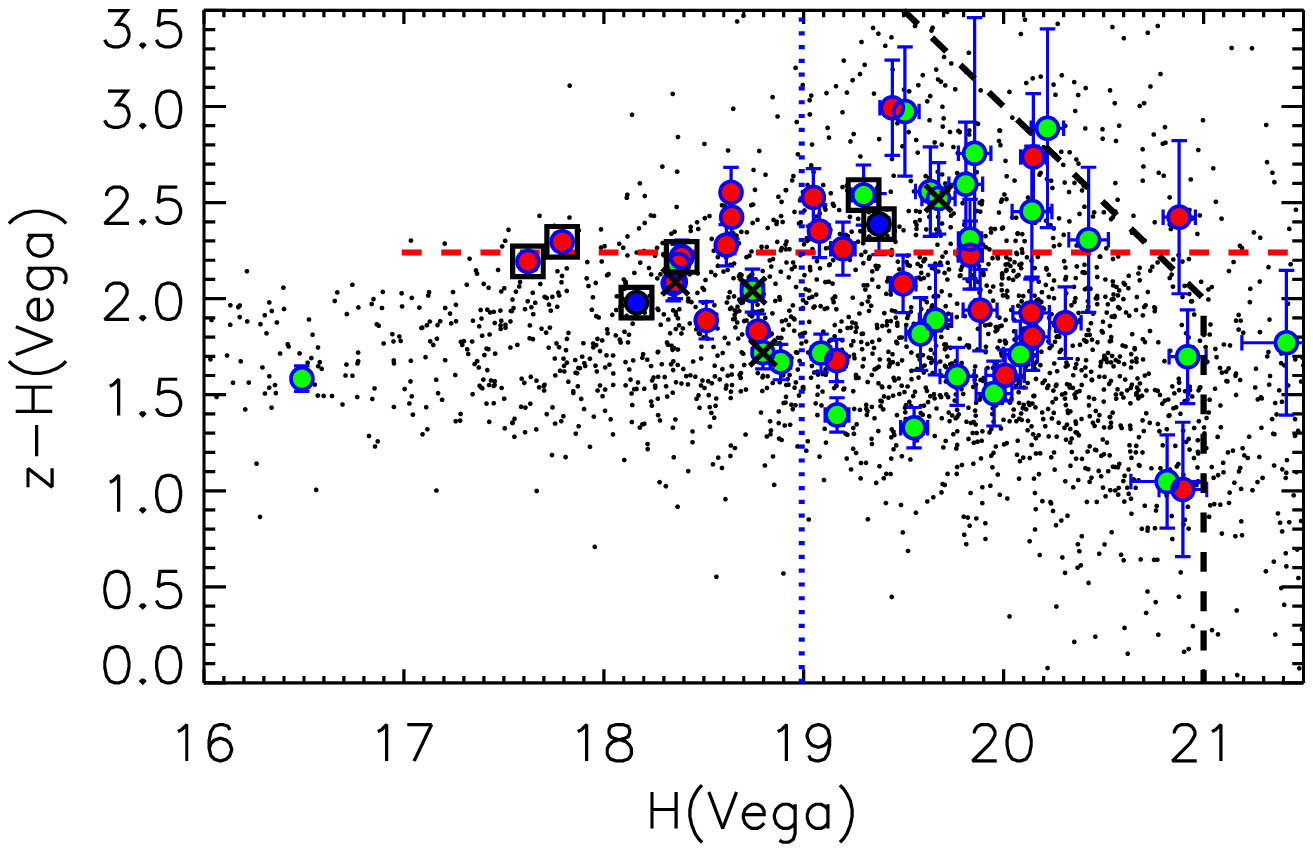} 
\includegraphics[angle=0,clip,width=0.496\textwidth]{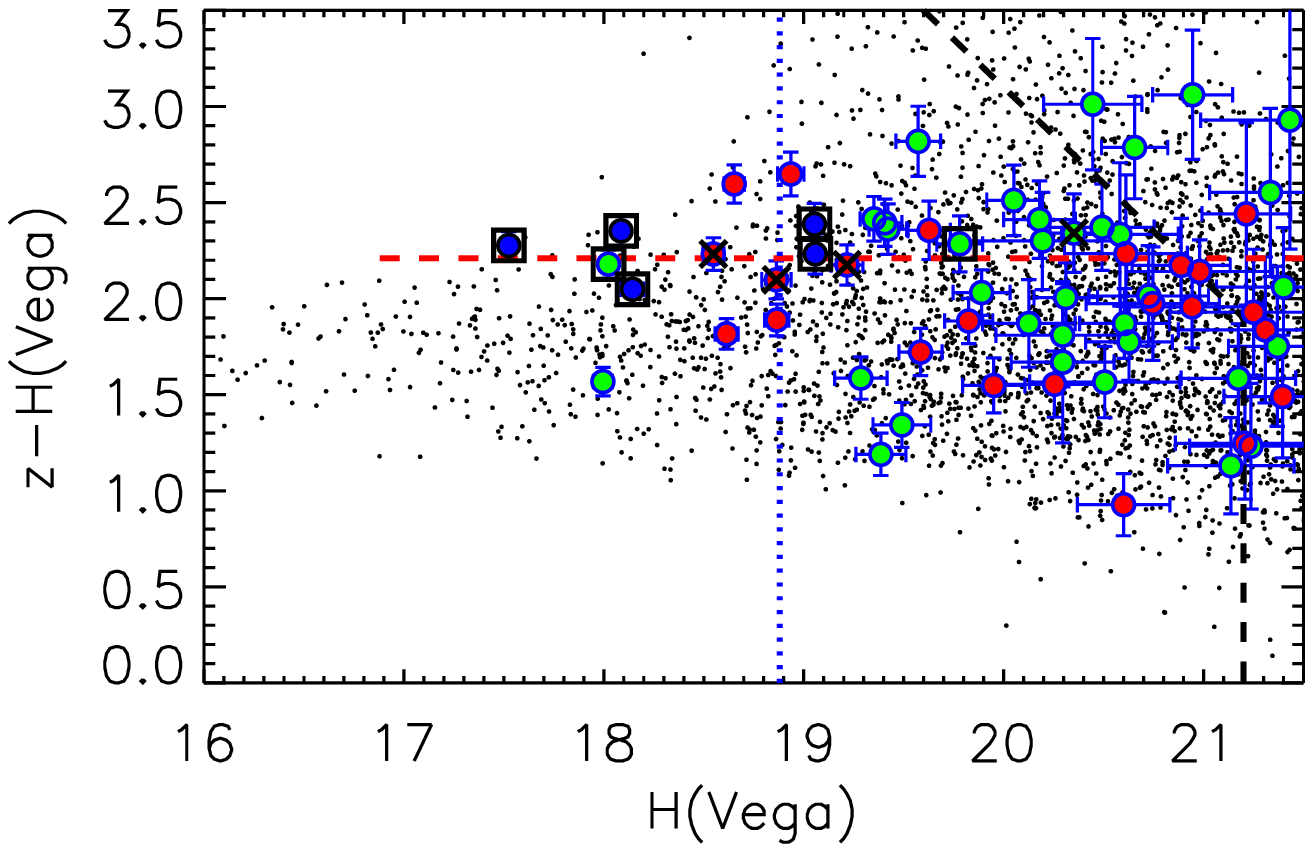} 
\includegraphics[angle=0,clip,width=0.495\textwidth]{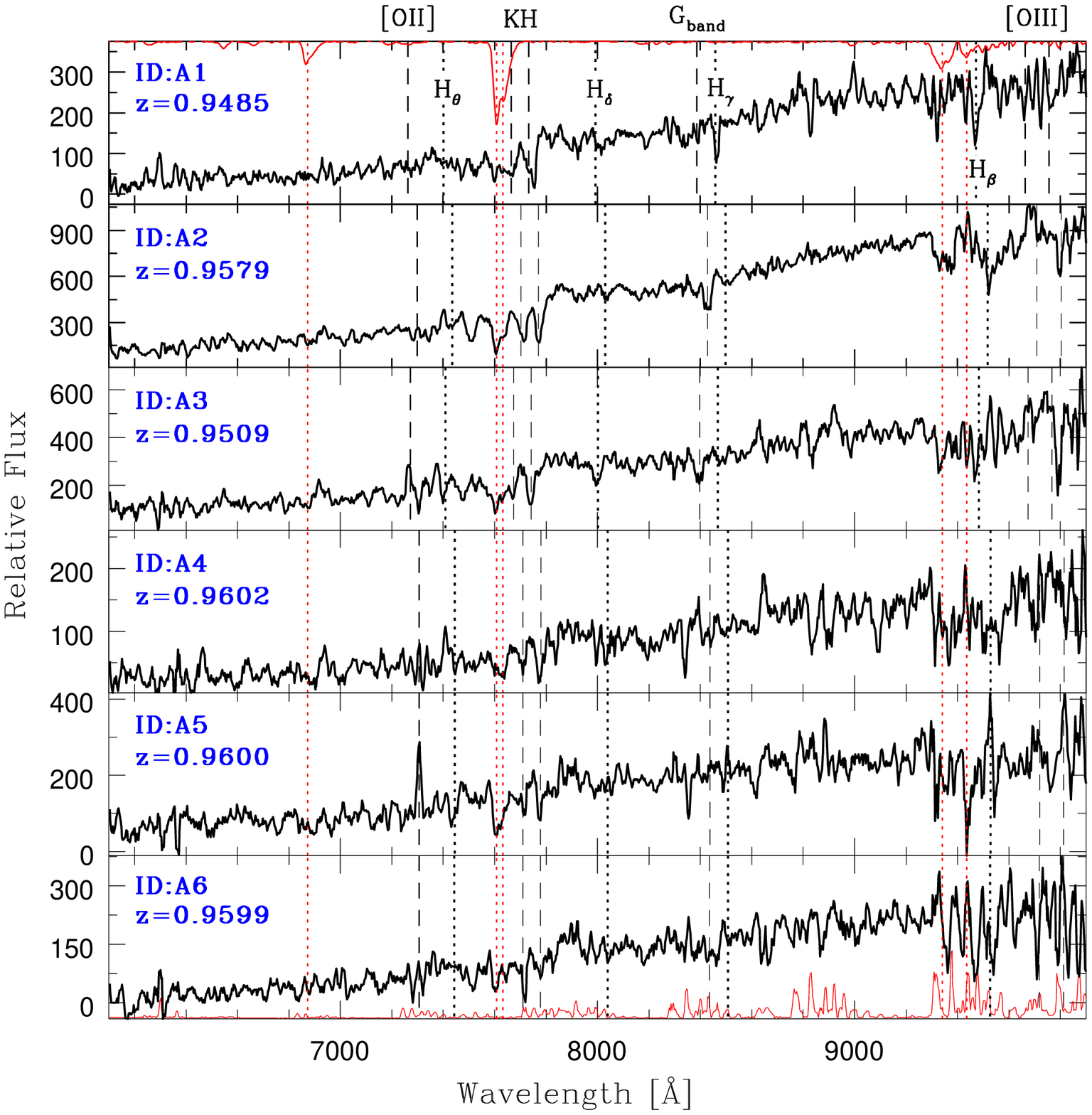} 
\includegraphics[angle=0,clip,width=0.495\textwidth]{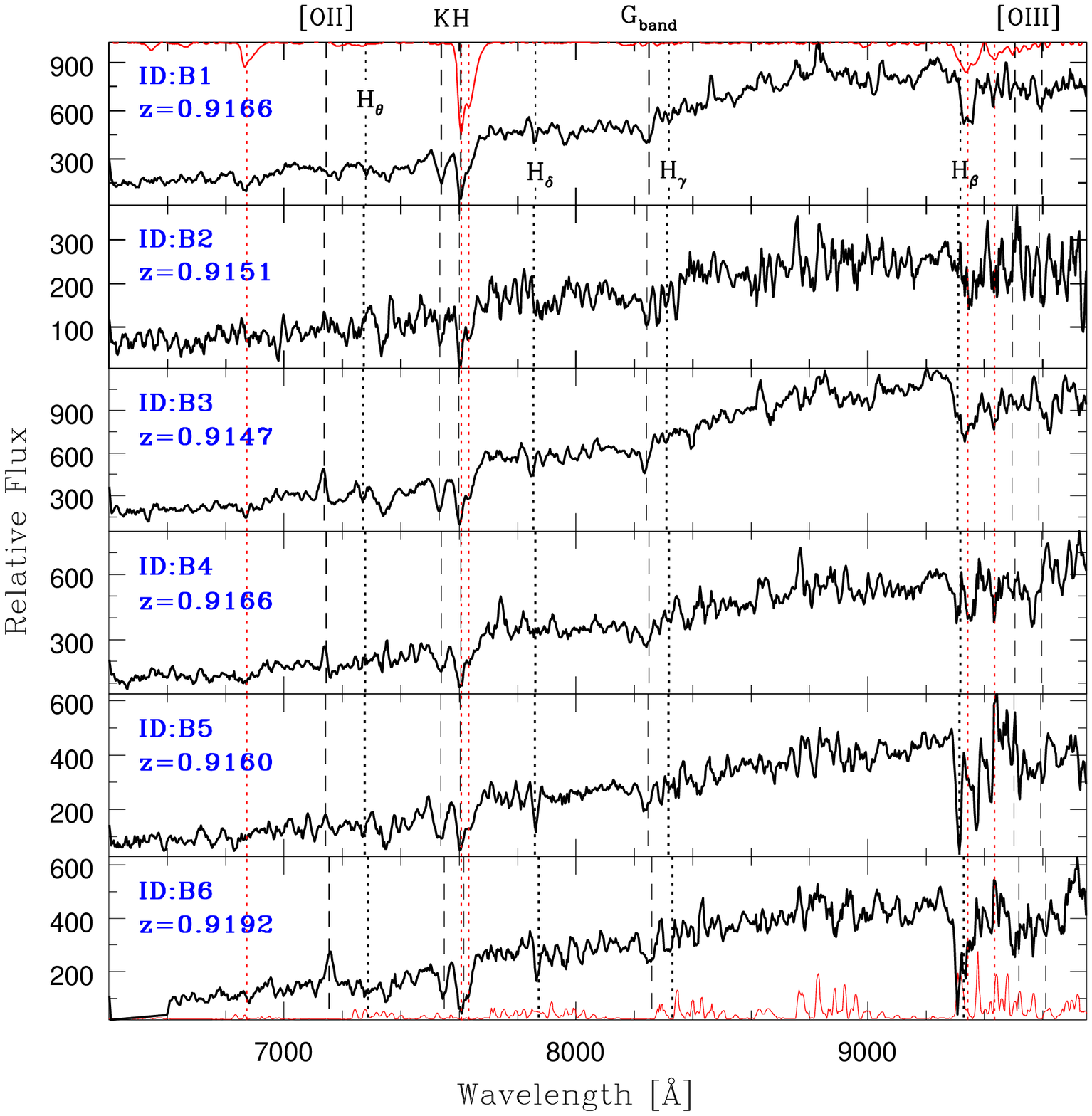} 
\vspace{-4ex}
\caption{Physical properties of the ICM and galaxy populations of the clusters of Fig.\,\ref{fig_012_077com_Opt}. {\em Top:} 
Growth curve of the extended X-ray emission measured for the PN (blue) and MOS detectors (red) 
in the 0.5-2\,keV band. Poisson errors plus 5\% background  uncertainties  are displayed by the dashed lines, the vertical solid (dotted) lines depict the R$_{500}$ (plateau level) radii. 
{\em Center:} z$-$H versus  H CMDs of the cluster fields with galaxies within 30\arcsec \,(60\arcsec) from the X-ray centroid marked in red (green), and spectroscopic members (black squares) at r$>$1\arcmin \ shown in blue. Black lines indicate the 50\% completeness limits, blue lines the H* magnitude at the cluster redshift, and the red dashed lines the expected color of a $z_f$=5 SSP model.
{\em Bottom:} Member galaxy spectra with  indicated redshifted spectral features, IDs correspond to Table\,\ref{tab_SpecRedshifts_012_077com}. Atmospheric absorption (top) and emission (bottom) features are overplotted in red. 
}
\label{fig_012_077com_CMDs}       
\end{figure}

\clearpage

For the determination of accurate X-ray parameters for the system XMMU\,J0035.8-4312/SpARCS\,J003550-431224 we followed the approach of Sect.\,\ref{s3_Xcharacterization} and measure a soft-band luminosity of L$^{0.5-2\,\mathrm{keV}}_{\mathrm{X,500}}\!\simeq\!(0.74\pm 0.22)\times 10^{44}$\,erg\,s$^{-1}$ and a bolometric energy output of L$^{\mathrm{bol}}_{\mathrm{X,500}}\!\simeq\!(1.8\pm 0.5)\times 10^{44}$\,erg\,s$^{-1}$.
With approximately 370 source counts in the soft band the  cluster signal is sufficient to  additionally 
allow a direct spectroscopic ICM temperature determination of  T$_{\mathrm{X}}\!\simeq\!4.5_{-2}^{+4}$\,keV using a local background extraction region close to the cluster  (see Table\,\ref{tab_XSparcs}, left column).

From the measurements of L$^{\mathrm{bol}}_{\mathrm{X,500}}$ and T$_{\mathrm{X}}$  we can derive total cluster mass estimates based on the latest M-L and M-T scaling relations (see Sect.\,\ref{s5_MassEstimates}). The  luminosity-based mass estimate M$^{\mathrm{L_X}}_{200}\simeq 1.7_{-0.5}^{+0.6}\times 10^{14}$\,M$_{\sun}$ and the independent temperature-based one  M$^{\mathrm{T_X}}_{200}\simeq 2.3_{-1.5}^{+4.3}\times 10^{14}$\,M$_{\sun}$ are fully consistent, indicating that the observed merging process does not have a significant influence on the system's location on the L-T relation compared to relaxed clusters. We can hence establish a robust X-ray-based total mass estimate for  XMMU\,J0035.8-4312/SpARCS\,J003550-431224 of 
 M$_{200}\!\simeq\!2\times 10^{14}$\,M$_{\sun}$ ($\pm$40\%), which places the system 
 in the medium mass category for distant clusters discussed in Sect.\,\ref{s5_MassEstimates}. However, these new X-ray mass estimates suggest that the original velocity dispersion-based dynamical mass estimate of  \citet{Wilson2009a} of  M$^{\sigma_r}_{200}\!\simeq\!(9.6 \pm 6.2)\times 10^{14}$\,M$_{\sun}$ is biased high by about a factor of 4-5 as a result  of the presented  evidence for major merging activity preferentially along the line-of-sight.

\subsection{The cluster XDCP\,J0027.2+1714 at $z$=0.959}
\label{s4_012com}

We now present results of the newly confirmed cluster XDCP\,J0027.2+1714 at a redshift of $z$=0.959.
The extended, very significant X-ray source associated with the cluster was detected in the XMM-{\it Newton} field with OBSID 0050140201 and an effective clean exposure time of 41.8\,ksec at an off-axis angle of 11.1\arcmin \ (see Tables\,\ref{tab_XSparcs}\,\&\,\ref{tab_masterlist_Xray}).
The X-ray surface brightness distribution of the system is more compact compared to SpARCS\,J003550-431224 
with a core radius (for $\beta$=2/3) of 
r$_{\mathrm{c}}\!\simeq\!15\arcsec$ ($\simeq$110\,kpc)   and a mostly regular morphology featuring an elongation in the SE-NW direction as shown in Fig.\,\ref{fig_012_077com_Opt} (left panels). The growth curve analysis (Sect.\,\ref{s3_Xcharacterization}) for the source yielded an unabsorbed soft-band flux of 
f$^{0.5-2\,\mathrm{keV}}_{\mathrm{X,500}}\!=\!(0.94\pm0.13)\times10^{-14}$erg\,s$^{-1}$cm$^{-2}$ as shown in the upper left panel of Fig.\,\ref{fig_012_077com_CMDs}.

Imaging follow-up observations (Sect.\,\ref{s3_FollowUpImaging}) in the  H- (50\,min) and z-band (34\,min)   took place  at the Calar Alto 3.5m telescope with the OMEGA2000 NIR camera on 3/4 January 2006 in moderate observing conditions, supplemented by short z-band calibration snapshot observations in photometric conditions on 30 October 2006. The final deep image stacks have an on-frame measured seeing of 1.59\arcsec \ (1.91\arcsec) in H (z) and limiting 50\% completeness Vega magnitudes of  H$_{\mathrm{lim}}\!\simeq\!21$\,mag and z$_{\mathrm{lim}}\!\simeq\!22.7$\,mag, respectively. 
The rich and centrally concentrated galaxy population of XDCP\,J0027.2+1714 is clearly visible in the central and top left panels of Fig.\,\ref{fig_012_077com_Opt}.

The z$-$H versus H color-magnitude diagram is displayed in the central left panel of Fig.\,\ref{fig_012_077com_CMDs} with the bright end of the observed \reds at the expected z$-$H SSP model color of $\simeq$2.24\,mag ($z_f$$=$$5$, red dashed line). Towards fainter magnitudes (H$\ga$19\,mag) the color uncertainties become significant due to the poor  seeing conditions  of 1.9\arcsec  \ for the PSF matched photometry resulting in a broadening of the observed cluster red-sequence.  In order to evaluate the spatial overdensity of galaxies close to the expected \reds color, the color cut 2.0$\le$z$-$H$\le$2.74 is considered, corresponding to the color interval spanning the range 0.2\,mag bluer than the  $z_f\!=\!3$ SSP model to   0.5\,mag redder than the  $z_f\!=\!5$ expectation. This choice of the color interval  is motivated by the decreasing density of background galaxies and the increasing photometric uncertainties towards redder colors in the CMD relative to the expected location of the red-sequence. 
 The spatial distribution of these color selected galaxies is shown in the lower left panel of Fig.\,\ref{fig_012_077com_Opt} (red circles), where the logarithmically spaced red isodensity contours mark  densities  of 6, 9, 13, 20, 29 galaxies per  arcmin$^2$ with a background of  (3.1$\pm$0.4)\,arcmin$^{-2}$. The main red galaxy density concentration coincides with the X-ray emission peak within a few arcseconds, indicating a dynamically evolved main cluster. A secondary galaxy density peak towards the East with a superimposed X-ray point source suggests an infalling structure on the cluster outskirts.


Spectroscopic observations of the cluster environment with VLT/FORS\,2 (Sect.\,\ref{s3_Spectroscopy}) were performed on 6 September 2010  in 1\arcsec \ seeing conditions for a total net exposure of 1.5\,h (run ID: 085.A-0647). Six secure spectroscopic cluster members with a median redshift of $z$=0.959 could be identified, whose locations are marked by circles in the top and bottom panels of Fig.\,\ref{fig_012_077com_Opt}. The spectra are shown in cluster-centric distance order in the bottom panel of Fig.\,\ref{fig_012_077com_CMDs} and properties of the individual member galaxies are given in Table\,\ref{tab_SpecRedshifts_012_077com} (IDs A1-A6). The spectroscopically confirmed BCG (ID A1) is located in close proximity to  the X-ray centroid 
at a projected distance of about 53\,kpc and features a passive spectrum without detectable emission lines. However,  the large observed rest-frame velocity offset of about $-$1600\,km/s relative to the median system redshift suggests that the BCG has not yet settled down to the bottom of the cluster potential well. 
The member galaxies with IDs A3 and A4 at projected distances of 210 and 340\,kpc show weak traces of \OII \ emission. Moreover, galaxy A5 at d$^{\mathrm{center}}\!\simeq\!520$\,kpc exhibits very significant  \OII \ line emission with an equivalent width of about 46\AA.
All three galaxies (A3-A5) are close to or redder than the expected SSP model 
color, which could point towards dusty star formation activity \citep[e.g.][]{Pierini2005a}.

Based on the spectroscopic system redshift, the cluster's soft-band X-ray luminosity can be determined as  L$^{0.5-2\,\mathrm{keV}}_{\mathrm{X,500}}\!=\!(0.40\pm0.06)\times10^{44}$\,erg\,s$^{-1}$ or in terms of the total bolometric energy output   L$^{\mathrm{bol}}_{\mathrm{X,500}}\!\simeq\!(1.0\pm0.1)\times 10^{44}$\,erg\,s$^{-1}$. Applying the scaling relation of  Sect.\,\ref{s5_MassEstimates} 
yields a total mass estimate for the cluster  XDCP\,J0027.2+1714 of 
M$^{\mathrm{L_X}}_{200}\simeq 1.6_{-0.4}^{+0.5}\times 10^{14}$\,M$_{\sun}$.





\begin{table}
\caption{Spectroscopic member galaxies of  XDCP\,J0027.2+1714 at $z\!=\!0.959$ (A, top) and XDCP\,J0338.5+0029  $z\!=\!0.916$ (B, bottom). The table lists for each member galaxy the identification number used in Fig.\,\ref{fig_012_077com_Opt} (bottom),  coordinates, total H-band magnitude, z$-$H color, projected cluster-centric distance d$_{\mathrm{cen}}$, spectroscopic redshift  $z_{\mathrm{spec}}$, and its uncertainty  $\sigma_z$. }\label{tab_SpecRedshifts_012_077com}
\begin{indented}
\item[]\begin{tabular}{@{}llllllll} 
\br
ID & RA     & DEC     & H & z$-$H  &d$_{\mathrm{cen}}$  & $z_{\mathrm{spec}}$  & $\sigma_z$ \\
   &  J2000 &  J2000  &  Vega mag  & Vega mag  &  arcsec             &         &              \\


\mr
A1 &  6.80767 & 17.24345  & 17.62  & 2.19   &  6.7 & 0.9485 &  0.0003  \\
A2 &  6.80770 & 17.24796  & 18.39  & 2.22   & 17.5 & 0.9579 & 0.0002   \\
A3 &  6.81179 & 17.23640   & 17.79  & 2.30   & 26.8 & 0.9509  &  0.0003  \\
A4 &  6.82149 & 17.24001  & 19.30  & 2.54   & 42.5 & 0.9602 & 0.0004    \\
A5 &  6.81302 & 17.22560  &  19.38 & 2.39   & 65.4 & 0.9600 &  0.0002  \\
A6 &  6.82069 & 17.26749  &  18.16 & 1.98   & 94.3 & 0.9599 & 0.0003   \\

\br
B1 &  54.64069 & 0.48888  & 18.02 & 2.18   &  49.0  & 0.9166 & 0.0002  \\
B2 &  54.64089 & 0.48512  & 19.78 & 2.29   & 51.4   & 0.9151 &  0.0003  \\
B3 &  54.64640 & 0.48250  & 17.52 & 2.28   &  73.9  & 0.9147 &   0.0002 \\
B4 &  54.64936 & 0.48243  & 18.09 & 2.35   & 83.4   & 0.9166 &  0.0004  \\
B5 &  54.67213 & 0.47336  & 19.05 & 2.39   & 172   & 0.9160 &   0.0002   \\
B6 &  54.67971 & 0.47059  & 18.14 & 2.05   &  200  & 0.9192 &  0.0003  \\
B7 &  54.68905 & 0.45572  & 19.06 & 2.23   & 253   & 0.9157 & 0.0002   \\

\br
\end{tabular}
\end{indented}
\end{table}

\subsection{The merging system XDCP\,J0338.5+0029 at $z$=0.916}
\label{s4_077com}


The optical and X-ray properties of the system XDCP\,J0338.5+0029 at redshift $z$=0.916 are displayed in the right panels of Figs.\,\ref{fig_012_077com_Opt}\,\&\,\ref{fig_012_077com_CMDs}. The associated X-ray source was detected in the XMM-{\it Newton} field with OBSID 0036540101 with a clean effective exposure time of 18\,ksec at an off-axis angle of 8\arcmin. The  imaging follow-up in z- (53\,min) and H-band (50\,min) took place on  5 January 2006 in good but non-photometric observing conditions with 1.2\arcsec \ seeing during the same campaign as for the cluster in 
Sect.\,\ref{s4_012com}, complemented again by photometric z-band calibration snapshot observations on 30 October 2006. 
The final image stacks reach limiting 50\% completeness Vega magnitudes of  H$_{\mathrm{lim}}\!\simeq\!21.2$\,mag and z$_{\mathrm{lim}}\!\simeq\!23.1$\,mag from which the photometry for the color-magnitude diagram in Fig.\,\ref{fig_012_077com_CMDs} (right central panel) was extracted.
The spectroscopic VLT/FORS\,2 follow-up observations were performed on 9 November 2007 (run ID: 079.A-0634) under moderate 1.5\arcsec \ seeing conditions for a total net exposure time of 2.2\,h and yielded seven spectroscopic cluster members (Table\,\ref{tab_SpecRedshifts_012_077com}, IDs B1-B7).


The situation and configuration for the system XDCP\,J0338.5+0029  is more 
complex 
than for XDCP\,J0027.2+1714 in Sect.\,\ref{s4_012com}. The detected X-ray emission in the case of  XDCP\,J0338.5+0029 has a lower extent significance ($\sim$3.2\,$\sigma$) and an irregular 
morphology with extensions in three directions (see Fig.\,\ref{fig_012_077com_Opt}).
The X-ray centroid is located in close proximity to a chip gap of the PN detector, which had to be discarded for this reason for the growth-curve analysis shown in the top right panel of Figs.\,\ref{fig_012_077com_CMDs}. The results based on the two MOS detectors yielded a flux of f$^{0.5-2\,\mathrm{keV}}_{\mathrm{X,500}}\!=\!(2.5\pm0.7)\times10^{-14}$erg\,s$^{-1}$cm$^{-2}$, which could be biased high due to unresolved point source contributions within the analysis aperture. The derived luminosities  L$^{0.5-2\,\mathrm{keV}}_{\mathrm{X,500}}\!=\!(0.89\pm0.23)\times10^{44}$\,erg\,s$^{-1}$ and L$^{\mathrm{bol}}_{\mathrm{X,500}}\!\simeq\!(2.6\pm0.7)\times 10^{44}$\,erg\,s$^{-1}$ are hence to be interpreted as upper limits, as is the luminosity-based mass estimate of 
M$^{\mathrm{L_X}}_{200}\simeq 2.6_{-0.7}^{+0.8}\times 10^{14}$\,M$_{\sun}$ (Tables\,\ref{tab_XSparcs}\,\&\,\ref{tab_masterlist_Xray}).

Owing to the complex multi-extension X-ray morphology, the effective measured core radius of the X-ray surface brightness distribution is  quite extended ($\sim$190\,kpc), with a value in between the cases of  SpARCS\,J003550-431224 and XDCP\,J0027.2+1714 discussed in Sects.\,\ref{s4_Sparcs}\,\&\,\ref{s4_012com}. 
In order to obtain a better understanding of  the observed  X-ray morphology of XDCP\,J0338.5+0029, it is instructive to inspect the larger-scale environment in the lower right panel of Fig.\,\ref{fig_012_077com_Opt}, where the X-ray surface brightness contours (blue) are plotted together with the color selected galaxy density contours (red) and the distribution of individual red galaxies (red small circles). The latter were derived with the same color selection criterion relative to the SSP model predictions for the system redshift as in Sect.\,\ref{s4_012com}, i.e.~1.97$\le$z$-$H$\le$2.71 in this case, with contours level spanning the range 
9, 12, 16, 22, 30 galaxies per  arcmin$^2$ and a background level of  (5.6$\pm$1.3)\,arcmin$^{-2}$. This representation shows that the complex morphology is  also reflected in the red galaxy distribution with a main density extension to the East and to the North and a connecting pivot point close to the centroid of the detected extended X-ray source. Both Northern and Eastern extensions of the galaxy 
distribution are still within the estimated projected cluster radius of R$_{\mathrm{X,200}}\!\simeq\!940$\,kpc$\simeq$2.0\arcmin \ based on the X-ray mass estimate. Very weak extended X-ray emission 
seems to be present for the main Eastern extension centered at an approximate distance of 1.5\arcmin \ from the main X-ray source centroid, while the X-ray emission towards the Northern extension is dominated by several point sources.  

This configuration suggests ongoing merging activity of at least three main components, similar to the discussed situation of SpARCS\,J003550-431224, with the difference that the bulk motions of the components are along the plane of the sky rather than in the radial direction. A merging configuration very close to the plane of the sky is also supported by the spectroscopic members of the systems, which all exhibit small rest-frame velocity offsets from the median redshifts of $\le$500\,km/s  (Table\,\ref{tab_SpecRedshifts_012_077com}, IDs B1-B7). Four of these members are located in the Eastern extension within 90\arcsec \  from the main X-ray centroid, three others were found beyond the nominal cluster radius towards the same direction. 
The tentatively identified BCG is the galaxy with ID B3 at a projected distance of about 570\,kpc from the determined X-ray center, which is likely part of the infalling Eastern structure. The currently available spectroscopy and confirmed cluster memberships are biased towards this Eastern extension since the MXU mask was centered on this structure to also incorporate the close-by system  XDCP\,J0338.7+0030 (Pierini et al, subm) on the same mask.
Although spectroscopic members in the immediate vicinity of the central X-ray centroid location are currently lacking, a chance superposition of the spectroscopically confirmed red galaxy component with the identified main and Eastern extended X-ray structures seems very unlikely (see Sect.\,\ref{s3_XSurveyArea}), which is also supported by the consistent CMD colors throughout the different components of the cluster environment.
All spectroscopic members are located close to the expected z$-$H SSP model color in the CMD, although most spectra (B2-B6) show indications of weak \OII \ emission pointing towards some ongoing star formation activity (Fig.\,\ref{fig_012_077com_CMDs}, right center and bottom panels), including the tentatively identified off-center BCG. Future multi-wavelength studies of this system will enable a more detailed characterization of this complex but intriguing system.


\begin{table}
\caption{Properties of the galaxy clusters SpARCS\,J003550-431224,  XDCP\,J0027.2+1714, and XDCP\,J0338.5+0029. The specified coordinates refer to the center of the detected X-ray emission. The given core radii r$_{\mathrm{c}}$ and cluster radii R$_{\mathrm{X,200}}$   are approximate values.}\label{tab_XSparcs}
\begin{indented}
\item[]\begin{tabular}{@{}lllll} 
\br
Property & SpARCS\,J0035.8-43  &  XDCP\,J0027.2+17  & XDCP\,J0338.5+00      &     Unit  \\

\mr
RA      & 00:35:50.1  & 00:27:14.3 	 & 03:38:30.5 	  &    \\ 
DEC   & -43:12:10.3  & +17:14:36.3 &  +00:29:20.2 &    \\
$z$     & 1.335 & 0.959 & 0.916  &    \\ 
d$^{\mathrm{center}}_{\mathrm{BCG}}$   & 13 (109) & 6.7 (53) & 73 (573)  & arcsec (kpc)    \\ 
{\tt DET\,ML}     & 76 & 54 & 15  &    \\ 
{\tt EXT\,ML}     & 56 & 24 &  6.6 &    \\ 
r$_{\mathrm{c}}$ ($\beta$=2/3)      & 31 (260) & 15 (110) &  24 (190)  &  arcsec (kpc)  \\
f$^{0.5-2\,\mathrm{keV}}_{\mathrm{X,500}}$     &  0.80$\pm 0.24$ & 0.94$\pm$0.13 & 2.5$\pm$0.7  &  $10^{-14}$erg\,s$^{-1}$cm$^{-2}$  \\ 
L$^{0.5-2\,\mathrm{keV}}_{\mathrm{X,500}}$    & 0.74$\pm 0.22$  & 0.40$\pm$0.06 & 0.89$\pm$0.23  &  $10^{44}$\,erg\,s$^{-1}$  \\ 
L$^{\mathrm{bol}}_{\mathrm{X,500}}$      & 1.8$\pm 0.5$ & 1.0$\pm$0.1 & 2.6$\pm$0.7  &  $10^{44}$\,erg\,s$^{-1}$  \\ 
T$_{\mathrm{X}}$         &  4.5$_{-2}^{+4}$  &  NA & NA  & keV   \\   
R$_{\mathrm{X,200}}$   &  700  & 770 &  940 & kpc   \\ 
 $M^{L_X}_{200}$    &  1.7$_{-0.5}^{+0.6}$ & 1.6$_{-0.4}^{+0.5}$ &  2.6$_{-0.7}^{+0.8}$  &   $10^{14}$\,M$_{\sun}$   \\ 

\br
\end{tabular}
\end{indented}
\end{table}

\subsection{Comparison of cluster configurations}
\label{s4_ClusterComp}

Galaxy clusters observed in the first half of cosmic time can be expected to be actively accreting mass from the surrounding large-scale structure and to exhibit a larger fraction of major merger events caught-in-the-act as part of the hierarchical structure growth process. The three systems presented in this section (see Table\,\ref{tab_XSparcs})  span a wide range of X-ray morphologies and dynamical states: (i) the multi-peaked X-ray emission of SpARCS\,J003550-431224 owing to major merger activity along the line-of-sight (Sect.\,\ref{s4_Sparcs}), (ii) the mostly regular X-ray properties of XDCP\,J0027.2+1714 (Sect.\,\ref{s4_012com}), and (iii) the irregular X-ray morphology and multi-component merging system XDCP\,J0338.5+0029 with bulk flow motions close to the plane of the sky  (Sect.\,\ref{s4_077com}).
 
From an observational point of view, the identification of the latter class (iii), i.e.~merger configurations close to the plane of the sky, is the most challenging at $z\!>\!0.9$ since the X-ray emission may be very irregular and the peaks of the galaxy component and the ICM emission may be spatially separated 
\citep[e.g.][]{Clowe2006a}. Radial merger configurations as in (i), on the other hand, are the easiest class for the observational cluster identification since the projected galaxy density in the vicinity of the extended X-ray emission is heavily boosted, which is also the case for the weak and strong lensing signals for follow-up studies. However, even at  $z\!>\!0.9$ the occurrence of major merging events as presented in  Sects.\,\ref{s4_Sparcs}\,\&\,\ref{s4_012com} is quite low based on the comparison to the observed X-ray morphologies of the full distant cluster sample shown in Fig.\,\ref{fig_Gallery} and discussed in Sect.\,\ref{s5_Xmorphologies}. In this respect, SpARCS\,J003550-431224  and XDCP\,J0338.5+0029 exhibit  the most extreme multi-peak/irregular X-ray morphologies among the whole comparison sample of 22 systems presented in the next section.

\section{The XDCP sample of 22 X-ray luminous galaxy clusters at $\bf z\!>\!0.9$ }
\label{c5_Sample}

Combining the data presented on the three systems in Sect.\,\ref{c4_Results} with previously published results, we can now complete the compilation of 
the largest sample of spectroscopically confirmed X-ray luminous galaxy clusters at  $z\!>\!0.9$ to date and discuss some first statistical characteristics of the  high-$z$ cluster population. This first XDCP distant cluster sample of 22 systems 
increases the size 
of homogeneously selected X-ray clusters in this redshift range by more than a factor of four \citep[e.g.][]{Rosati2000a,Adami2011a} and is still significantly larger in the targeted redshift regime than the recent first data release of the XMM Cluster Survey (XCS) based on the full XMM-{\it Newton} archive \citep{Mehrtens2011a}.

\subsection{The distant cluster sample}
\label{s5_SampleCharacter}

Tables\,\ref{tab_masterlist_Opt}\,\&\,\ref{tab_masterlist_Xray} list the optical and X-ray properties of the present XDCP galaxy cluster sample at  $z\!>\!0.9$. Both tables start with the cluster IDs and the system redshifts for easier cross-referencing of objects. 20 clusters have related publications in the literature (or are submitted/in preparation) that are listed in the last column (21) of Table\,\ref{tab_masterlist_Xray}. Five clusters (C04, C07, C08, C15, C16) were spectroscopically confirmed by other projects, from which the official name of the first publication is listed in column (3).

The stated coordinates in Table\,\ref{tab_masterlist_Opt} (4\,\&\,5) refer to the X-ray centroid of the detected extended X-ray sources, from which all projected cluster-centric distances are measured, e.g.~in columns (8) and (9). 
The given X-ray centroid position is  the first moment (i.e.~the `center-of-mass') of the extended  X-ray emission  as measured during the maximum likelihood source evaluation procedure discussed in Sect.\,\ref{s3_Xdetection}.
The number of spectroscopic cluster members and the follow-up imaging technique and color (see Sect.\,\ref{s3_FollowUpImaging}) are given in columns (6) and (7). The cluster-centric BCG offsets (8) will be further analyzed in Sect.\,\ref{s5_BCGoffsets}, and Sect.\,\ref{s5_RadioProp} discusses the statistics of nearby 1.4\,GHz radio sources listed in column (9). Total mass estimates are either X-ray luminosity based (10) as discussed in Sect.\,\ref{s5_MassEstimates}, or derived from other methods, where available, in column (11).

\begin{figure}[h]
   \centering
    \includegraphics[width=1.04\textwidth, clip=true]{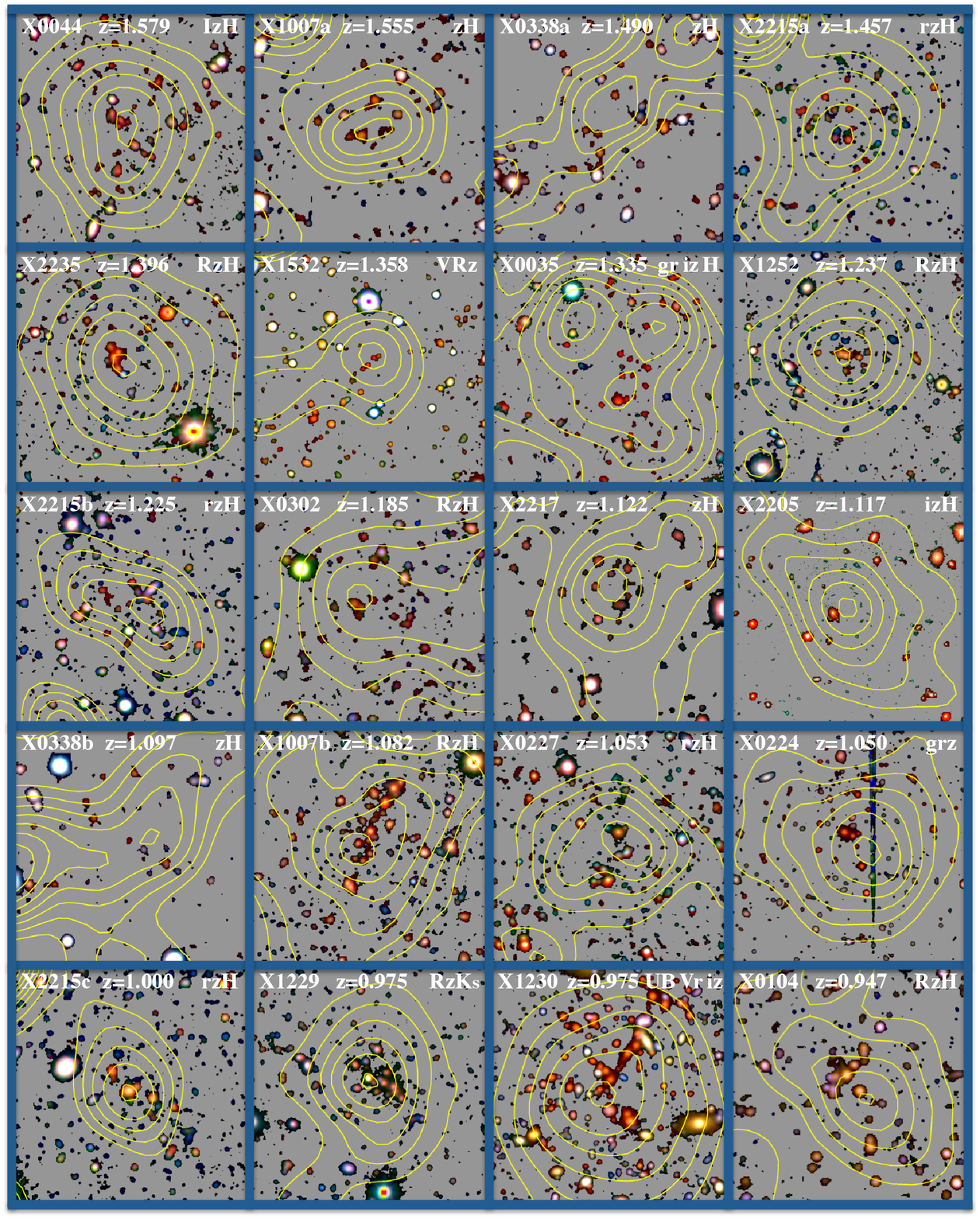}
      \caption{XDCP gallery of the 20 X-ray luminous galaxy clusters at  $z\!>\!0.9$  presented in Table\,\ref{tab_masterlist_Opt} that were not already shown in Fig.\,\ref{fig_012_077com_Opt}. 
Logarithmically spaced XMM-{\it Newton} X-ray surface brightness contours are overlaid in yellow.  Each cluster image is centered on the X-ray centroid location and has a sidelength of 1.5\arcmin$\times$1.5\arcmin \ with the black background remapped to gray scale for contrast enhancement. The top of the panels lists the cluster acronym (see Table\,\ref{tab_masterlist_Xray}), the system redshift, and the bands used for the shown color composite.
      }
         \label{fig_Gallery}
\end{figure}

Table\,\ref{tab_masterlist_Xray} focusses on the X-ray properties of the systems starting with the XMM-{\it Newton} serendipitous source name (12), an acronym (13) used for Fig.\,\ref{fig_Gallery}, and the physical X-ray source parameters (see Sect.\,\ref{s3_Xcharacterization}) soft-band 0.5-2\,keV flux (14), bolometric luminosity (15), and X-ray temperature (16).  The  XMM-{\it Newton}  field observation identifier of the detected serendipitous X-ray source is listed in (17), the effective clean exposure time (ECT, see Sect.\,\ref{s3_Xprocessing}) of the field is given in column (18), and the off-axis angle of the source is stated in (19).


Column (20) in Table\,\ref{tab_masterlist_Xray} lists an overall X-ray quality (XFl), which summarizes the confidence that the detected extended X-ray emission of the source originates predominantly from thermal emission of the ICM. This flag takes into account all presently available information on the source to assign a confidence class based on (i) the original source detection parameters (Sect.\,\ref{s3_Xdetection}), (ii) the more detailed source characterization (Sect.\,\ref{s3_Xcharacterization}), (iii) the imaging data information to check for potentially contaminating objects (Sect.\,\ref{s3_FollowUpImaging}), and (iv) the optical spectra of central sources to probe for AGN signatures (Sect.\,\ref{s3_Spectroscopy}). This evaluation yielded for 17 clusters (77\%) a secure (+++) X-ray 
quality flag, implying a high confidence 
($>$98\%) that the source emission is dominated by thermal ICM emission.

\pagebreak

\begin{landscape}
\begin{table*}
\caption{General properties of the 22 XDCP galaxy clusters at  $z\!>\!0.9$ presented in this work.
The table lists a cluster identification number (column 1), the system redshift (2),  the official cluster name (3), X-ray centroid coordinates (4+5), the number of secure (tentative) spectroscopic members (6), and the imaging color used for the photometric identification (7). The projected cluster-centric distance of the BCGs is given in column (8) with (t) marking tentative identifications, and column (9) lists the closest 1.4\,GHz radio source within a radius of 2\arcmin. X-ray luminosity-based total mass estimates are provided in column (10). 
Other mass estimates from the referenced publications in column (21) of Table\,\ref{tab_masterlist_Xray}, where available, are listed in (11) with the method indicated (T: X-ray temperature-based, HE: hydrostatic equilibrium method, WL: weak lensing, M$_g$: gas mass-based). The entries `lit.' refer to literature references of other projects listed 
in Table\,\ref{tab_masterlist_Xray}.

}\label{tab_masterlist_Opt}
\begin{indented}
\item[]\begin{tabular}{@{}lllllllllll} 
\br          
ID & $z$ & Official Name &  RA & DEC & Specs & Follow-up  &  d$^{\mathrm{center}}_{\mathrm{BCG}}$ & S$_{\mathrm{1.4\,GHz}}$  &    M$^{\mathrm{L_X}}_{200}$  & M$_{200}$  \\
 &  &   & J2000 & J2000 & \# (ten.) & color &  kpc & mJy (d$_\mathrm{cen}$) & $10^{14}$\,M$_{\sun}$ & $10^{14}$\,M$_{\sun}$    \\
(1)  &  (2) &  (3)  & (4)  & (5)  & (6)  & (7)  & (8)  & (9)  & (10)  & (11)   \\
\mr
C01 &	1.579 &		XDCP\,J0044.0-2033 &		00:44:05.2 &	-20:33:59.7 & 3 &   I$-$H&  73 &3.2\,(0.6\arcmin) & 2.9$_{-0.8}^{+1.1}$ &  \\
C02 &	1.555 &		XDCP\,J1007.3+1237 &	 	10:07:21.6 & 	+12:37:54.3 & 3 (1)& z$-$H & 36& 2.2\,(0.1\arcmin)& 1.7$_{-0.5}^{+0.7}$ &  \\
C03 &	1.490 &		XDCP\,J0338.8+0021 &		03:38:49.5 &	+00:21:08.1 & 7 (1) & z$-$H& 176& - & 1.2$_{-0.5}^{+0.6}$ &  \\
C04 &	1.457 &		XMMXCS\,J2215.9-1738 &	22:15:58.5 &	-17:38:05.8 &  lit.& R$-$z, z$-$H& 300(t)& 3.3\,(1.8\arcmin) & 1.8$_{-0.5}^{+0.7}$ & 1.9$_{-0.8}^{+0.6}$ (T)	  \\
C05 &	1.396 &		XDCP\,J2235.3-2557 &		22:35:20.4 &	-25:57:43.2 & 30& R$-$z & 31& - & 4.1$_{-1.0}^{+1.5}$ & 6.6$_{-1.0}^{+1.0}$ (HE/WL)  \\
C06 &	1.358 &		XDCP\,J1532.2-0837 &		15:32:13.2 &	-08:37:01.4 & 3 &R$-$z & 46(t)& - &  1.1$_{-0.3}^{+0.4}$ &  \\
C07 &	1.335 &		SpARCS\,J0035.8-4312 &	00:35:50.1 &	-43:12:10.3 & lit. & grizH  & 109 & 0.2\,(0.2\arcmin) & 1.7$_{-0.5}^{+0.6}$ &  2.3$_{-1.5}^{+4.3}$  (T) \\
C08 &	1.237 &		RDCS\,J1252.9-2927 &		12:52:54.5 &	-29:27:18.0 &  lit. & lit., R$-$z& 11 & 15.3\,(0.8\arcmin) & 3.7$_{-0.9}^{+1.2}$ & 2.9$_{-0.5}^{+0.5}$ (HE)  \\ 
C09 &	1.227 &		XDCP\,J2215.9-1751 &		22:15:56.9 &	-17:51:40.9 &  7 (5) &R$-$z, z$-$H& 57(t) & 3.1\,(0.8\arcmin) &  1.0$_{-0.2}^{+0.3}$ & 0.7$_{-0.2}^{+0.2}$ (T) \\
C10 &	1.185 &		XDCP\,J0302.1-0001 &		03:02:11.9 &	-00:01:34.3 &  6 &z$-$H & 47 & - & 2.1$_{-0.5}^{+0.7}$ &  \\
C11 &	1.122 &		XDCP\,J2217.3+1417 &		22:17:20.8 &	+14:17:54.6 & 7 (3) & z$-$H& 35(t)& 18.0\,(0.2\arcmin)  & 1.8$_{-0.5}^{+0.6}$ & \\
C12 &	1.117 &		XDCP\,J2205.8-0159 &		22:05:50.3 &	-01:59:27.4 &  3 & R$-$z, z$-$H & 57 & -  & 1.8$_{-0.4}^{+0.6}$ & \\
C13 &	1.097 &		XDCP\,J0338.7+0030 &		03:38:44.2 &	+00:30:01.8 & 4 & z$-$H& 347(t)& -&  1.5$_{-0.4}^{+0.5}$ & \\
C14 &	1.082 &		XDCP\,J1007.8+1258 &		10:07:50.5 &	+12:58:18.1 &  19 &R$-$z & 199& 3.6\,(0.8\arcmin)  & 1.7$_{-0.4}^{+0.5}$ &  \\
C15 &	1.053 &		XLSS\,J0227.1-0418 &		02:27:09.2 &	-04:18:00.9 &  lit.& z$-$H& 113(t)& - & 2.0$_{-0.5}^{+0.6}$&  \\
C16 &	1.050 &		XLSS\,J0224.0-0413 &		02:24:04.1 &	-04:13:31.7 & lit. & lit.  & 44& 0.1\,(0.9\arcmin) &   3.3$_{-0.8}^{+1.0}$ &  2.0$_{-0.5}^{+1.4}$ (HE)\\
C17 &	1.000 &		XDCP\,J2215.9-1740 &		22:15:57.5 &	-17:40:25.6 &  10 (2) & R$-$z, z$-$H & 20 & 14.4\,(0.7\arcmin) & 1.1$_{-0.3}^{+0.3}$ & 0.8$_{-0.2}^{+0.2}$ (T) \\
C18 &	0.975 &		XDCP\,J1229.4+0151 &		12:29:29.2 &	+01:51:31.6 & 27 & R$-$z& 8(t)&  - & 4.8$_{-1.2}^{+1.5}$	 & 5.1$_{-1.5}^{+1.6}$ (T) \\
C19 &	0.975 &		XDCP\,J1230.2+1339 &		12:30:16.9 &	+13:39:04.3 & 65 (20) & R$-$z& 134& 1.7\,(0.3\arcmin)   & 4.1$_{-1.0}^{+1.2}$ & 4.2$_{-0.8}^{+0.8}$ (T/M$_{\mathrm{g}}$/WL)\\
C20 &  	0.959 &   	        XDCP\,J0027.2+1714 &         	00:27:14.3 &  	+17:14:36.3 & 6 & z$-$H & 53 &  3.3\,(1.7\arcmin) &1.6$_{-0.4}^{+0.5}$ &  \\
C21 &	0.947 &		XDCP\,J0104.3-0630 &		01:04:22.3 &	-06:30:03.1 &  7 (8) & R$-$z, z$-$H & 30& 11.9\,(0.0\arcmin)  & 2.1$_{-0.5}^{+0.6}$ &  \\
C22 &	0.916 &		XDCP\,J0338.5+0029 &		03:38:30.5 &	+00:29:20.2 & 7  & z$-$H& 573(t)& - & 2.6$_{-0.7}^{+0.8}$  &  \\

\br
\end{tabular}
\end{indented}
\end{table*}
\end{landscape}

\begin{landscape}
\begin{table*}
\caption{Continuation of Table\,\ref{tab_masterlist_Opt} focused on the X-ray properties of the clusters.  The  XMM-{\it Newton} source name is listed in (12), an acronym form in (13), the 0.5-2\,keV soft-band X-ray flux inside the R$_{500}$ aperture in (14), the bolometric cluster luminosity in (15), and the spectroscopic X-ray temperature in (16), where feasible. The XMM-{\it Newton} detection field is listed in (17), the corresponding effective clean time (ECT) of the field in (18), the source off-axis angle in (19), an overall X-ray quality flag (XFl) in (20), and relevant literature references  to the cluster in (21). 

}\label{tab_masterlist_Xray}
\begin{indented}
\item[]\begin{tabular}{@{}lllllllllllllllll} 
\br          
ID & $z$ &  XMM Source Name & Acron. &  f$^{0.5-2}_{\mathrm{X,500}}$/$10^{-14}$ & L$^{\mathrm{bol}}_{\mathrm{X,500}}$/$10^{44}$ &  T$_{\mathrm{X}}$ &  OBSID & ECT & $\Theta_{\mathrm{off}}$ & XFl & References$^a$\\
 &  &   & & erg\,s$^{-1}$cm$^{-2}$ & erg\,s$^{-1}$ &  keV &  & ksec & \arcmin &  & \\
(1)  &  (2) &  (12)  & (13)  & (14)  & (15)  & (16)  & (17)  & (18)  & (19)  & (20) & (21)  \\
\mr
C01 &	1.579 &		XMMU\,J0044.0-2033 &  X0044  & 1.6$\pm$0.3 & 6.1$\pm$1.0& NA & 0042340201 & 8.5 & 10.8 & +++ & Sa11\\
C02 &	1.555 &		XMMU\,J1007.3+1237 &  X1007a & 0.56$\pm$0.11& 2.1$\pm$0.4 & NA & 0140550601 & 19.4 & 10.7 & ++  & Fa11a \\
C03 &	1.490 &		XMMU\,J0338.8+0021 &  X0338a  & 0.30$\pm$0.18 & 1.1$\pm$0.6 & NA & 0036540101 & 18.0  & 5.6 & + & Na11 \\
C04 &	1.457 &		XMMU\,J2215.9-1738 &  X2215a & 1.1$\pm$0.1 & 2.2$\pm$0.3 & 4.1$_{-0.9}^{+0.6}$ & 0106660101 & 51.7 & 9.3 & +++& St06,Hi07/9/10,B10\\
C05 &	1.396 &		XMMU\,J2235.3-2557 &  X2235  & 3.2$\pm$0.1& 10.0$\pm$0.8 & 8.6$_{-1.2}^{+1.3}$ & 0111790101& 13.6 & 8.3 & +++& Mu05,Ro09,J09,S10 \\
C06 &	1.358 &		XMMU\,J1532.2-0837 &  X1532  & 0.29$\pm$0.11 & 0.78$\pm$0.30& NA & 0100240801 & 22.4 & 5.7 & + & Su11 \\
C07 &	1.335 &		XMMU\,J0035.8-4312 &  X0035 &  0.80$\pm 0.24$ & 1.8$\pm 0.5$  & 4.5$_{-2}^{+4}$ & 0148960101 & 47.2 & 6.3 & +++ & Wi09, this work \\
C08 &	1.237 &		XMMU\,J1252.9-2927 &  X1252  & 3.0$\pm$0.4 & 6.8$\pm$1.1 &  6.0$_{0.5}^{+0.7}$ & 0111020201 & 6.5& 14.0 & +++ & Ro04,De07\\
C09 &	1.227 &		XMMU\,J2215.9-1751 &  X2215b & 0.37$\pm$0.04 & 0.55$\pm$0.07  & 2.0$_{-0.2}^{+0.2}$  & 0106660601 & 82.2 & 9.8 & +++ & dHo11,B10 \\
C10 &	1.185 &		XMMU\,J0302.1-0001 &  X0302  & 1.2$\pm$0.1  & 2.2$\pm$0.3 & NA & 0041170101 & 40.9 & 10.7 & +++ & Su11\\
C11 &	1.122 &		XMMU\,J2217.3+1417 &  X2217   &1.0$\pm$0.2 & 1.6$\pm$0.4& NA & 0103660301 & 10.3 & 3.6 & +++ & Fa11c \\
C12 &	1.117 &		XMMU\,J2205.8-0159 &  X2205 & 0.95$\pm$0.15 & 1.5$\pm$0.2& NA & 0012440301 & 24.9& 10.5 & +++ & Da09,Fa11c \\
C13 &	1.097 &		XMMU\,J0338.7+0030 &  X0338b  & 0.71$\pm$0.23& 1.1$\pm$0.3 & NA &  0036540101 & 18.0  & 9.6 & +  & Pi11\\
C14 &	1.082 &		XMMU\,J1007.8+1258 &  X1007b & 0.82$\pm$0.19 & 1.3$\pm$0.3& 5.7$_{-2.2}^{+\infty}$ & 0140550601 & 19.4 & 11.7 & +++& Sc10 \\
C15 &	1.053 &		XMMU\,J0227.1-0418 &  X0227  & 1.1$\pm$0.1 & 1.9$\pm$0.2& 3.7$_{-1.0}^{+1.5}$ & 0112680101& 22.7 & 7.7& +++& An05,Pa07,Ad10 \\
C16 &	1.050 &		XMMU\,J0224.0-0413 &  X0224  & 3.1$\pm$0.2 & 4.6$\pm$0.4&  3.4$_{-0.2}^{+0.3}$ &  0112680301 & 19.2 & 8.9 & +++ &  Pa07,Ma08,Ad10 \\
C17 &	1.000 &		XMMU\,J2215.9-1740 &  X2215c &  0.54$\pm$0.05 &  0.47$\pm$0.05 &  2.1$_{-0.2}^{+0.2}$ & 0106660101 & 51.7 & 7.7 & +++ & dHo11  \\
C18 &	0.975 &		XMMU\,J1229.4+0151 &  X1229  & 6.0$\pm$0.9 & 8.8$\pm$1.5 &  6.4$_{-0.6}^{+0.7}$ & 0126700201 & 8.7 & 12.7 & +++ & Sa09\\
C19 &	0.975 &		XMMU\,J1230.2+1339 &  X1230   & 5.1$\pm$0.5 & 6.5$\pm$0.7 & 5.3$_{-0.6}^{+0.7}$& 0112552101 & 10.3 & 4.3 & +++&  Fa11b/11d, Le11 \\
C20 &  	0.959 &   	        XMMU\,J0027.2+1714 &  X0027  & 0.94$\pm$0.13 & 1.0$\pm$0.1 & NA & 0050140201 & 41.8 & 11.1 & +++ & this work\\
C21 &	0.947 &		XMMU\,J0104.3-0630 &  X0104 &  1.7$\pm$0.3&  1.7$\pm$0.4 & NA & 0112650401 & 18.4 & 5.5 & +++& Fa08 \\
C22 &	0.916 &		XMMU\,J0338.5+0029 &  X0338c  & 2.5$\pm$0.7 & 2.6$\pm$0.7 & NA &  0036540101 & 18.0 & 8.0 & ++ & this work\\

\br
\end{tabular}
$^a$ Sa11: \citet{Santos2011a};
Fa11a: Fassbender et al. (\citeyear{Fassbender2011a}a);
Na11: \citet{Nastasi2011a};
St06: \citet{Stanford2006a};
Hi07/9/10: \citet{Hilton2007a,Hilton2009a,Hilton2010a};
B10: \citet{Bielby2010a};
Mu05: \citet{Mullis2005a};
Ro09: \citet{Rosati2009a}; 
J09: \citet{Jee2009a};
S10: \citet{Strazzullo2010a};
Su11: \citet{Suhada2011a};
Wi09: \citet{Wilson2009a};
Ro04: \citet{Rosati2004a};
De07:\citet{Demarco2007a};
dHo11: de Hoon et al. (in prep.);
Fa11c: Fassbender et al. (in prep.),
Da09: \citet{Dawson2009a};
Pi11:  Pierini et al. (subm.);
Sc10: \citet{Schwope2010a};
An05:\citet{Andreon2005a};
Pa07:\citet{Pacaud2007a};
Ad10: \citet{Adami2011a};
Ma08:\citet{Maughan2008a};
Sa09: \citet{Santos2009a};
Fa11b/11d: Fassbender et al. (\citeyear{Fassbender2011b}b, in prep.);
Le11: \citet{Lerchster2011a};
Fa08: \citet{RF2008b}
\end{indented}
\end{table*}
\end{landscape}

\clearpage

Two clusters (9\%) have assigned intermediate (++) X-ray confidence flags, with a non-negligible probability of up to 20\% that non-thermal emission processes may be major contributors ($\ga$50\%) to the detected X-ray flux. System C02 at  $z\!=\!1.555$  (Fassbender et al. \citeyear{Fassbender2011a}a) with an extent significance of $\sim$4\,$\sigma$ hosts a radio galaxy in the center which could emit non-thermal X-rays, and C22 at $z\!=\!0.916$ with its complex configuration was discussed in Sect.\,\ref{s4_077com}.

A lower X-ray confidence flag (+) was given to three systems (14\%), where the probability of possible predominant non-thermal X-ray emission appears to be at levels $>$20\%. These sources (C3, C06, C13) are from the supplementary X-ray sample  (Sect.\,\ref{s3_Xdetection}) and were originally selected with extent significances of   2-3\,$\sigma$, i.e.~very close to the detection threshold. All three systems feature a red galaxy population peaked within 30\arcsec \ from the X-ray centroid, of which 3-7 galaxies are spectroscopically confirmed  members \citep[][ Pierini et al., subm.]{Suhada2011a,Nastasi2011a}, 
in the case of XDCP\,J1532.2-0837 (C06) with signatures of some central AGN activity. The properties of such low-mass ($M_{200}\!\la\!1.5\times 10^{14}$\,M$_{\sun}$), high redshift ($z\!\ga\!1.1$) systems and their X-ray point source contents are currently unexplored territory and will require further investigations.

\subsection{Redshift distribution and mass estimates}
\label{s5_RedshiftSampling}
\label{s5_MassEstimates}

\begin{figure}[t]
   \centering
    \includegraphics[width=11cm, clip=true]{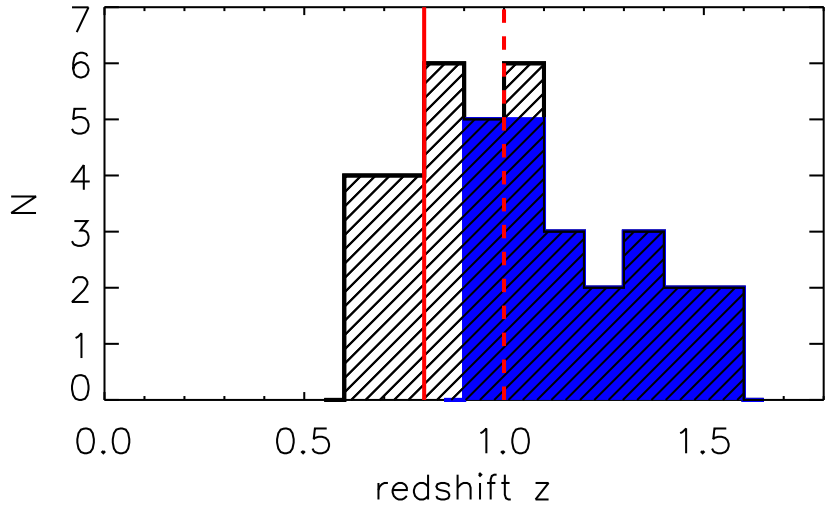}
      \vspace{-1ex}
      \caption{Redshift histogram of  all currently spectroscopically confirmed XDCP galaxy clusters (black hashed) and the sample presented in this work (blue). The solid (dashed) vertical line marks the redshift $z\!=\!0.8$ ($z\!=\!1$). }
         \label{fig_zHistogram}
\end{figure}

The histogram in Fig.\,\ref{fig_zHistogram} (blue shaded region) displays the redshift distribution of the  clusters presented in this work as well as the full current XDCP sample (black hashed) for comparison.  With 17 systems at $z\!\ge\!1$ and 7 clusters at $z\!>\!1.3$, this sample provides an almost homogeneous redshift coverage up to $z\!\sim\!1.6$.



\begin{figure}[t]
   \centering
    \includegraphics[width=11cm, clip=true]{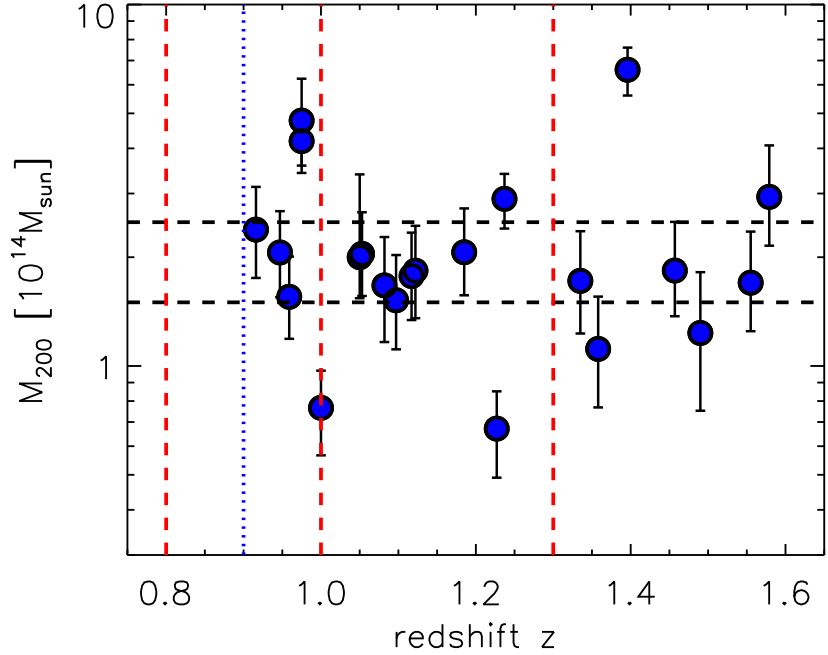}
       \vspace{-1ex}
      \caption{XDCP clusters in the mass versus redshift plane. The mass estimates with the lowest uncertainty for each galaxy cluster were used according to Table\,\ref{tab_masterlist_Opt} (10\,\&\,11). Future studies can investigate the properties of the galaxy cluster population in at least three redshift bins at $z\!>\!0.8$ (vertical dashed lines) and three mass bins (horizontal dashed lines).
      }
         \label{fig_MassRedshift}
\end{figure}


Advances in the empirical calibration of local X-ray scaling relations  \citep[e.g.][]{Pratt2009a} and their redshift evolution (Reichert et al. 2011)
allow now to obtain robust mass estimates based on the X-ray luminosity and the ICM temperature all the way to the highest accessible redshifts. 
Here we make use of the latest empirically calibrated  M-L and M-T relations by \citet{Reichert2011a} with the explicit forms

\begin{eqnarray}
\label{e5_ScalingRelations_L}
M^{\mathrm{L_X}}_{500} = (1.64 \pm 0.07)\cdot  \left( \frac{L^{\mathrm{bol}}_{\mathrm{X,500}}}{10^{44}\,\mathrm{erg\,s}^{-1}} \right)^{0.52 \pm 0.03} \cdot E(z)^{-0.90^{+0.35}_{-0.15}} \times 10^{14} \,\mathrm{M_{\sun}} \ \ \mathrm{and} \\
\label{e5_ScalingRelations_T}
M^{\mathrm{T_X}}_{500} = (0.291 \pm 0.031)\cdot  \left( \frac{T_{\mathrm{X}}}{1\,\mathrm{keV}} \right)^{1.62 \pm 0.08} \cdot E(z)^{-1.04 \pm 0.07} \times 10^{14} \,\mathrm{M_{\sun}} \ ,
\end{eqnarray}

\noindent where E(z)=H(z)/H$_0$ is the cosmic evolution factor of the Hubble expansion.
These relations provide the best current constraints on the redshift evolution factors and their uncertainties, which in the case of the  M-L relations is significantly slower than the self-similar model predictions (e.g.~Kaiser \citeyear{Kaiser1986a}; B\"ohringer et al., in prep.). Since the evolution factors for the relevant redshift regime $0.9\!<\!z\!\la\!1.6$ cover the range E(z)$\simeq$1.7-2.4, the uncertainty in the exponent of the $E(z)$-term dominates the error budget for luminosity-based mass estimates at  high-$z$ together with the effect of intrinsic scatter, which is currently only quantified at low-redshifts  \citep[][]{Pratt2009a}.
The considered error budget hence includes this (local) intrinsic scatter, the redshift evolution uncertainty including sample bias effects, the errors in normalization and slope of the relation, and the measurement uncertainties in L$_X$ (T$_X$). As a last step, M$_{500}$ values are scaled to total mass estimates  M$_{200}\!\simeq\!(1.54\pm0.06)\cdot M_{500}$  by assuming an NFW mass profile with concentration parameters $c\!=\!3.0\pm 0.5$ matched to our redshift and mass range following \citet{Duffy2008a}.


The resulting luminosity-based total cluster mass estimates are listed in column (10) of Table\,\ref{tab_masterlist_Opt} for each system.    
ICM temperature-based mass estimates according to (\ref{e5_ScalingRelations_T}) are given in column (11) with a (T) label, whenever meaningful T$_X$ constraints are available. In several cases (C05, C08, C16, C19) more accurate mass estimates are available and listed in (11), which are 
mostly based on the standard hydrostatic equilibrium (HE) method, weak lensing (WL) measurements, or combinations thereof.  

Figure\,\ref{fig_MassRedshift} shows the XDCP cluster mass estimates with the lowest uncertainty as a function of the system redshift. 
The characteristic median mass of the sample is M$_{200}\!\simeq\!2 \times 10^{14}$\,M$_{\sun}$, with a mass range spanning approximately 0.7-7$\times 10^{14}$\,M$_{\sun}$. 
The distribution shows a fairly homogeneous and unbiased mass sampling with indications of an  increasing lower mass cut with redshift as expected. 
The achieved coverage of the mass-redshift plane will allow future investigations of the distant galaxy cluster population properties in at least three redshift (vertical dashed lines) and mass bins (horizontal dashed lines).
The latter mass bins allow an approximate distinction of 
the three classes of massive distant X-ray  clusters with M$_{200}\!>\!2.5 \times 10^{14}$\,M$_{\sun}$, medium mass objects at  $1.5 \times 10^{14}$\,M$_{\sun}$$\!<\!$
M$_{200}\!\la\!2.5 \times 10^{14}$\,M$_{\sun}$, and low mass systems with M$_{200}\!\la\!1.5 \times 10^{14}$\,M$_{\sun}$.

\subsection{X-ray morphologies}
\label{s5_Xmorphologies}

Figure\,\ref{fig_Gallery} displays an optical/NIR-X-ray gallery of all systems, in additions to the two new clusters presented in Fig.\,\ref{fig_012_077com_Opt} (central panels). All image sizes are 1.5\arcmin$\times$1.5\arcmin, corresponding to physical length scales of $\sim$700-760\,kpc at  $z\!\simeq$0.9-1.6.
The cluster acronym (column (13) in Table\,\ref{tab_masterlist_Xray}), the system redshift, and the filter bands used for the color  
image are listed in the top part of each panel.  Most of the optical and near-infrared images for the color composites originate from our designated XDCP follow-up imaging campaigns (see Sect.\,\ref{s3_FollowUpImaging} and references in Table\,\ref{tab_masterlist_Xray}), complemented by some images from the public CFHT data archive  
\citep{Gwyn2008a}. Logarithmically spaced XMM-{\it Newton} X-ray surface brightness contours are overlaid for each system, with optimized adjusted levels for each source to allow a fair representation of the underlying X-ray morphology. 

This X-ray surface brightness morphology is generally closely linked to the dynamical state of the systems \citep[e.g.][]{HxB2010a, Mohr1993a}.
Although the presented distant clusters do not constitute a representative sample and the signal strength is very limited, a rough qualitative morphological classification can provide some first clues on the typical  high-$z$ cluster X-ray appearance within the limitations of the XMM-{\it Newton} resolution capabilities. As the simplest qualitative classification, we can consider the following four categories: regular morphology (R), mostly regular but with a clear elongation axis   (R-), intermediate states (0), and multi-peaked/irregular (M/I) morphologies. Such a scheme yields roughly 4/22 (18\%) regular systems (C05, C08, C16, C19), 12 (55\%) mostly regular morphologies (C01, C02, C04, C06, C10, C11, C12, C14, C17, C19, C20, C21), 4 (18\%) intermediate state systems (C03, C09, C13, C15), and the 2 (9\%) multi-peaked/irregular X-ray morphologies (C07, C22) discussed in Sects.\,\ref{s4_Sparcs}\,\&\,\ref{s4_077com}.

The majority of the systems (16/22  or 73\%) hence exhibit at least a mostly regular X-ray morphology (R or R-), which can be interpreted as a first indication for advanced evolutionary states. The four most regular (R) systems are located in the top half of the mass range (M$_{200}\!\ga\!2 \times 10^{14}$\,M$_{\sun}$), feature BCGs close to the X-ray centroid, and  show very evolved galaxy populations  \citep[e.g.][]{Strazzullo2010a,Rettura2010a,Santos2009a}. The more elongated X-ray structure of the mostly regular (R-) cluster category, on the other hand, may indicate the major matter accretion axis or minor merging activity   
(e.g.~Fassbender et al. \citeyear{Fassbender2011b}b),
 while the multi-peaked/irregular (M/I) systems suggest ongoing major mergers (Sects.\,\ref{s4_Sparcs}\,\&\,\ref{s4_077com}).

\subsection{BCG offsets and luminosity gaps}
\label{s5_BCGoffsets}

\begin{figure}[t]
   \centering
    \includegraphics[width=11cm, clip=true]{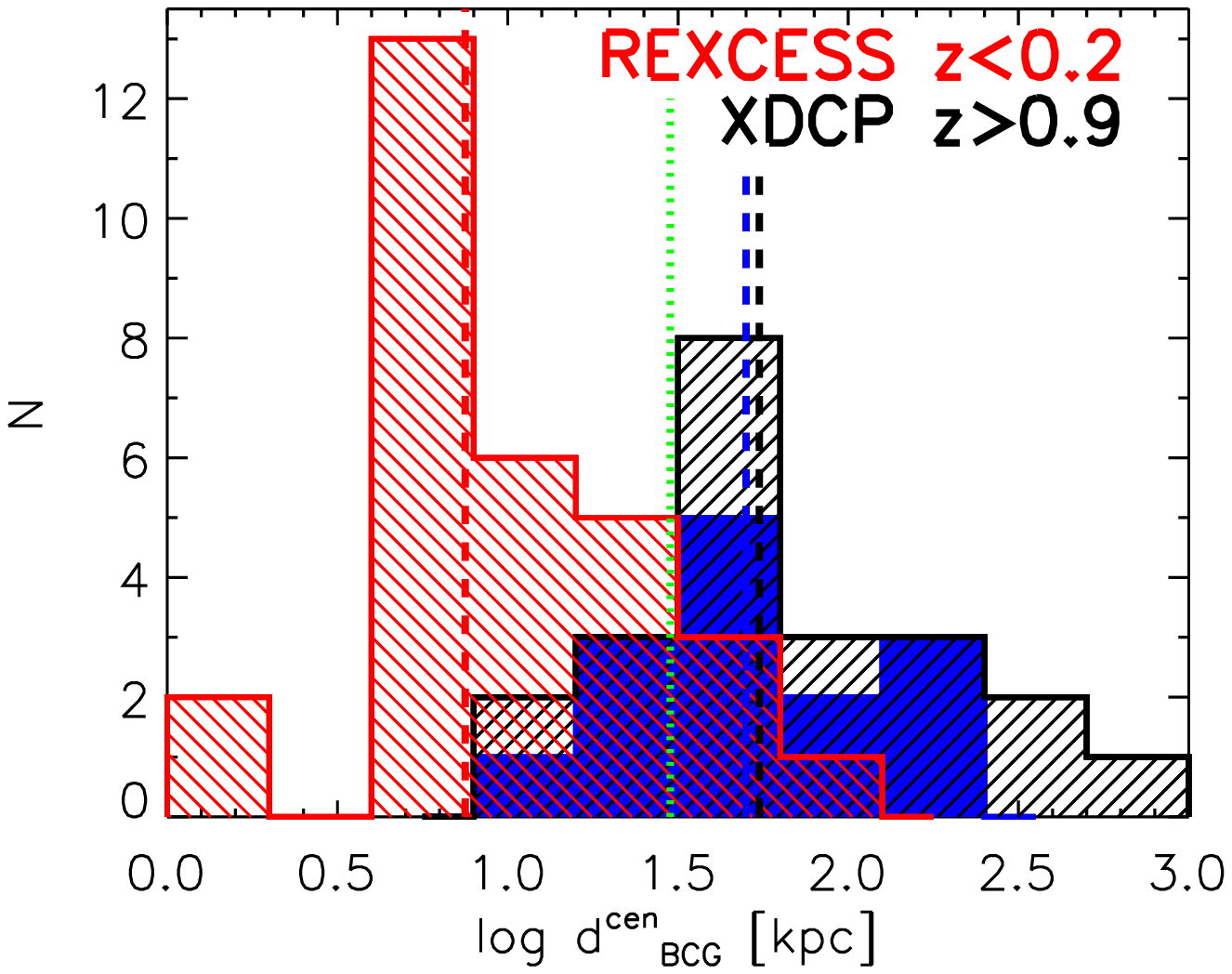}
  \includegraphics[width=11cm, clip=true]{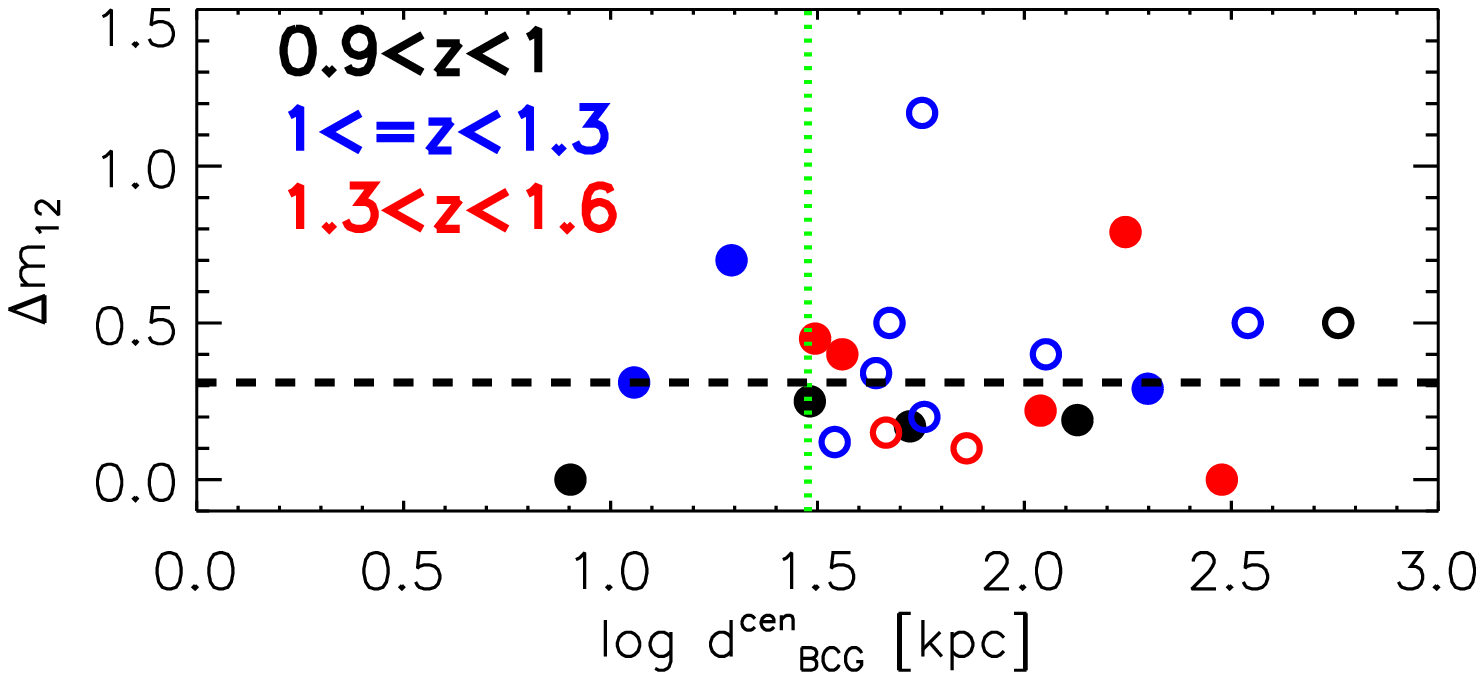}
      \caption{Properties of brightest cluster galaxies in the XDCP sample. {\em Top panel:}
      Comparison of the BCG offsets from the X-ray centroid for the low-z REXCESS sample (red histogram) and the $z\!>\!0.9$ XDCP sample (black histogram), where secure BCG identifications for the  high-$z$ clusters are indicated by the blue background color. The median XDCP cluster centroid offsets  for all BCGs (secure identifications) of 55\,kpc (50\,kpc) are marked by the black (blue) dashed vertical lines, whereas the green dotted line depicts the average measurement uncertainty. 
 The median BCG offset for the REXCESS sample of 7.5\,kpc is indicated by the red dashed line for reference.
 {\em Bottom panel:}  Magnitude difference $\Delta m_{12}$ between the first- and second-ranked cluster galaxies 
 as a function of the BCG centroid offset. 
 Clusters belonging to different redshifts bins are marked by different colors. Filled (open) symbols indicate  secure (tentative) identifications of the two top-ranked galaxies. The median magnitude gap of 0.31\,mag for the full sample is marked by the horizontal black line, the green vertical line marks the centroid measurement uncertainty as above.
  }
         \label{fig_BCG_Offsets}
\end{figure}

Another indicator for the dynamical state of a system at low redshifts is the location of the BCG with respect to the X-ray centroid position 
\citep[e.g.][]{Haarsma2010a,Smith2010a,Sanderson2009a}. Studies of the representative REXCESS reference sample by \citet{Haarsma2010a} show that $\sim$80\% of the local ($z\!<\!0.2$) clusters host a central dominant brightest cluster galaxy within 20\,kpc of the X-ray peak, 
with a median offset for the full population of 7.5\,kpc (red histogram and red dashed line in the top panel of Fig.\,\ref{fig_BCG_Offsets}).

The situation is clearly different at $z\!>\!0.9$, where a central dominant BCG coincident with the X-ray centroid is more an exception than the rule, as is evident from 
Fig.\,\ref{fig_Gallery}. The X-ray centroid position, as the first moment of the surface brightness distribution of the extended cluster emission, is generally robustly 
 determined with XMM-{\it Newton}, 
 even in the low-count regime at  high-$z$, with an average statistical positional uncertainty of 3\arcsec. 
 The (Gaussian) combination of this statistical error with the average systematic absolute astrometric offset of  1\arcsec \  \citep[e.g.][]{Watson2009a} leads to an average total X-ray centroid uncertainty of 25-28\,kpc in the targeted redshift regime.   
For this work, we conservatively assume a total positional error radius of the X-ray centroid determination of 30\,kpc (green dotted line in Fig.\,\ref{fig_BCG_Offsets}), which would result in an observed median BCG offset for the REXCESS sample of 26.8$\pm$4.4\,kpc based on 1000 Monte Carlo realizations with random offset directions.

 The unambiguous identification of the BCG 
 can be a challenging task at  high-$z$ for a significant fraction of non-trivial cases. 
Owing to the standard paradigm of hierarchical built-up of  BCGs  \citep[e.g.][]{DeLucia2007a}, the  high-$z$ progenitors of present day centrally dominant galaxies may not necessarily be the brightest galaxies at any redshift and may have migrated long distances within the larger scale cluster environment. 
Related to this, three main issues for the observational identification process arise in practice: (i) clusters may host several top ranked galaxies with similar absolute magnitudes or mass \citep[e.g. X1229, ][]{Santos2009a}; (ii) the brightest galaxy of the cluster environment may still be outside the formal R$_{200}$ radius (e.g.~X2217,  Fassbender et al., in prep.); and (iii) off-center BCG candidates may still lack the spectroscopic membership confirmation or not all brighter galaxies at lower cluster-centric distance have been spectroscopically excluded as members  (e.g.~X0338c,  Sect.\,\ref{s4_077com}).
Such ambiguous cases  (8/22) are flagged as tentative (t) BCG identifications in column (8) of Table\,\ref{tab_masterlist_Opt}, where the projected  cluster-centric BCG distances are listed.

\pagebreak

The black hashed histogram in  Fig.\,\ref{fig_BCG_Offsets} (top panel) shows the observed distribution of BCG offsets for the full $z\!>\!0.9$ sample as a function of cluster-centric distance, whereas the blue shaded regions indicate the 14 secure BCG identifications. This distribution does not peak 
at small cluster-centric distances 
as the local reference sample (red hashed), but rather exhibits a median offset of 55\,kpc (50\,kpc for the secure BCGs) from the X-ray centroid, with a wing extending towards large cluster-centric distances. 
Considering the discussed measurement uncertainty, the offsets of only  7/22 systems of the sample with observed d$^{\mathrm{center}}_{\mathrm{BCG}}$$<$40\,kpc are 
statistically consistent with harboring a central BCG at offsets of $\la$20\,kpc  
\citep[see e.g.~the case of X2235, ][]{Rosati2009a}. The determined median BCG offset of d$^{\mathrm{center}}_{\mathrm{BCG}}$$\sim$50\,kpc, on the other hand, is robust 
and basically unaffected 
by centroid uncertainties since it is governed by the largest half of the distribution of cluster-centric distances.

At lookback times of 7.3-9.5\,Gyrs  for the present sample, the observed BCG population has hence generally not yet reached the bottom of the cluster potential well  (see e.g.~the case of X1230, Fassbender et al.~\citeyear{Fassbender2011b}b), but is rather still caught in the process of inward migration via dynamical friction. A first hint for a further redshift evolution of the BCG offsets can be obtained by considering the redshift bins of Fig.\,\ref{fig_MassRedshift}, which yields median projected cluster-centric BCG distances of $\sim$52\,kpc for the first two bins at  $z\!\le\!1.3$ and $\sim$73\,kpc for the seven highest-z systems at  $z\!>\!1.3$. 

As a second straightforward test concerning the position and role of BCGs in high-$z$ clusters, we can consider the luminosity gap $\Delta m_{12}$ between the first- and second-ranked galaxies. This statistic was studied by \citet{Smith2010a} for a sample of massive (M$_{200}$$\sim$$10^{15}$\,M$_{\sun}$), low-redshift ($0.15\!\le\!z\!\le\!0.3$) clusters and was found to correlate tightly with the dynamical state of the systems, e.g.~large $\Delta m_{12}$ generally imply small amounts of substructure and cuspy gas density profiles. The median luminosity gap for this local reference sample is measured to be $\Delta m_{12,{\mathrm{med}}}\!\simeq\!0.67$\,mag and the fraction of clusters with very dominant BCGs with $\Delta m_{12}\!>\!1$\,mag is about 37\%.

The bottom panel of Fig.\,\ref{fig_BCG_Offsets} shows the luminosity gaps $\Delta m_{12}$ of the XDCP $z\!>\!0.9$ cluster sample as a function of the cluster-centric BCG offsets. The $\Delta m_{12}$ measurements were obtained in the reddest (K$_s$, H, or z) optical/NIR band available (see column 21 in Table\,\ref{tab_masterlist_Xray} for references). The color coding groups the systems into the different redshift bins, whereas open circles indicate tentative identifications of the first- and/or second ranked galaxies as discussed above. Clear trends of $\Delta m_{12}$ with either the BCG offset or as a function of redshift are not obvious in the bottom panel of Fig.\,\ref{fig_BCG_Offsets}. However, the statistics of the $\Delta m_{12}$ distribution reveals again marked evolutionary differences compared to the low-$z$ reference sample. The measured  median luminosity offset of the high-$z$ clusters is found to be   $\Delta m_{12,{\mathrm{med}}}\!\simeq\!0.31$\,mag ($\Delta m_{12,{\mathrm{med}}}\!\simeq\!0.28$\,mag for secure identifications) and the sample  only contains one candidate system (e.g.~$\la$5\%) with a very dominant $\Delta m_{12}\!>\!1$\,mag BCG, XDCP\,J2205.8-0159 at $z\!=\!1.117$ (C12 in Table\,\ref{tab_masterlist_Opt}, Fassbender et al., in prep.). The population of $z\!>\!0.9$ BCGs is hence significantly less dominant compared to the ones observed in their more  massive successor systems at $z\!<\!0.3$.  In particular, we note that our current sample based on the discussed follow-up strategy (Sects.\,\ref{s3_FollowUpImaging}\,\&\,\ref{s3_Spectroscopy})
does not include any candidates that would qualify them as fossil groups with 
$\Delta m_{12}\!\ge\!2$\,mag \citep[e.g.][]{Jones2003a}.
We conclude that   BCGs at $0.9\!<\!z\!\la\!1.6$ are generally still  observed at an earlier phase of their evolutionary track 
on the way to 
 their typical central cluster position in  low-$z$ systems and their dominance with respect to second-ranked galaxies.



\subsection{Radio properties}
\label{s5_RadioProp}

The statistics of radio sources associated with  high-$z$ galaxy clusters is of prime importance for ongoing 
Sunyaev-Zeldovich effect (SZE) surveys \citep[e.g.][]{Williamson2011a,Marriage2010a,Planck2011a}. Radio emitting sources at the cluster locations pose the main source of potential contamination for SZE selected cluster samples, since these sources can (partially) fill in the SZE decrement signal and hence lead to an underestimation of cluster counts and mass estimates.  While detailed radio source studies in clusters in the local Universe \citep[e.g.][]{Lin2007a,Best2007a,Mittal2009a} and at moderate redshifts \citep{Sommer2011a} are now available, robust statistics for the $z\!>\!0.9$ cluster population have not been accessible so far or are limited to the galaxy group regime \citep{Smolcic2011a}. 

We queried the NASA Extragalactic Database 
for 1.4\,GHz radio sources within 2\arcmin \ ($\sim$1\,Mpc) from the X-ray centroids. The 1.4\,GHz radio flux densities of the closest sources with the range of 0.1-18\,mJy and their cluster-centric distances are listed in column (9) of Table\,\ref{tab_masterlist_Opt}. For 13 clusters (59\%) at least one 1.4\,GHz radio source was found, most of which (10) from the NVSS survey \citep{Condon1998a} and three sources (C07, C16, C19) at lower flux densities observed with  ATCA  \citep{Middelberg2008a},  VLA-VIRMOS \citep{Bondi2003a}, and the
 VLA FIRST survey \citep{White1997a,Becker2003a}.

The 1.4\,GHz NRAO VLA Sky Survey (NVSS) covers the full XDCP distant cluster sample (except C07) at a completeness limit of $\sim$2.5\,mJy (45\arcsec \  FWHM resolution)  and hence allows a first evaluation of the frequency of cluster-associated radio sources at bright flux densities ($\ga$2\,mJy). The average surface density of NVSS sources amounts to 53.4 radio sources per square degree or 1 source per 67.4 square arcminutes, corresponding to an expectation rate for random radio sources within an area of radius 0.5\arcmin/1\arcmin/2\arcmin \ of 1.2\%/4.7\%/18.6\%. 

The observed number of   3/8/10  NVSS radio sources at radii within 0.5\arcmin/1\arcmin/2\arcmin \ from the centers of the 22 distant XDCP clusters is to be compared to the background expectation of 0.3/1.0/4.1 random sources within these apertures. This yields a background-corrected expectation value of 6-7 cluster-associated NVSS radio sources equivalent to a `radio active' cluster fraction of about 30\% and a preferred location within 1\arcmin \ ($\sim$500\,kpc) from the X-ray center. The radio flux density of these sources spans a range of 2.2-18\,mJy with a median value of 3.5\,mJy. A trend of the cluster-associated radio source fraction with redshift is not apparent over the three probed redshifts bins with the current statistics. In terms of cluster mass bins, there is a hint that intermediate mass systems (M$_{200}\!\sim\!2 \times 10^{14}$\,M$_{\sun}$) may be preferred environments for cluster-associated radio sources with an observed  fraction of approximately 50\%.

With the assumption of a typical spectral index of $\alpha\!\simeq\!-0.8$ \citep[e.g.][]{Miley2008a}, the NVSS flux limit translates into 
increasing minimal  radio powers of  $P_{1.4\,\mathrm{GHz}}\!\ga\!(0.8$-$3)\!\times\!10^{25}$ W\,Hz$^{-1}$ for the detection of cluster-associated radio sources in the probed redshift range $0.9\!<\!z\!\la\!1.6$. A comparison to results obtained at lower redshifts is hence only possible for the most luminous radio source bin with  $P_{1.4\,\mathrm{GHz}}\!\ga\!10^{25}$ W\,Hz$^{-1}$, for which a low-$z$ fraction of central radio sources of $\sim$6\% were determined by \citet{Lin2007a} and \citet{Best2007a}, while the HIFLUGCS sample of \citet{Mittal2009a} contains a fraction of $\sim$12\%. The derived value of about 30\% for the high-$z$ sample thus suggests an increase of the fraction of very luminous cluster-associated radio sources by about a factor of 2.5-5. 

An upper limit of  $P_{1.4\,\mathrm{GHz}}\!\la\!1.2\!\times\!10^{26}$ W\,Hz$^{-1}$ for the potentially most luminous radio sources in the sample can be derived from the observed flux densities at the locations of clusters C08 and C11 (see Table\,\ref{tab_masterlist_Opt}). These maximal flux densities are expected to drop by a factor of  $\sim$40 to $\la$0.5\,mJy  when extrapolated to 150\,GHz using the assumed spectral index. The observed radio sources in our sample would thus only have  a small impact on the detection efficiency of massive clusters with SZE surveys for the assumed extrapolation\footnote[1]{This extrapolation by more than a factor of 100 in frequency using $\alpha\!\simeq\!-0.8$ may not be valid for the full radio source population. Individual AGN with shallower (or even rising) spectral slopes may contribute significantly more than the estimated upper limit.}, with a  maximum radio source flux contribution at   150\,GHz of $\la$10\%   at the typical cluster detection limit of e.g. the South Pole Telescope \citep{Carlstrom2011a}.



\subsection{The $z\!\ge\!1.5$ galaxy cluster frontier}
\label{s5_HizFrontier}

As the final point to be addressed in this section, we have a closer look at the current galaxy cluster redshift frontier at $z\!\ga\!1.5$ and the state of the galaxy populations in theses systems. 
A detailed study of the very massive cluster XDCP\,J2235.3-2557 (C05) by \citet{Strazzullo2010a} revealed a very evolved central galaxy population with very little star formation activity and a fully formed, tight red-sequence. 
The intermediate mass system XMMXCS\,J2215.9-1738 (C04), on the other hand, features very active star formation activity down to central cluster regions  
\citep{Hilton2010a,Hayashi2010a}, which was also reported for a system at $z\!=\!1.62$ in the group regime  (M$_{200}\!<\!10^{14}$\,M$_{\sun}$) by  \citet{Tran2010a}.

\begin{figure}[t]
\centering
\includegraphics[angle=0,clip,width=0.88\textwidth]{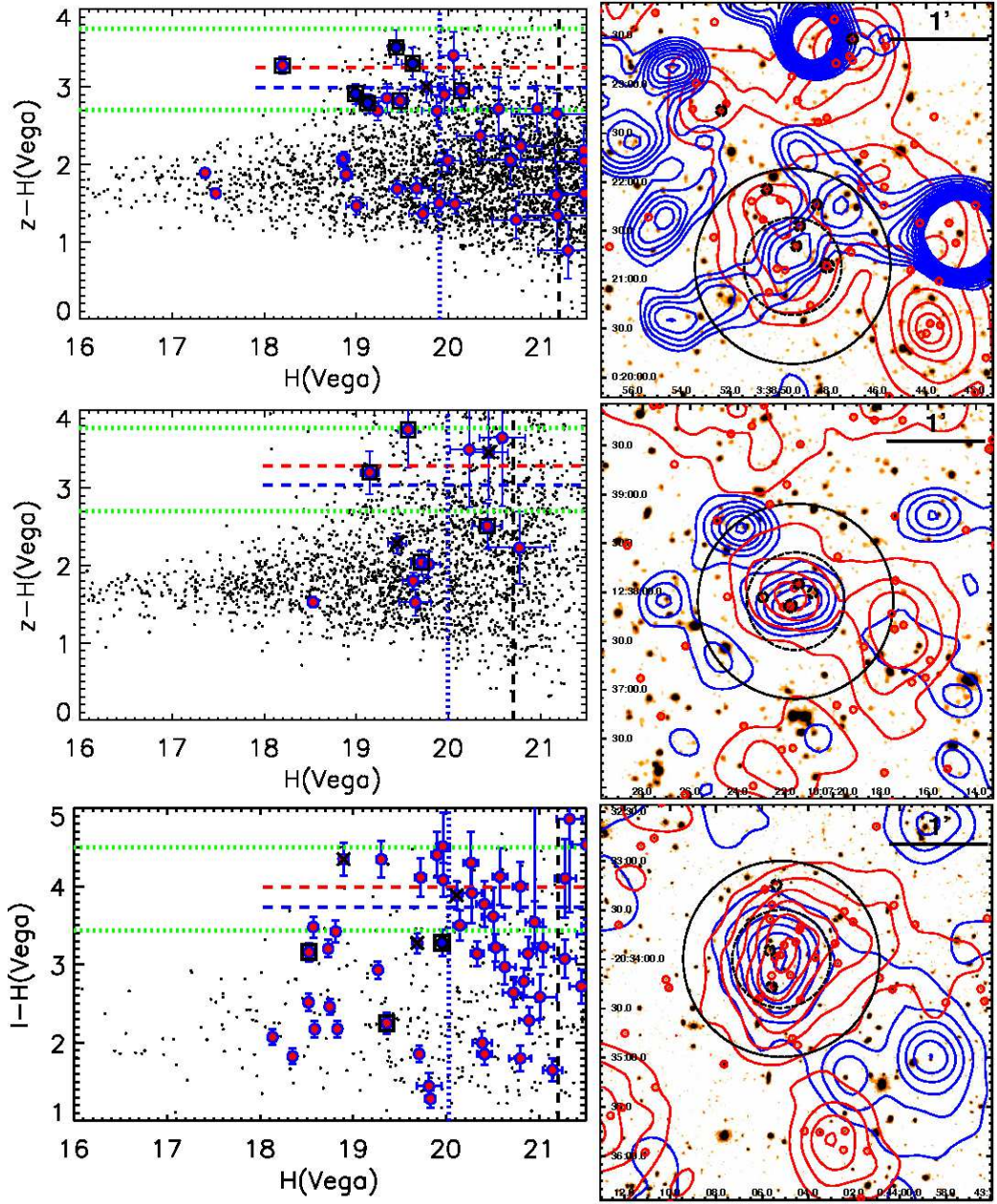}           
\vspace{-1ex}
\caption{Comparison of the presently three most distant clusters in the XDCP sample:  XDCP\,J0338.8+0021 at $z\!=\!1.490$ (top), XDCP\,J1007.3+1237  at $z\!=\!1.555$ (center), and XDCP\,J0044.0-2033  at $z\!=\!1.579$ (bottom). The left column shows the color-magnitude diagrams with red circles indicating galaxies within 40\arcsec \ from the X-ray center, blue symbols representing spectroscopic members at $r\!>\!40\arcsec$, and black dots all other objects in the FoV. Spectroscopic members are marked by open squares, the 50\% completeness limits by the vertical dashed black lines, and  the apparent characteristic H-band magnitudes H* at the 
cluster redshifts by the vertical dotted blue line. Horizontal blue ($z_f\!=\!3$) and red ($z_f\!=\!5$) dashed lines indicate SSP solar metallicity  model predictions for different stellar formation redshifts   $z_f$, and the dotted lines confine the applied color cuts for the red galaxy densities. 
The corresponding 4\arcmin$\times$4\arcmin \ ($\sim$2$\times$2\,Mpc) \ H-band  cluster environments are shown on the right-hand side with XMM-{\it Newton} X-ray contours overlaid in blue.  Large black circles indicate the 60\arcsec and 30\arcsec \ radii around the X-ray center, small black circles mark spectroscopic members, and red circles represent  color selected red objects with corresponding logarithmically spaced red density contours with levels of 
3.3, 5.2, 8.0, 13, 20, 30  
galaxies per arcmin$^2$ covering the significance range of 2-29\,$\sigma$ above the background.  
}
\label{fig_HizFrontierClusters}       
\end{figure}

The three top ranked clusters from Table\,\ref{tab_masterlist_Opt} at redshifts of 1.490, 1.555, and 1.579 are presently the most distant, spectroscopically confirmed, X-ray luminous systems known in the cluster regime at M$_{200}\!\ga\!10^{14}$\,M$_{\sun}$. The systems
XDCP\,J0044.0-2033 (C01, Santos et al., \citeyear{Santos2011a}),  XDCP\,J1007.3+1237 (C02, Fassbender et al. \citeyear{Fassbender2011a}a), and
XDCP\,J0338.8+0021 (C03, Nastasi et al., \citeyear{Nastasi2011a}) are hence good test cases to probe cluster environments at lookback times beyond 9.2\,Gyr.
Figure\,\ref{fig_HizFrontierClusters} (left panels) shows the color-magnitude diagrams of the three systems based on the initial two-band imaging data. Galaxies within 40\arcsec \ from the X-ray centroid position are indicated in red and spectroscopic members are marked by square boxes.
Simple stellar population model predictions for stellar formation redshifts of 5 (3) are displayed by red (blue) dashed lines and green dotted lines confine the applied color cuts for each system spanning the color range between 0.3\,mag bluer than the $z_f\!=\!3$ SSP model to 0.5\,mag redder than the $z_f\!=\!5$ value. These color selected galaxies are shown in the H-band  images in the right panels (red circles), which display the  4\arcmin$\times$4\arcmin \ ($\sim$2$\times$2\,Mpc) \  cluster environments with the red galaxy iso-density contours and the X-ray surface brightness contours in blue. 
The background galaxy density for the applied color selection of very red galaxies is low with a value of about 1.4$\pm$1.0 arcmin$^{-2}$.

\clearpage

The presently highest redshift 
XDCP cluster XDCP\,J0044.0-2033 at $z\!=\!1.579$ (bottom panels) is a remarkable system for this cosmic 
epoch with its high X-ray luminosity and corresponding high mass estimate (Tables\,\ref{tab_masterlist_Opt}\,\&\,\ref{tab_masterlist_Xray}). This massive structure is also reflected in the rich galaxy population (see Fig.\,\ref{fig_Gallery}, top left panel), which marks a $\ga$29\,$\sigma$ galaxy overdensity centered on the X-ray emission. Both the galaxy distribution and X-ray emission are elongated along the NS direction, which may reflect the main cluster assembly axis.  The CMD is well populated in the applied color cut region with the main noteworthy feature that the brightest central galaxies, including the spectroscopically confirmed BCG candidate, are all significantly bluer than the expected \reds color.  

The second ranked system XDCP\,J1007.3+1237  at $z\!=\!1.555$ (central panels) is an intermediate mass cluster with a central, red, radio-loud BCG. The 
currently available imaging depth is 0.5\,mag shallower compared to the other two fields, implying that only the bright end of the underlying galaxy population is presently accessible. Two spectroscopic members close to the characteristic magnitude H* and within a projected distance of 200\,kpc  from the X-ray centroid are blue and feature prominent 
\OII \ emission lines, which provide evidence for strong starburst activity in these massive galaxies. 

The seven spectroscopic members of the lower mass system XDCP\,J0338.8+0021 at $z\!=\!1.490$ (top panels), on the other hand, show very little signs of star formation out to beyond the nominal R$_{200}$ radius. The BCG is a red, merging, off-center galaxy at the expected SSP color, while other galaxies along the apparently well populated \reds seem to show an increased spread in color, compared to lower-z clusters. Towards fainter magnitude (H$\sim$21\,mag) several central galaxies are just below the applied color cut, with the effect that the otherwise central galaxy density peak \citep{Nastasi2011a} is now shifted Northward of the X-ray centroid position. This system features the widest spatial distribution of red galaxy overdensities, spread over almost the full $\sim$2$\times$2\,Mpc region displayed in the right panel, which may be an indication that we are observing a young cluster environment. 

Although no clear simple picture for the general state of galaxy populations in $z\!\ga\!1.5$ cluster environments  is evident yet, it is apparent that dramatic changes do occur once lookback times of $\ga$9.2\,Gyr  are probed. As the 
 observed star formation activity proceeds towards the highest galaxy masses and the densest core environments at these epochs, the cluster \reds seems to gradually lose its universal, well defined form characteristic for clusters up to about 9\,Gyr in lookback time.

\subsection{Outlook and prospects}
\label{s5_Prospect}




 With the recent observational advances to push the high redshift cluster frontier to $z\!\ga\!1.5$, we are now closing in on the formation epoch of these most massive collapsed structures in the Universe. Key questions on the formation and evolution of the hot intracluster medium and the galaxy populations in the densest environment can be addressed observationally with upcoming deep multi-wavelength follow-up data to allow a more detailed physical characterization of the different cluster components and their mutual interactions.   


Besides the aspect of reaching out to redshifts of $z$$\sim$$1.6$, other key features of the presented XDCP distant cluster sample is the almost homogeneous coverage of the targeted redshift baseline and the wide cluster mass interval probed, which spans the rich group to the massive cluster regimes. 
Future distant galaxy cluster population studies can thus connect the well studied redshift regime at   $z\!<\!0.8$ to the $z\!\ga\!1.5$ frontier in order to continuously trace cluster evolution as a function of redshift {\em and} total system mass. 
The presented sample with 22 distant test objects is clearly a key step forward to achieve these goals. However, the spectroscopic follow-up of all  high-$z$ XDCP candidate clusters  is still ongoing with good prospects to double the number of the present sample over the next few years.

\section{Summary and conclusions}
\label{c7_Summary}

We presented a description of the survey strategy of the XMM-{\it Newton} Distant Cluster Project to detect, identify, and study X-ray luminous galaxy clusters at $z\!>\!0.8$.
All clusters are X-ray selected as extended sources in deep archival XMM-{\it Newton} data and are hence unbiased with respect to their galaxy populations. We provided an overview of the  X-ray data processing of the 469 survey fields and discussed the detection capabilities of XMM-{\it Newton} concerning faint extended X-ray sources down to soft-band flux levels of $<$$10^{-14}$erg\,s$^{-1}$cm$^{-2}$.

We discussed different imaging techniques for the efficient follow-up and photometric identification of distant cluster candidates. In particular we compared the efficacy of two-band imaging strategies based on the R$-$z and z$-$H colors using 20 spectroscopically confirmed reference clusters in each case. We applied a robust prescription to blindly measure the characteristic color of red cluster galaxies and its uncertainty for candidate systems of unknown redshifts to be compared with simple stellar population galaxy evolution model predictions. We confirmed the general expectations on the redshift accuracy performance of 
the R$-$z color, which yields accurate  estimates at $z\!<\!0.9$ and allows the photometric identification of distant clusters at $0.9\!<\!z\!\la\!1.4$ albeit with significantly increasing redshift uncertainties. In this high-$z$ range the z$-$H color provides more reliable redshift estimates owing to its steep redshift dependence which also allows robust cluster identifications out to  $z\!\ga\!1.5$. The empirically calibrated redshift evolution models for the  R$-$z and z$-$H colors are provided in table format as part of the online material.  

We outlined the spectroscopic cluster confirmation process with VLT/FORS\,2 and our applied observational galaxy cluster definition based on (i) the detected extended X-ray emission, (ii) a coincident red galaxy population, and (iii) a minimum of three associated concordant spectroscopic member redshifts. 

We discussed the X-ray properties of the previously identified rich cluster SpARCS\,J003550-431224 at $z$=1.335 with a bolometric luminosity  of  L$^{\mathrm{bol}}_{\mathrm{X,500}}\!\simeq\!(1.8\pm 0.5)\times 10^{44}$\,erg\,s$^{-1}$, an ICM temperature of T$_{\mathrm{X}}\!\simeq\!4.5_{-2}^{+4}$\,keV, and a derived consistent mass estimate from both measurements of about M$_{200}\!\simeq\!2 \times 10^{14}$\,M$_{\sun}$ ($\pm$40\%), which is significantly lower than the previously reported velocity dispersion based mass. 
The cluster features a very extended (r$_{\mathrm{c}}\!\simeq\!260$\,kpc),  multi-peaked X-ray morphology, which in conjunction with the bimodal redshift distribution provides evidence for a major merger configuration close to the line-of-sight.

We presented X-ray and optical properties of the two newly identified systems XDCP\,J0027.2+1714 at $z$=0.959 and XDCP\,J0338.5+0029 at $z$=0.916. 
For XDCP\,J0027.2+1714 we measured  L$^{\mathrm{bol}}_{\mathrm{X,500}}\!\simeq\!(1.0\pm0.1)\times 10^{44}$\,erg\,s$^{-1}$ with a corresponding mass estimate of M$^{\mathrm{L_X}}_{200}\simeq 1.6_{-0.4}^{+0.5}\times 10^{14}$\,M$_{\sun}$. The X-ray morphology is elongated, but mostly regular with a coincident rich red galaxy population and a central BCG with a significant rest-frame velocity offset of $-$1600\,km/s.
The system XDCP\,J0338.5+0029 shows evidence for major merging activity along the plane of the sky based on the observed complex X-ray and red galaxy density morphologies with separated centers and a very narrow redshift interval of the spectroscopic members. 
The derived X-ray luminosity of  L$^{\mathrm{bol}}_{\mathrm{X,500}}\!\simeq\!(2.6\pm0.7)\times 10^{44}$\,erg\,s$^{-1}$  and the mass estimate of M$^{\mathrm{L_X}}_{200}\simeq 2.6_{-0.7}^{+0.8}\times 10^{14}$\,M$_{\sun}$ for the system are to be considered as 
upper limits due to potential unresolved point source contributions to the flux measurements.  

These new systems together with the previously published ones constitute the largest sample of X-ray selected distant galaxy clusters to date. In total, we presented X-ray and optical properties for 22 X-ray luminous systems at $z\!>\!0.9$, with an almost homogeneous redshift coverage all the way to  $z\!\sim\!1.6$. The sample has a median total cluster mass of 
 M$_{200}\!\simeq\!2 \times 10^{14}$\,M$_{\sun}$  and spans  a mass range of approximately 0.7-7$\times 10^{14}$\,M$_{\sun}$.
A first qualitative (non re-presentative) assessment of X-ray morphologies of the sample showed that the majority of the systems ($>$70\%) exhibit at least a mostly regular morphology, albeit predominantly ($\sim$55\%) with clear indications for an elongation along one axis.

We investigated the distribution of cluster-centric offsets of the brightest cluster galaxies from the X-ray centroid locations. In contrast to local clusters of which $\sim80$\% harbor a dominant BCG within 20\,kpc from the X-ray center, the brightest galaxies of the majority of the  $z\!>\!0.9$ clusters show significant offsets from their X-ray centers and are less dominant with respect to the second-ranked galaxy. We find a median cluster-centric BCG offset for the sample of $\sim$50\,kpc, with a significant tail towards large projected off-center distances (i.e.~$>$100\,kpc) for about one third of the systems. 
The median observed luminosity gap between the first- and second-ranked galaxy for the high-$z$ cluster sample is $\Delta m_{12,\mathrm{med}}\!\simeq\!0.3$\,mag and the fraction of systems with very dominant BCGs ($\Delta m_{12}\!>\!1$) is $\la$5\%, compared to $\Delta m_{12,\mathrm{med}}\!\simeq\!0.67$\,mag  and a fraction of 37\% of BCGs with $\Delta m_{12}\!>\!1$ in the $z\!<\!0.3$ reference sample.
These findings provide evidence that the BCGs in distant clusters observed at lookback times of 7.3-9.5\,Gyr have generally not yet fully migrated towards the centers of the systems' gravitational potential wells and have yet to establish 
their local luminosity dominance with respect to the non-BCG galaxy populations in clusters.

For 13/22 cluster locations (59\%)  we found the presence of a 1.4\,GHz radio source within 2\arcmin \ from the X-ray centers, of which 10/22 (45\%) are NVSS sources with flux density levels of $>$2\,mJy. Statistically accounting for random superpositions of radio sources with cluster positions results in the estimate that $\sim$30\% of the systems host a cluster-associated NVSS 1.4\,GHz  radio source with  flux densities in the range of 2.2-18\,mJy, predominantly at locations within 1\arcmin \ (i.e.~$\la$500\,kpc) from the center. With the current statistics, no change of the radio-loud cluster fraction with redshift over the probed interval is evident, while the data suggest an increase of the fraction of very luminous cluster-associated radio sources by about a factor of 2.5-5 relative to low-$z$ systems. 

As a final point, we focussed on the galaxy populations of the most distant $z\!\ga\!1.5$ systems, which currently constitutes the redshift frontier for bona fide  
$\ga$$10^{14}\,\mathrm{M_{\sun}}$ clusters. Although red galaxy populations close the predicted SSP model colors are already present in these systems, drastic changes at the massive end of the galaxy populations are evident compared to the evolved, tight red-sequences observed in massive clusters at 
$z\!\la\!1.4$. These observed changes in the three most distant XDCP systems include (i) significantly bluer colors than the \reds for the brightest galaxies (C01), (ii) starburst activity for central massive galaxies (C02),  and (iii) an apparent observed increase in the \reds scatter (C03). Even though no clear picture on the evolution of the galaxy populations in these densest cluster environments is established yet at lookback times of $\ga$9.2\,Gyr, the available observations provide evidence that the well-defined characteristic cluster red-sequences lose their universal form and start to dissolve once redshifts of  $z\!\ga\!1.5$ are probed.

The presented sample of 22 $z\!>\!0.9$ X-ray luminous galaxy clusters is a first step forward to allow redshift and mass dependent galaxy cluster population studies
that continuously connect the formation epoch of massive systems at $z\!>\!1.5$ to the well studied regime in the second half of cosmic time at $z\!<\!0.8$ and to trace the evolution of the different cluster components in the hot and cold phases.

\ack
We thank the anonymous referee and Christophe Adami for insightful comments that helped to improve the clarity of the paper.
This research was supported by the DFG cluster of excellence `Origin and Structure of the Universe' (www.universe-cluster.de), 
 by the DFG under grants Schw536/24-1,
Schw 536/24-2, BO 702/16-3, and the German DLR under grant 50 QR 0802. 
RF acknowledges the hospitality of the Department of Astronomy and Astrophysics at Pontificia Universidad  Cat\'olica de Chile. 
HQ thanks the FONDAP Centro de Astrofisica for
partial support. 
We acknowledge the excellent support provided by Calar Alto and VLT staff in carrying out the service observations. 
The XMM-{\it Newton} project is an ESA Science Mission with instruments and contributions directly funded by ESA Member
States and the USA (NASA). 
This research has made use of the NASA/IPAC Extragalactic
Database (NED) which is operated by the Jet Propulsion Laboratory, California Institute of Technology, under contract
with the National Aeronautics and Space Administration.


\bibliographystyle{jphysicsB} 
\bibliography{../BIB/RF_BIB_11}

\end{document}